\documentclass[10pt,aps,physrev,reprint,floatfix,superscriptaddress,longbibliography]{revtex4-1}
\pdfoutput=1
\usepackage{graphicx,latexsym}
\usepackage{dcolumn,amsfonts}
\usepackage{amssymb,amsmath,bm}

\usepackage[breaklinks=true,pdfencoding=auto]{hyperref}

\usepackage{natbib}
\hypersetup{
  colorlinks   = true, 
  urlcolor     = blue, 
  linkcolor    = blue, 
  citecolor    = red 
}

\usepackage{color}
\usepackage{ulem}

\newcommand{\VG}{\textcolor{blue}}

\begin{document}

\title{Self-induction and magnetic effects in
       electron transport through a photon cavity}

\author{Vidar Gudmundsson}
\email{vidar@hi.is}
\affiliation{Science Institute, University of Iceland, Dunhaga 3, IS-107 Reykjavik, Iceland}
\author{Nzar Rauf Abdullah}
\affiliation{Physics Department, College of Science, 
             University of Sulaimani, Kurdistan Region, Iraq}
\affiliation{Computer Engineering Department, College of Engineering, Komar University 
             of Science and Technology, Sulaimani 46001, Kurdistan Region, Iraq}
\author{Chi-Shung Tang}
\email{cstang@nuu.edu.tw}
\affiliation{Department of Mechanical Engineering, National United University, Miaoli 36003, Taiwan}
\author{Andrei Manolescu}
\email{manoles@ru.is}
\affiliation{School of Science and Engineering, Reykjavik University, Menntavegur 
             1, IS-101 Reykjavik, Iceland}
\author{Valeriu Moldoveanu}
\email{valim@infim.ro}
\affiliation{National Institute of Materials Physics, PO Box MG-7, Bucharest-Magurele, Romania}

%

\begin{abstract}
	We explore higher order dynamical effects in the transport through
a two-dimensional nanoscale electron system embedded in a three-dimensional
far-infrared photon cavity. The nanoscale system is considered to be a
short quantum wire with a single circular quantum dot defined in a GaAs heterostructure.
The whole system, the external leads and the central system are placed in a
constant perpendicular magnetic field. The Coulomb interaction of the electrons,
the para- and diamagnetic electron-photon interactions are all treated by
a numerically exact diagonalization using step-wise truncations of the appropriate
many-body Fock spaces. We focus on the difference in
transport properties between a description within an electric dipole approximation
and a description including all higher order terms in a single photon mode model.
We find small effects mostly caused by an electrical quadrupole and a magnetic dipole
terms that depend strongly on the polarization of the cavity field with respect to
the transport direction and the photon energy. When the polarization is aligned along 
the transport direction we find indications of a weak self-induction that we 
analyze and compare to the classical counterpart, and the self-energy contribution of
high-order interaction terms to the states the electrons cascade through on their
way through the system. Like expected the electron-photon interaction is well 
described in the dipole approximation when it is augmented by the lowest order 
diamagnetic part for a nanoscale system in a cavity in an external magnetic field.
\end{abstract}

\maketitle
%
%

\section{Introduction}
The non-perturbative coupling of two-dimensional (2D)
electrons in a magnetic field in a heterostructure with high-quality-factor terahertz 
or far-infrared photons has been achieved \cite{Zhang1005:2016}. The coupling of electronic
systems in circuit QED planar microwave cavities to external leads gives the hope that
this will also be accomplished for the terahertz systems 
\cite{PhysRevX.6.021014,0953-8984-29-43-433002,Delbecq11:01,2017arXiv170401961L,PhysRevX.7.011030}.  
Several groups have modeled various aspects of the transport of electrons through
nanoscale systems placed in photon cavities 
\cite{PhysRevLett.116.113601,PhysRevB.92.165403,Gudmundsson16:AdP_10,2017arXiv170300803H,PhysRevB.97.195442},
just to mention few. Like for most physical phenomena, different modeling approaches have 
been applied ranging from non-equilibrium Green functions \cite{PhysRevB.94.035434,PhysRevLett.119.223601,PhysRevB.99.035129}
to master equations of various types \cite{PhysRevB.90.125402,PhysRevB.90.085416,Entropy19:731}. The modeling efforts have not been straight
forward, and some authors have emphasized the role of the geometrical shape of the systems \cite{Gudmundsson16:AdP_10},
while other have explored how appropriate the use of a single cavity mode is \cite{SanchezMunoz2018}, with respect to causality,
or if relativistic corrections have to be taken into account \cite{Lopp_2018}, 
besides the fundamental issues on the derivation of an appropriate 
description when using a master equation approach \cite{PhysRev.129.2342,PhysRevA.84.043832}. 

Various approaches have been used for the matter-photon interactions in 
different systems of spins, electrons, or atoms placed in a cavity. 
Naturally, most can be referred back to an electrical dipole interaction,
expressed either in terms of a space integral over the inner product of the vector 
potential $\mathbf{A}$ and the charge current density $\mathbf{j}$, or the 
integral of the inner product of the electrical field and the position operator
\cite{Scully97,Ford97:377}.
Some models have included the much weaker diamagnetic interaction, that is often
referred to as the $A$-square term 
\cite{PhysRevLett.35.432,doi:10.1002/andp.201700334,Nataf2010,PhysRevA.84.012510}.

Analytical or high-order numerical methods lead to results of high accuracy
within the electrical dipole interaction approximation, relying or 
not \cite{Feranchuk96:4035,PhysRevLett.98.013601} on the 
rotating wave approximation \cite{Scully97,Ford97:377}.

The dipole approximation is known to be valid if the wavelength of the electromagnetic
field is usually much larger than the size of the matter system. Magnetic
effects in the interaction will only be evident as the variation of the 
electromagnetic vector potential is taken into account within the matter
or the electron system. They are thus bound to be tiny in most cases.

Earlier, we have shown that high order transitions can be
important when describing the approach of systems to a steady state in which otherwise
low order forbidden transitions play a role \cite{doi:10.1002/andp.201900306}.
This has lead us to the questions: What about the higher order electric and 
magnetic terms of the para- and diamagnetic electron-photon interactions?
Do they lead to effects, that among others, can be understood as self-induction
in time-dependent electron transport through a system with a nonzero spatial 
extension in a photon cavity?

To explain our ideas, let us consider the following set-up:
Initially, the external leads are coupled to 
an empty central system, i.\ e.\ neither an electron nor a photon are present, 
the initial system is in its vacuum state. The coupling to the external leads offers 
only electrons to the system, no photons. We consider an external photon reservoir 
coupled to the cavity to be at zero temperature, so no photons enter the system from
the photon reservoir. The electron-photon coupling in the central systems turns all 
many-body states into either pure photon states, or electrons states dressed 
with cavity photons. Thus, after the coupling to the leads the photon expectation value in the 
cavity increases and can become large depending on which dressed electron
states are in the bias window defined by the external leads. In this set-up the photon 
content of the central system will reach a maximum at some intermediate time and 
decrease as the system reaches a steady state.

In a classical system this emergence of an electromagnetic component in transport
could lead to the consideration of inductive effects if higher order terms were considered
for the electron-photon interaction.  
For a ``few-body'' quantum system one expects the identification of such inductive
effects probably to be obscured by the quantum details of countably few transport paths
or transitions through the system. Comparison with classical circuits is further hindered by the simplicity of
the approach to view few-body high frequency quantum circuits using classical terms as
``lumped elements'', with separable capacitive and inductive elements.

In quantum electrodynamics (QED) all higher order terms of an electron-photon interaction,
describing processes during which an electron interacts with its own electromagnetic field,
are included in the self-energy of the electron. Here, in a confined central system the discrete 
electron states all acquire a higher-order self-energy that depends on the geometry of both the 
electron states and the photon field in the cavity in addition to other properties the states
have.

The paper is organized as follows: Section \ref{Model} gives a short description
of the model with the time-independent properties in the Subsection \ref{Time-ind},
but the time-dependent transport in Subsection \ref{Tran-prop}.
Results are presented in Section \ref{Results}, with general quantities outlined 
in subsection \ref{Trans-res}, while differences between the models with different 
electron-photon interactions are displayed in Subsection \ref{Dyn-diff}. 
Conclusions are drawn in Section \ref{Conclusions}.

\section{Model}
\label{Model}
\begin{figure}[htb]
	\centerline{\includegraphics[width=0.48\textwidth]{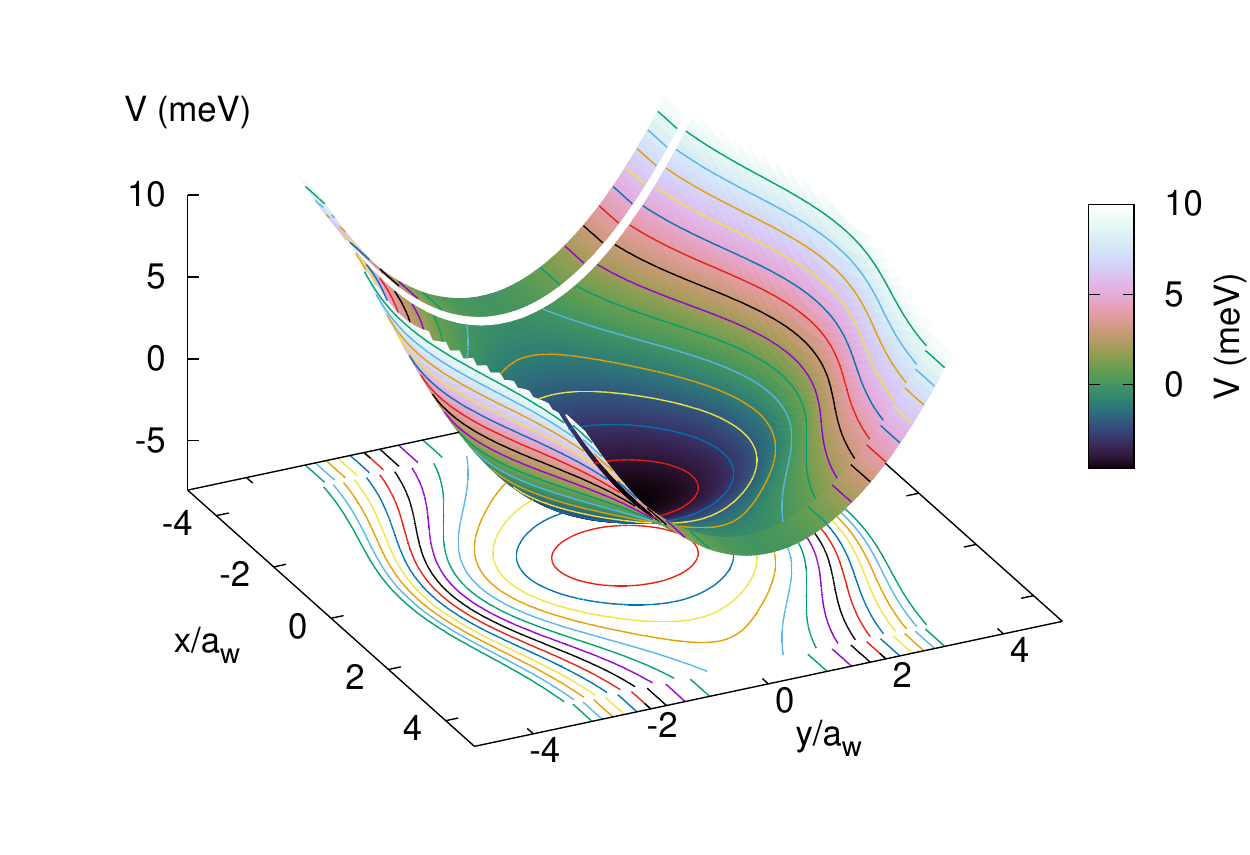}}
	\caption{The potential, $V$, defining the shape of the central 
	         system, a circular quantum dot embedded in a short quantum wire.
	         The length of the wire is 180 nm, extending from 
	         $x/a_w\approx -4.3$ to $x/a_w\approx +4.3$, where 
	         $a_w\approx 20.75$ nm is the effective magnetic length
	         for the parabolic confinement in the $y$-direction
	         with characteristic energy $\hbar\Omega_w=2.0$ meV.
	         The tiny gaps at either end of the short wire indicate
	         the onset of the external leads with same confinement
	         in the $y$-direction.}
	\label{Fig01}
\end{figure}

We model a short quantum wire of length $L_x=180$ nm in the $x$-direction 
with a parabolic confinement with characteristic energy $\hbar\Omega_0=2.0$ meV
in the $y$-direction. We assume a hard wall confinement at the ends in the 
$x$-direction and a constant external magnetic field $\mathbf{B}=B\hat{\mathbf{z}}$
with strength $B=1.0$ T perpendicular to the two-dimensional quantum wire assumed 
to be formed in a GaAs heterostructure with effective mass $m^*=0.067m_e$ dielectric constant
$\kappa_e=12.4$, and $g^*=-0.44$. Embedded in the short wire is a quantum dot potential as is 
shown in Fig.\ \ref{Fig01}
The short quantum wire with the dot is defined by the potential
\begin{align}\label{V-pot}
      V(x,y) =& \theta\left(\frac{L_x}{2}-|x|\right)
      \left[\frac{1}{2}m^*\Omega_0^2y^2 -eV_\mathrm{g} \right. \\ 
      &+ \left. V_d\sum_{i=1}^2 \exp{\left\{-\beta^2 x^2-\beta^2(y-y_{0i})^2\right\}} 
      \right] ,\nonumber
\end{align}
formed by two strongly overlapping Gaussian 
dips in order to get an almost circular shape in the parabolic confinement. 
The parameters are $V_d = -6.0$ meV, $\beta = 0.018$ nm$^{-1}$, defining their depth and extent, 
$y_{01}=-48$ nm, and $y_{02}=+48$ nm defining their location. $V_\mathrm{g}$ is a
plunger gate voltage used to raise or lower certain states into the bias window. Here, it will 
be set to $V_\mathrm{g}=2.47$ meV.

\subsection{Time-independent properties}
\label{Time-ind}
The Hamiltonian of the central system is 
\begin{align}\label{HS}
      H_\mathrm{S} =& \int d^2r\; \psi^\dagger (\mathbf{r})\left\{\frac{\pi^2}{2m^*}+
        V(\mathbf{r})\right\}\psi (\mathbf{r}) \nonumber\\
        +&H_\mathrm{EM} + H_\mathrm{Coul}+H_\mathrm{Z}\nonumber\\ 
        +&\frac{1}{c}\int d^2r\;\mathbf{j}(\mathbf{r})\cdot\mathbf{A}_\gamma
        +\frac{e^2}{2m^*c^2}\int d^2r\;\rho(\mathbf{r}) A_\gamma^2,
\end{align}
in terms of the fermionic field operators $\psi$ and $\psi^\dagger$ for electrons,
the probability density $\rho = \psi^\dagger\psi$, and the charge current density
$\mathbf{j} = -e\{\psi^\dagger\bm{\pi}\psi + \bm{\pi}^*\psi^\dagger\psi\}/(2m^*)$,
with ${\bm{\pi}}=(\mathbf{p}+e\mathbf{A}_{\mathrm{ext}}/c)$,
where $\mathbf{A}_{\mathrm{ext}}=(-By,0,0)$ is the vector potential defining the external
constant magnetic field $\mathbf{B}$. The external magnetic field together with the parabolic
confinement energy $\hbar\Omega_0$ define the effective confinement energy  
$\hbar\Omega_w=\hbar({\omega_c^2+\Omega_0^2})^{1/2}$ and the effective magnetic length
$a_w=(\hbar /(m^*\Omega_w))^{1/2}$, where $\omega_c=(eB_{\mathrm{ext}})/(m^*c)$
is the cyclotron frequency. We will explore properties of the central system for 
$B=1.0$ T, such that $a_w\approx 20.75$ nm and $\hbar\Omega_w\approx 2.642$ meV.
$H_\mathrm{EM}=\hbar\omega a^\dagger a$ is the Hamiltonian for the single-mode cavity
with energy $\hbar\omega$, $H_\mathrm{Z}$ is the Zeeman term for the electrons, 
and $H_\mathrm{Coul}$ is the Coulomb interaction of the electrons with a kernel
\begin{equation}
      V_{\mathrm{Coul}}(\mathbf{r}-\mathbf{r}') = \frac{e^2}{\kappa_e\sqrt{|\mathbf{r}-\mathbf{r}'|^2+\eta_c^2}},
\label{VCoul}
\end{equation}
where a small regularization parameter $\eta_c/a_w=3\times 10^{-7}$ has been used.
The electron-photon interactions terms in the 3rd line of Eq.\ (\ref{HS}) are the paramagnetic-
and the diamagnetic parts, respectively.

\begin{figure}[htb]
	\centerline{\includegraphics[width=0.38\textwidth]{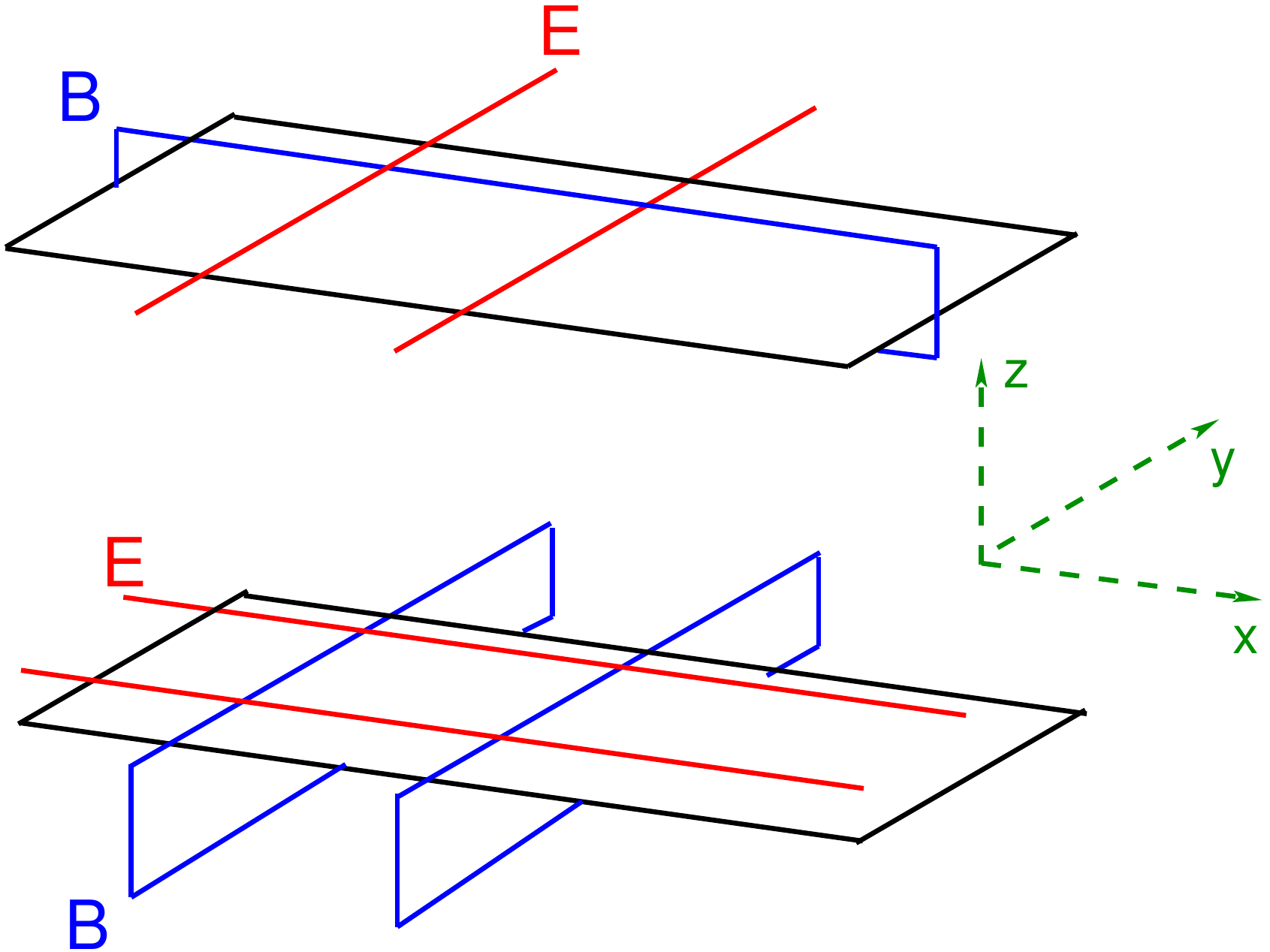}}
	\caption{A schematic diagram indicating the directions of the
	         electric ($E$, red) and the magnetic ($B$, blue) fields 
	         of the cavity photon field for $y$-polarization (upper)
	         and $x$-polarization (lower). The black horizontal strip
	         represents the plane of the two-dimensional electron gas.}
	\label{Fig02}
\end{figure}

We model a rectangular photon-cavity with dimensions $a_\mathrm{c}\times b_\mathrm{c}\times d_\mathrm{c}$,
using the Coulomb gauge for the quantized vector potential $\mathbf{A}_\gamma$ 
for the single-mode photon
field of the cavity. The electric field of the cavity photons is aligned to the 
transport in the $x$-direction (with the unit vector $\mathbf{e}_x$) in the TE$_{011}$ mode, 
and perpendicular to it (defined by the unit vector $\mathbf{e}_y$) in the TE$_{101}$ mode. 
The general configuration of the electric and magnetic components of the cavity field are
schematically displayed in Fig.\ \ref{Fig02} for the two linear polarizations.
The quantized vector potential for the two polarizations for the cavity photon field is expressed 
as \cite{Gudmundsson19:10,doi:10.1002/andp.201900306}
\begin{equation}
      \mathbf{A}_{\gamma i} (\mathbf{r})=\hat{\mathbf{e}}_i {\cal A}\left\{a+a^{\dagger}\right\}
      u_i(z)
\label{Cav-A}
\end{equation}
where $i=x$ or $y$ labels the direction of the polarization and
\begin{align}
\label{u-iz}
      u_x(z) &= \cos{\left(\frac{\pi y}{b_\mathrm{c}}\right)}\cos{\left(\frac{\pi z}{d_\mathrm{c}}\right)} 
              \quad\mbox{and}\nonumber,\\
      u_y(z) &= \cos{\left(\frac{\pi x}{a_\mathrm{c}}\right)}\cos{\left(\frac{\pi z}{d_\mathrm{c}}\right)}.     
\end{align}
$a$ is the annihilation operator for a cavity photon,
and $a^\dagger$ is the corresponding creation operator.
The magnitude of the vector potential, ${\cal A}$,
and the electron-photon coupling constant are related by $g_\mathrm{EM} = e{\cal A}\Omega_wa_w/c$.
For the cavity with photon energy $\hbar\omega =2.63$ meV the ratio of the size of the 
cavity to the length of the central system, $a_c/L_x$ or $b_c/L_x$, is 26.40, and for 
the energy $\hbar\omega =0.98$ meV it is 70.7, so as expected the higher-order terms
will be small. Simplistically, one might expect the higher photon energy leads to
increased higher order effects, but this view will be challenged below.

For an easier comparison of the electron-photon interaction used in other models   
we use the vector potential (\ref{Cav-A}) and the field operators for the electrons
to rewrite the Hamiltonian for the electron-photon interactions as \cite{doi:10.1002/andp.201900306}
\begin{align}
\label{H-e-EM-q}
      H_\mathrm{e-EM}=&g_\mathrm{EM}\sum_{ij}d^\dagger_id_j g^p_{ij}\{a+a^\dagger \}\\ \nonumber
      +&\frac{g_\mathrm{EM}^2}{\hbar\Omega_w}\sum_{ij}d^\dagger_id_j g^d_{ij}
      \left[\left(a^\dagger a+\frac{1}{2}\right) +\frac{1}{2}(a^\dagger a^\dagger +aa)\right] ,
\end{align}
where $d_i$ is the annihilation operator for an electron in the single-particle state labeled
by $i$, and $d^\dagger_i$ is the corresponding creation operator.
The definition of the coupling matrices (in the $z=0$ plane) 
\cite{doi:10.1002/andp.201900306} in terms of the wavefunctions of the original
single-electron basis $\{|a\rangle \}$ is
\begin{equation}
      g_{ab}^{di} = \langle a|\{u_i(0)\}^2 |b\rangle ,
\label{g-d}
\end{equation}
with $i=x$ or $y$ for the diamagnetic electron-photon interaction. If the size of the 
cavity would be assumed to approach infinity (compared to $L_x$) $u_i(0)\rightarrow1$
and then $g_{ab}^d\rightarrow\delta_{a,b}$. The coupling matrix for the
paramagnetic interaction is written symbolically as
\begin{equation}
      g_{ab}^{pi} = \frac{a_w}{2\hbar} \langle a|(\hat{\mathbf{e}}_i\cdot\bm{\pi})\; 
                  u_i(0) |b\rangle  ,
\label{g-p} 
\end{equation}
where the basis $\{|a\rangle\}$ consists of the original single-electron states of the 
wire-dot system with no interaction to the photons.
As the counter rotating terms in the 
electron-photon interactions are important 
\cite{PhysRevA.96.033802,PhysRevA.96.063820,RevModPhys.91.025005}, 
especially for the self-energy of the 
dressed states \cite{doi:10.1002/andp.201900306} we do not apply the rotating 
wave approximation. The eigenstates of the central system described by the Hamiltonian 
(\ref{H-e-EM-q}) are denoted by $|\breve{\mu})$.

If in Eqs.\ (\ref{g-d}-\ref{g-p}) the spatial variation of the vector potential of the photon field, 
$\mathbf{A}_\gamma (\mathbf{r})$ over the electron system is neglected, we have a kind of a dipole
interaction, but including the diamagnetic electron-photon interaction. Note, that 
in this case we still use the exact numerical diagonalization that thus includes higher
order terms of the dipole origin into the self-energies of the dressed electron many-body states.
We will notice below that at least the lowest order of the diamagnetic interaction
is necessary due to the strong external magnetic field having orbital effects.

\subsection{Transport description}
\label{Tran-prop}
The external leads have the same parabolic confinement as the short quantum wire
and are subjects to the same perpendicular magnetic field. Their coupling to the 
central system at $t=0$ is described by the Hamiltonian
\begin{equation}
    H_T = \theta (t)\sum_{il} \int d  \mathbf{q} \left(T_{\mathbf{q}i}^l c_{\mathbf{q}l}^\dagger d_i +
          (T_{\mathbf{q}i}^l )^* d_i^\dagger c_{\mathbf{q}l}\right).
\label{H_T}
\end{equation} 
Electrons in the short wire are created (annihilated) by the operators $d_i^\dagger$ ($d_i$), and
in the leads by the operators $c_{\mathbf{q}l}^\dagger$ ($c_{\mathbf{q}l}$).
The quantum number $\mathbf{q}$ stands both for the continuous momenta in the leads and the appropriate
subband index.  
Due to the shape of the central system it is essential to account for the fact that states
in the leads and the central system couple differently well or badly depending on their energy
and the shape of their probability densities. Thus, the coupling matrix $T_{\mathbf{q}i}^l$ is 
calculated using the probability density of each single-electron state of the lead $l$ and the 
central electron system in the contact region that is defined to extend approximately one $a_w$ 
into each subsystem \cite{Gudmundsson09:113007,Moldoveanu09:073019,Gudmundsson12:1109.4728}. 
See also Appendix A in Ref.\ \cite{doi:10.1002/andp.201900306}. The coupling of the external
leads to the central system has an overall strength $g_0g_\mathrm{LR}a_w^{3/2}=0.101$ meV, when 
the overall dimensionless coupling constant $g_0=1.0$.

The time evolution of the central system under the influence of the reservoirs,
the external leads and the photon reservoir, is calculated using a Markovian master
equation for the reduced density operator $\rho_\mathrm{S}$ in the Liouville space of 
transitions \cite{Weidlich71:325,Nakano2010,Petrosky01032010}, that has been derived from 
a generalized master equation \cite{Zwanzig60:1338,Nakajima58:948}
in a many-body Fock space by applying a Markovian approximation and a vectorization of 
the matrices \cite{IMM2012-03274,2016arXiv161003223J,doi:10.1002/andp.201900306}.
The coupling of the cavity to the external photon reservoir is $\kappa = 1.0\times 10^{-5}$ meV,
and the average photon number in the reservoir is set by the parameter $\bar{n}_\mathrm{R}$
\cite{GUDMUNDSSON20181672}.

From the generalized master equation we derive the mean 
current \cite{Gudmundsson09:113007,Moldoveanu09:073019,Gudmundsson12:1109.4728,GUDMUNDSSON20181672} into the 
central system, $I_\mathrm{L}$, from the left lead (L), and the mean current out of the 
central system, $I_\mathrm{R}$, into the right lead (R). For comparison with simple
classical circuit it would be convenient to consider the mean net current through the 
central system. Due to the capacitance and charging of the central system this 
quantity is not well defined, and we use the mean current, $I=I_\mathrm{L}+I_\mathrm{R}$, as a 
measure of it. Conveniently, in the regime when $I_\mathrm{L}=I_\mathrm{R}$, $I_\mathrm{L}$ is double this
net current through the system, but the net current into the central system, $I_i=I_\mathrm{L}-I_\mathrm{R}$,
vanishes.

In addition to calculating the mean current, $I$, we evaluate the mean number of electrons
in the central system, $N_e$, the mean number of photons $N_\gamma$, the mean value of the 
$z$-component of the total spin $S_z$, and the R{\'e}niy-2 entropy of the central 
system \cite{2017arXiv170708946S,2018arXiv180608441B,2011arXiv1102.2098B}
\begin{equation}
      S=-k_B\ln{[Tr(\rho^2_\mathrm{S})]}.
\label{entropy}
\end{equation}
The mean entropy (\ref{entropy}) serves as a sensitive measure of changes in the central system,
and we only use it for that purpose here.

Information about details in the numerical computations is found in 
Appendix \ref{AppNumDetails}.

\section{Results}
\label{Results}
We first concentrate
on properties of the system with the full electron-photon interactions. Then we analyze the differences 
of these quantities between a system with the full electron-photon interactions
and a system where the dipole approxmation is considered.

\subsection{Transport results for numerically exact one-mode electron-photon interaction}
\label{Trans-res}
\begin{figure*}[htb]
	{\includegraphics[width=0.48\textwidth]{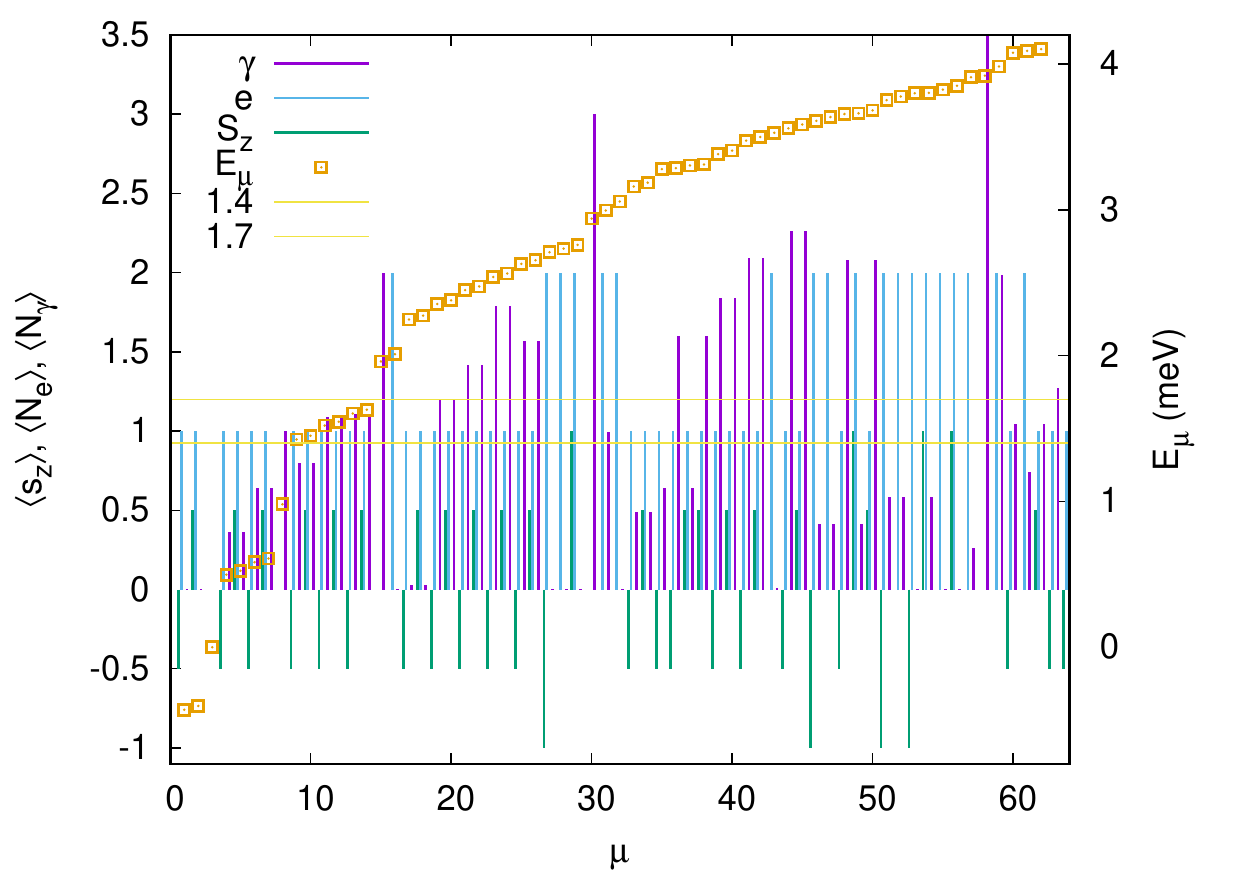}}
	{\includegraphics[width=0.48\textwidth]{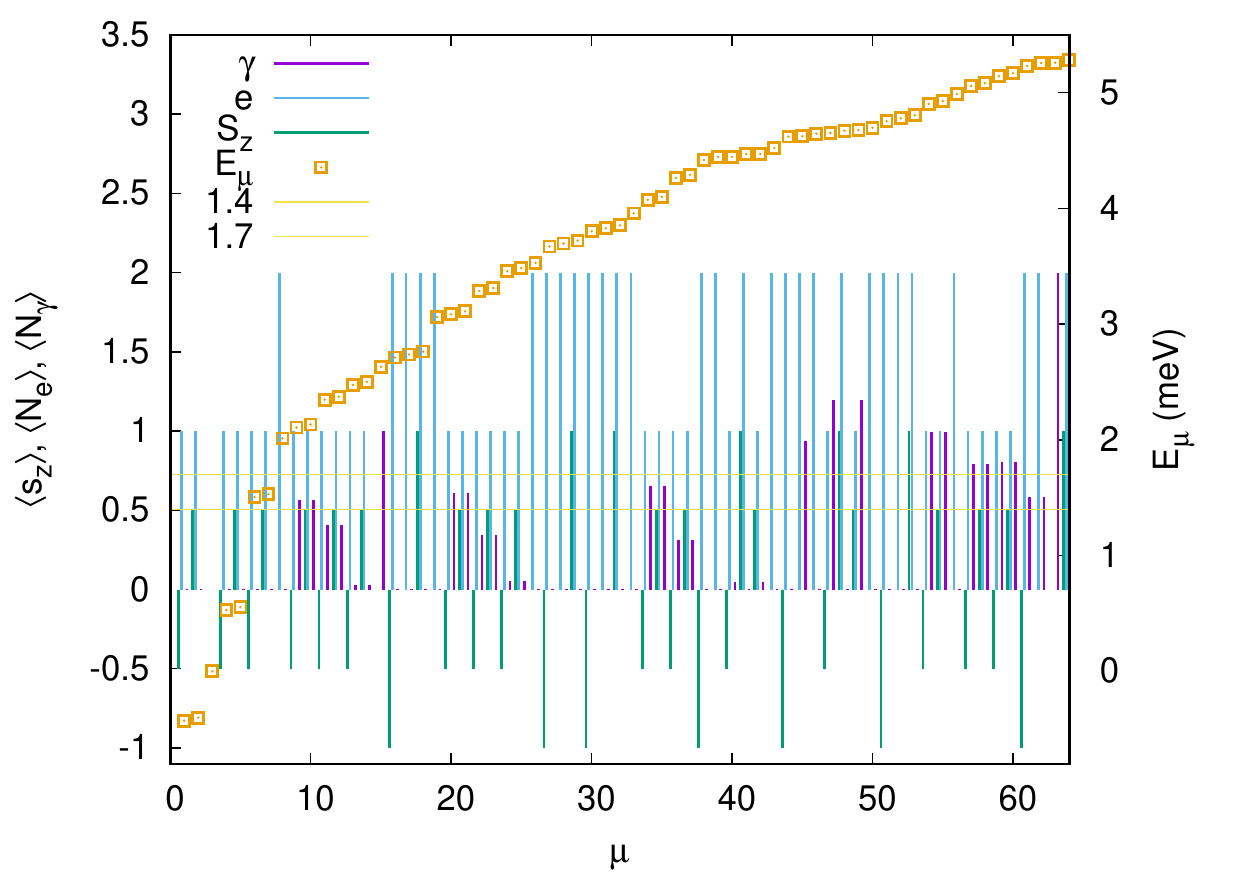}}\\
	{\includegraphics[width=0.48\textwidth]{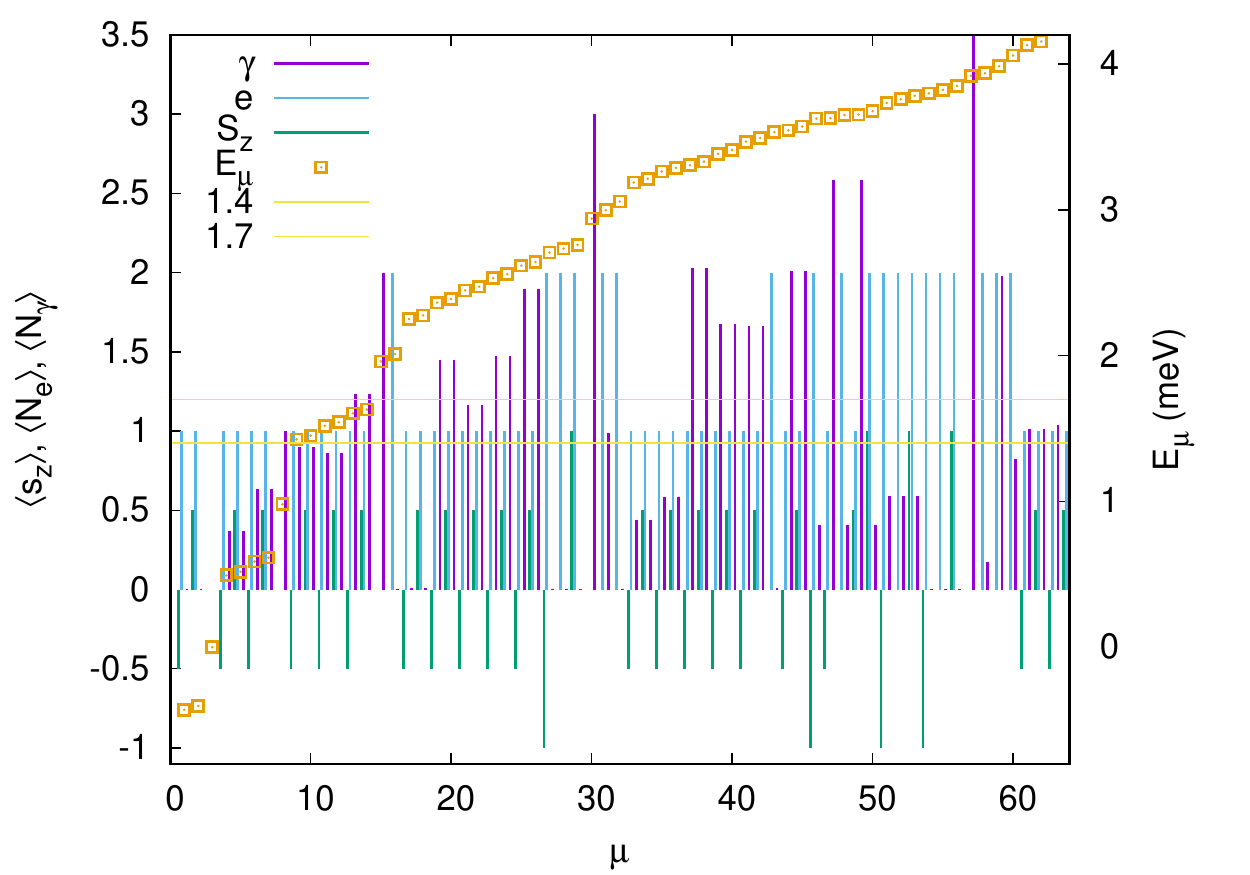}}
	{\includegraphics[width=0.48\textwidth]{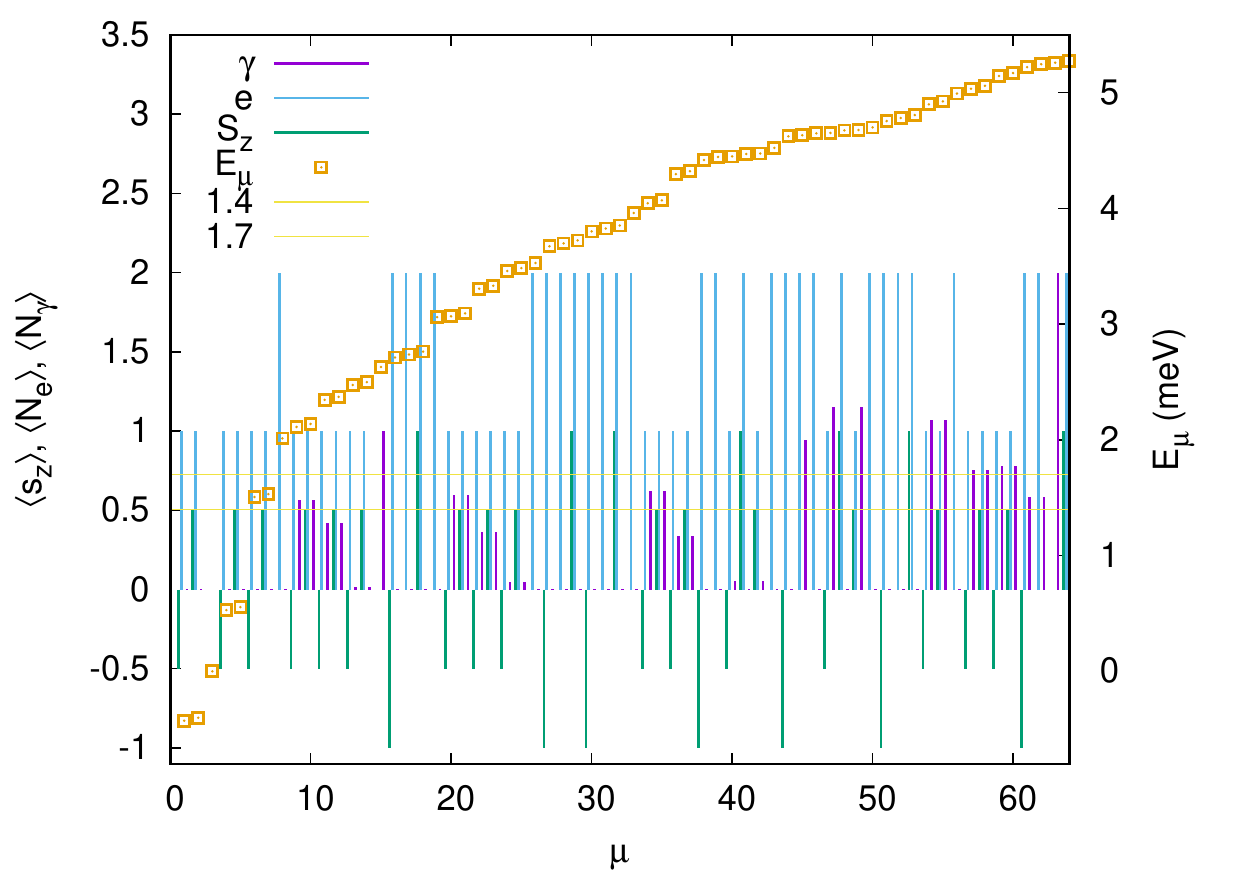}}
	\caption{The energy ($E$), the electron ($e$), the photon ($\gamma$),
	         expectation values for fully interacting state $|\breve{\mu})$,
	         together with its value of $S_z$, for $\hbar\omega = 0.98$ meV
             (left), $\hbar\omega = 2.63$ meV (right), $x$-polarization 
             (upper), and $y$-polarization (lower).	The two horizontal 
	         yellow lines indicate the chemical potentials of the left $\mu_\mathrm{L}$
	         and right $\mu_\mathrm{R}$ leads defining the bias window. 
	         $g_\mathrm{EM}=0.1$ meV, $B=1.0$ T,  
	         $-eV_\mathrm{g}=2.47$ meV, and $L_x=180$ nm.}
	\label{Fig03}
\end{figure*}
The external homogeneous magnetic field $B=1.0$ T. 
We select the chemical potentials of the left and right lead to be
$\mu_\mathrm{L} =1.7$ meV and $\mu_\mathrm{L} =1.4$ meV defining a bias window $\Delta\mu$ of 
0.3 meV. The plunger gate voltage $V_\mathrm{g}$ is set such that
$-eV_\mathrm{g}=2.47$ meV. Further more we will explore the transport
properties of the system for two values of the cavity photon energy,
$\hbar\omega = 0.98$, and 2.63 meV. The one-electron ground state of
the quantum dot (and the central system) is a typical circular symmetric
state with vanishing angular momentum, while the next one-electron states
in energy are typical circular ring states with unit angular momenta 
quantum numbers. In the Fock-Darwin energy spectrum for a 
circular symmetric parabolically confined quantum dot in a magnetic field
the radial wavefunctions are the same for the $M=\pm 1$ states \cite{Fock28:446}.
In our case they are very similar, but the higher once (in the bias window)
look slightly ellipical due to the opening of the dot into the short wire
in the $x$-direction.
In our quantum dot the energy difference between these two states is close to
the lower photon frequency, leading to a very special Rabi-splitting as
will be seen below. We select these circular states of a quantum dot in 
an external magnetic field to enhance possible self-induction in the system.
At the same time the one-electron ground state is in resonance with the 
first excited one-electron state.

As Figure \ref{Fig03} shows the one electron ground state $|\breve{1})$ and
its partner with opposite electron spin $|\breve{2})$ ({corresponding to the 
$M=0$ states of the Fock-Darwin spectrum}) are well below the 
bias window. For the case of $\hbar\omega = 2.63$ meV only the two spin
partners of the second excited one-particle states $|\breve{6})$ and 
$|\breve{7})$ ({corresponding to the higher $M=\pm 1$ Fock Darwin states, 
$|\breve{4})$ and $|\breve{5})$ correspond to the lower ones}) 
are within the bias window, but there is a Rabi-resonance 
between two spin partners of the ground state and states slightly above the 
bias window, such that the lower Rabi branch is either of the spin partners
$|\breve{9})$ or $|\breve{10})$, and the upper branch is $|\breve{12})$ or 
$|\breve{13})$. This can best be seen by the mean photon content of these dressed
one-electron states which is close to half a photon.

In this case the photon content of the dressed one-electron states in 
the bias window is small. When the photon energy is $\hbar\omega = 0.98$ meV
there is, on the other hand, a Rabi-resonance between the second and the third
excited one-electron states in the system on top of a resonance between the 
ground state and the first excited state. We thus end up with six one-electron
states in the bias window, all with a noninteger photon content. For this
case one would expect a faster charging of the system and it should likewise approach
the steady state as most transitions needed would be photon aided.
We also notice from Fig.\ \ref{Fig03} that in the case of $\hbar\omega = 0.98$ meV
(left subfigures) the photon content of the dressed states is more dependent on the 
polarization of the photon field than for $\hbar\omega = 2.63$ meV (right subfigures).

\begin{figure}[htb]
	\centerline{\includegraphics[width=0.48\textwidth]{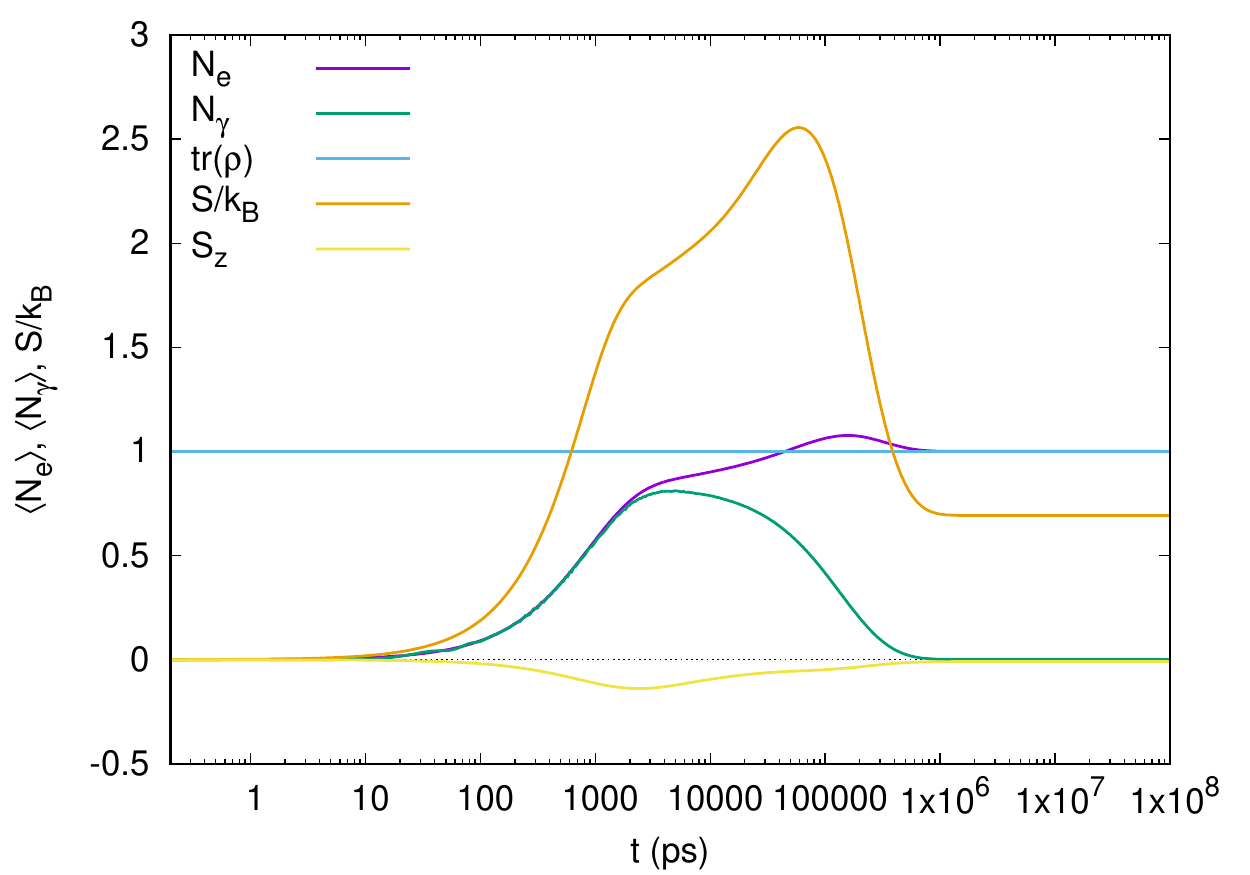}}
	\centerline{\includegraphics[width=0.48\textwidth]{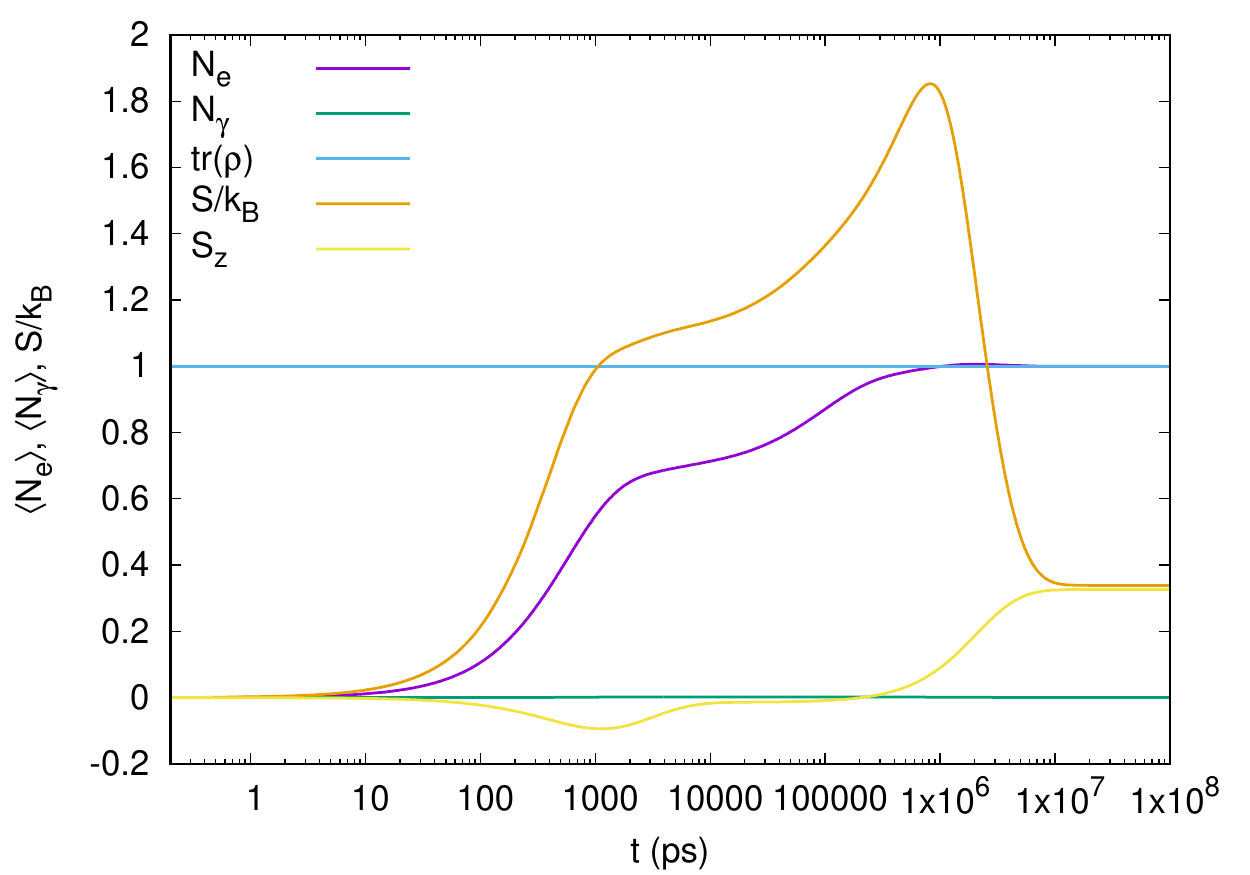}}
	\caption{The mean electron $N_e$ and photon $N_\gamma$ number,
	         the trace of the reduced density matrix, the mean
	         entropy $S$, and $z$-component of the spin for
	         photon energy $\hbar\omega = 0.98$ meV (upper),
	         and $\hbar\omega = 2.63$ meV (lower). $g_\mathrm{EM}=0.1$ meV,
	         $B=1.0$ T, $\bar{n}_\mathrm{R}=0$,
	         $\kappa = 1.0\times 10^{-5}$ meV, $-eV_\mathrm{g}=2.47$ meV,
	         $L_x=180$ nm, and $g_\mathrm{0}g_\mathrm{LR}a_w^{3/2}=0.101$ meV.}
	\label{Fig04}
\end{figure}

Indeed, in Fig.\ \ref{Fig04} we see this photon enhancement of all transitions
leading to a faster charge for $\hbar\omega = 0.98$ meV (upper)
than for $\hbar\omega = 2.63$ meV (lower).
This can be verified in Fig.\ \ref{Fig04} by the difference in the mean
photon content for either case with different photon energy. We notice that the steady
states for these two cases are different as can be seen by their difference in
the mean spin $z$-component. We come back to this below.

\begin{figure}[htb]
	\centerline{\includegraphics[width=0.48\textwidth]{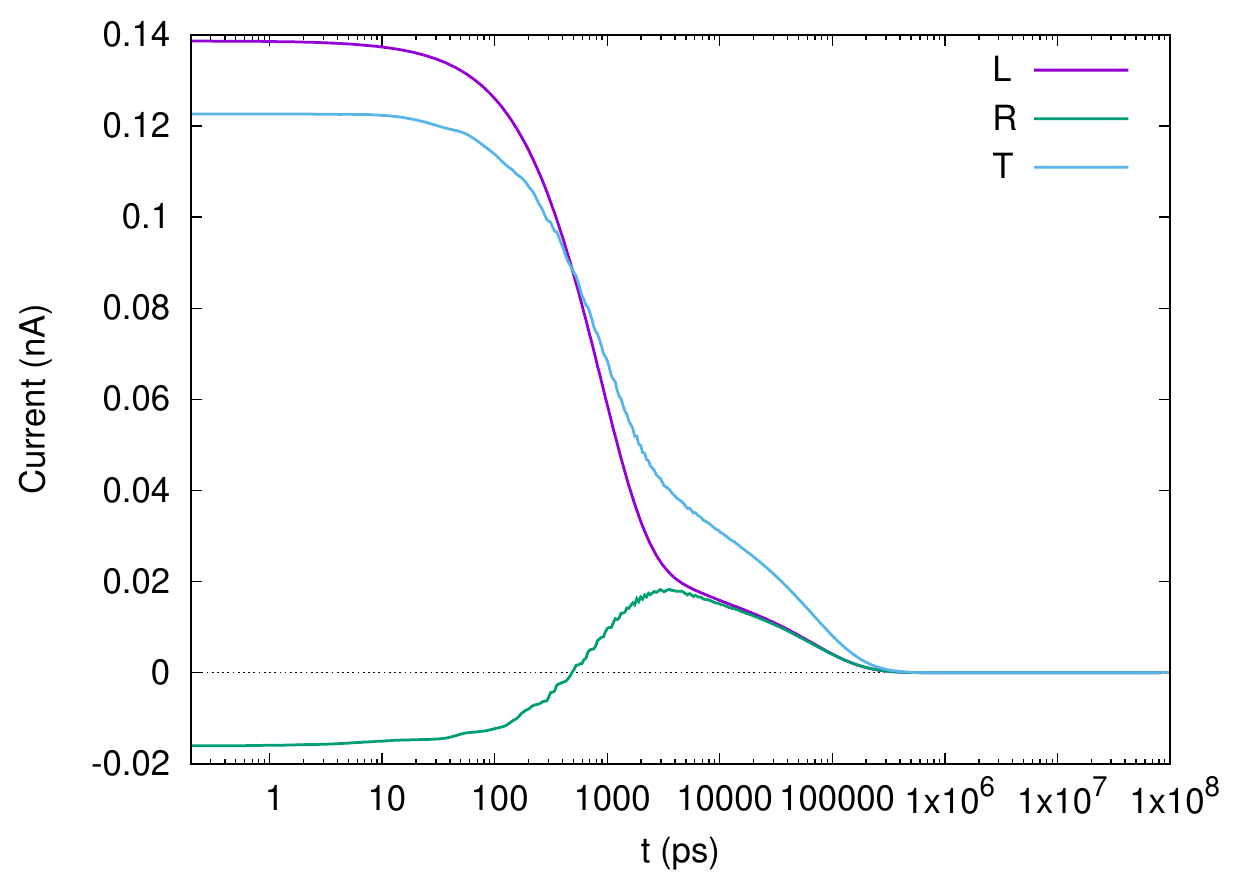}}
	\centerline{\includegraphics[width=0.48\textwidth]{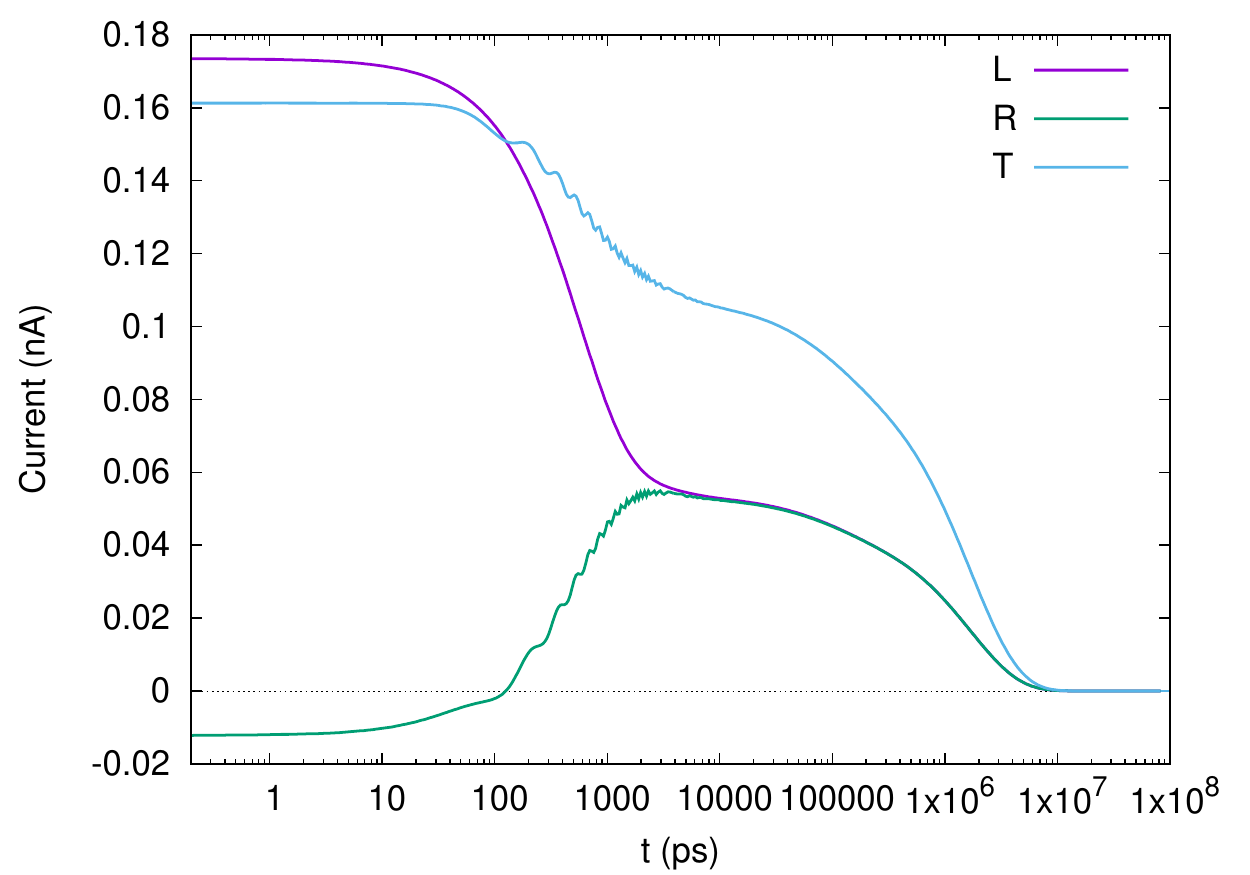}}
	\caption{The mean electron current from the left lead into the 
	         central system $I_\mathrm{L}$ (L), from the central system to the right
	         lead $I_\mathrm{R}$ (R), and the current $I=I_\mathrm{L}+I_\mathrm{R}$ (T), 
	         for photon energy $\hbar\omega = 0.98$ meV (upper),
	         and $\hbar\omega = 2.63$ meV (lower). $g_\mathrm{EM}=0.1$ meV,
	         $B=1.0$ T, $\bar{n}_\mathrm{R}=0$,
	         $\kappa = 1.0\times 10^{-5}$ meV, $-eV_\mathrm{g}=2.47$ meV,
	         $L_x=180$ nm, and $g_\mathrm{0}g_\mathrm{LR}a_w^{3/2}=0.101$ meV.}
	\label{Fig05}
\end{figure}

Fig.\ \ref{Fig05} displays the mean currents for either photon energy,
the current from the left lead into the central system, $I_\mathrm{L}$ labeled (L), the current from the
central system into the right lead, $I_\mathrm{R}$ labeled (R), and the current $I=I_\mathrm{L}+I_\mathrm{R}$ labeled (T).
As $\bar{n}_\mathrm{R}=0$ and there are several one-electron states below the bias 
window and the lowest two-electron states are above it, the system is in a Coulomb 
blockade in the steady state. If $\bar{n}_\mathrm{R}=1$, or higher, the current
will not vanish in the steady state.

\begin{figure}[htb]
	\centerline{\includegraphics[width=0.48\textwidth]{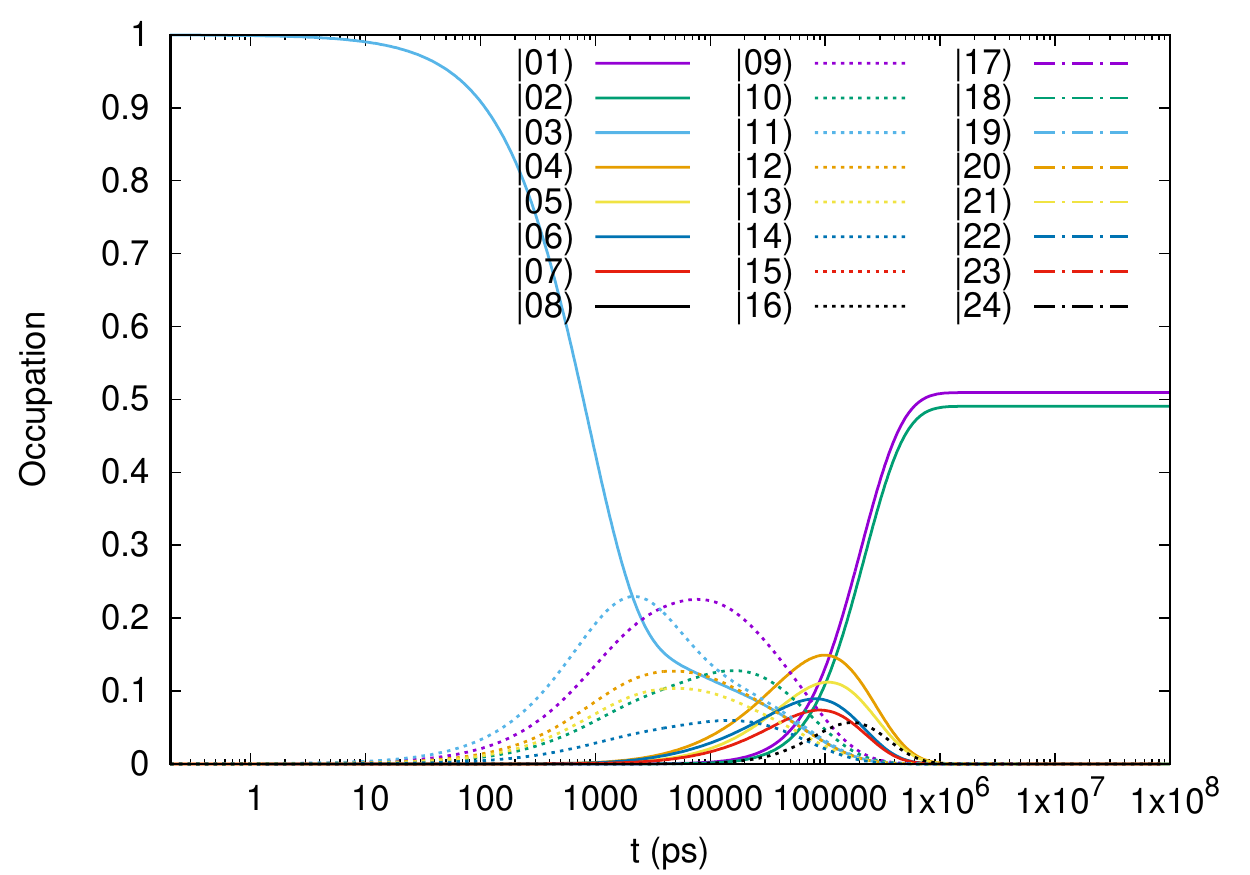}}
	\centerline{\includegraphics[width=0.48\textwidth]{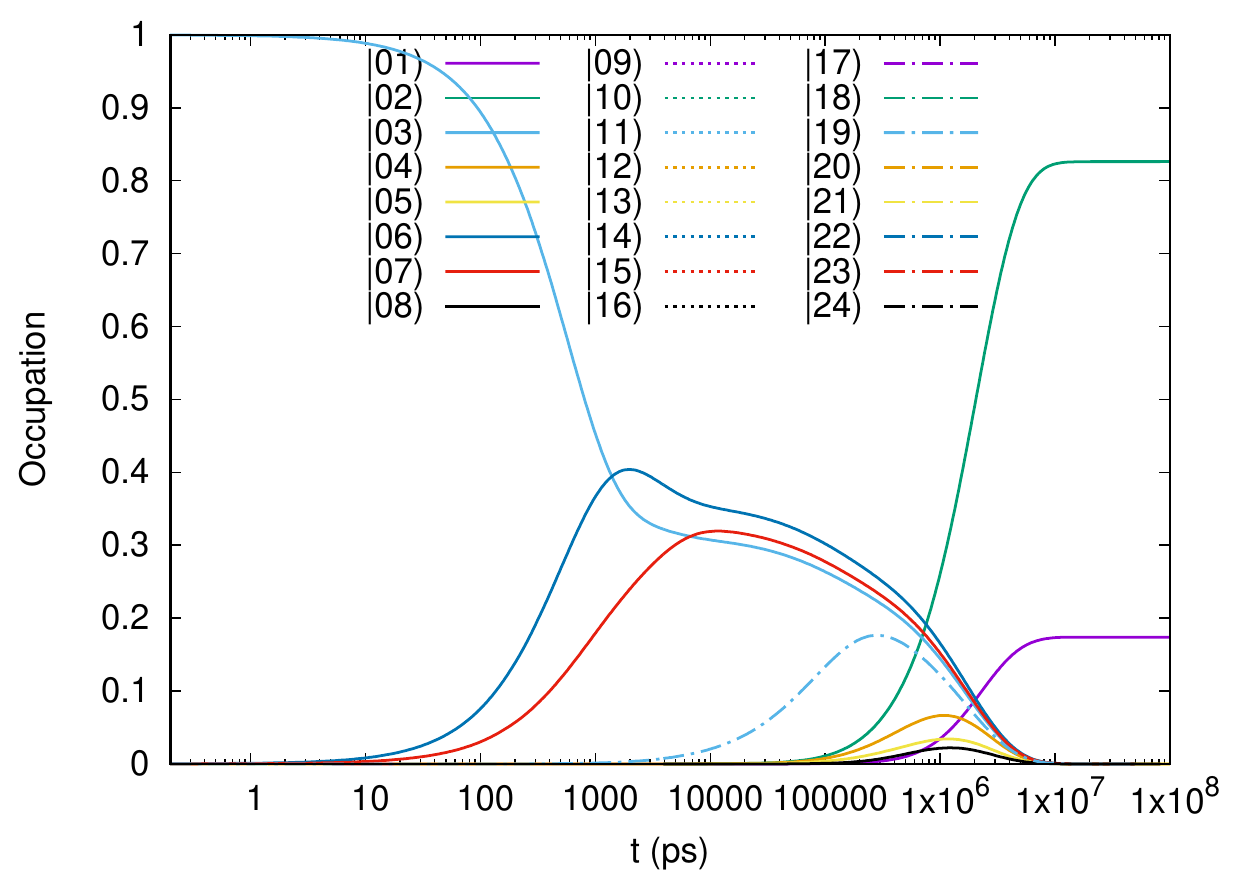}}
	\caption{The time-dependent occupation of selected many-body states $|\breve{\mu})$
	         for $\hbar\omega = 0.98$ meV (upper), and $\hbar\omega = 2.63$ meV (lower).
             $g_\mathrm{EM}=0.1$ meV, $B=1.0$ T, $\bar{n}_\mathrm{R}=0$,
             $-eV_\mathrm{g}=2.47$ meV, and $L_x=180$ nm, and 
             $g_\mathrm{0}g_\mathrm{LR}a_w^{3/2}=0.101$ meV.}
	\label{Fig06}
\end{figure}

Importantly, there is a time interval just when
the system is getting fully charged where the left and right currents are almost
the same. During this time interval the system is approaching the steady state 
through radiative or nonradiative transitions \cite{Gudmundsson16:AdP_10} 
as can be seen by the time dependent
occupations seen in Fig.\ \ref{Fig06}
By comparing the information in Fig.\ \ref{Fig06} and \ref{Fig03} we see that
on the way to the steady state the system probes the two-electron ground state,
but then relaxes finally into a combination of one-electron states made
up of contributions from both spin components of the one-electron ground state.
The route taken to the steady state depends strongly on the photon energy
as the electron-photon interaction makes different transitions available
to the system.

The results shown in Figs.\ \ref{Fig04} and \ref{Fig05} allow us to estimate 
two time scales relevant for classical circuits. First, using the size of the 
bias window, $\Delta\mu =0.3$ meV, and the average current outside the steady state,
$I=0.1$ nA, we arrive at the contact resistance $R\approx 3$ M$\Omega$. 
Secondly, from the geometry of the electron system we estimate its coefficient of inductance
to be $L\approx 36$ fH. Together these give the time scale of the inductance to be
$\tau_L\approx 1.2\times 10^{-20}$ s = 1.2$\times$10$^{-8}$ ps.
This shows plainly that we can not expect to observe this $\tau_L$ in the present system.

From the charging time $\tau_C\approx 10^6$ ps (see  Fig.\ \ref{Fig06}) and the contact 
resistance we estimate the capacity of the electron system to be approximately $C\approx 1$ pf. If our 
system could be compared to an RCL circuit this allows us to extract the resonance frequency of oscillations 
to be $\omega_0=1/\sqrt{LC}\approx 5.3$ (ps)$^{-1}$, or their energy $\hbar\omega_0\approx 3.5$ meV.
Interestingly, this resonance energy is inside the part of the energy spectrum we are exploring, 
but as we are dealing with few particles in a system with discrete energy
levels we can not expect to find this classical resonance energy in our quantum system.


Looking at Fig.\ \ref{Fig04} and \ref{Fig06} one has to wonder why there are
no Rabi-oscillations seen in the density or the occupation of states like has
been seen for different dot systems at lower magnetic field 
\cite{doi:10.1002/andp.201900306,Gudmundsson19:10}. We would expect them,
at least for the case of the photon energy $\hbar\omega = 0.98$ meV,
when there are Rabi-split states in the bias window.
The answer lies in the combination of high external magnetic field ($B=1.0$ T), 
and the simple circular dot potential in the central system. Like, before we
draw an analogy with the Fock-Darwin energy spectrum \cite{Fock28:446}, 
mentioned above, the lowest Rabi-resonance is between states with opposite angular momenta,
but having the same radial wavefunctions. The resonance thus leads to circular
oscillations in the local current density in the central system. These 
divergence-free oscillations can not lead to oscillations in charge density. We have thus
``transverse'' Rabi-oscillations that will be clear in the current, but not in 
the density. This behavior will even be clear on a higher resolution scale
introduced below.

This resonance between the lowest lying ring-formed states corresponding to the lowest
$M=\pm 1$ Fock-Darwin states would not appear in the lowest order dipole approximation,
and at the same time there is a resonance between the one-electron ground state
and the lower of the $M=\pm 1$ states that would be enabled in a dipole approximation.
The Rabi-resonance caused by the diamagnetic part of the electron-photon interactions
occurs even for constant vector potential inside the electron system, as it is 
facilitated by the external constant magnetic field, that leads to divergent-free rotating
currents in the system \cite{doi:10.1002/andp.201700334}.

\subsection{Exploring dynamical differences}
\label{Dyn-diff}
\begin{figure*}[htb]
	{\includegraphics[width=0.48\textwidth]{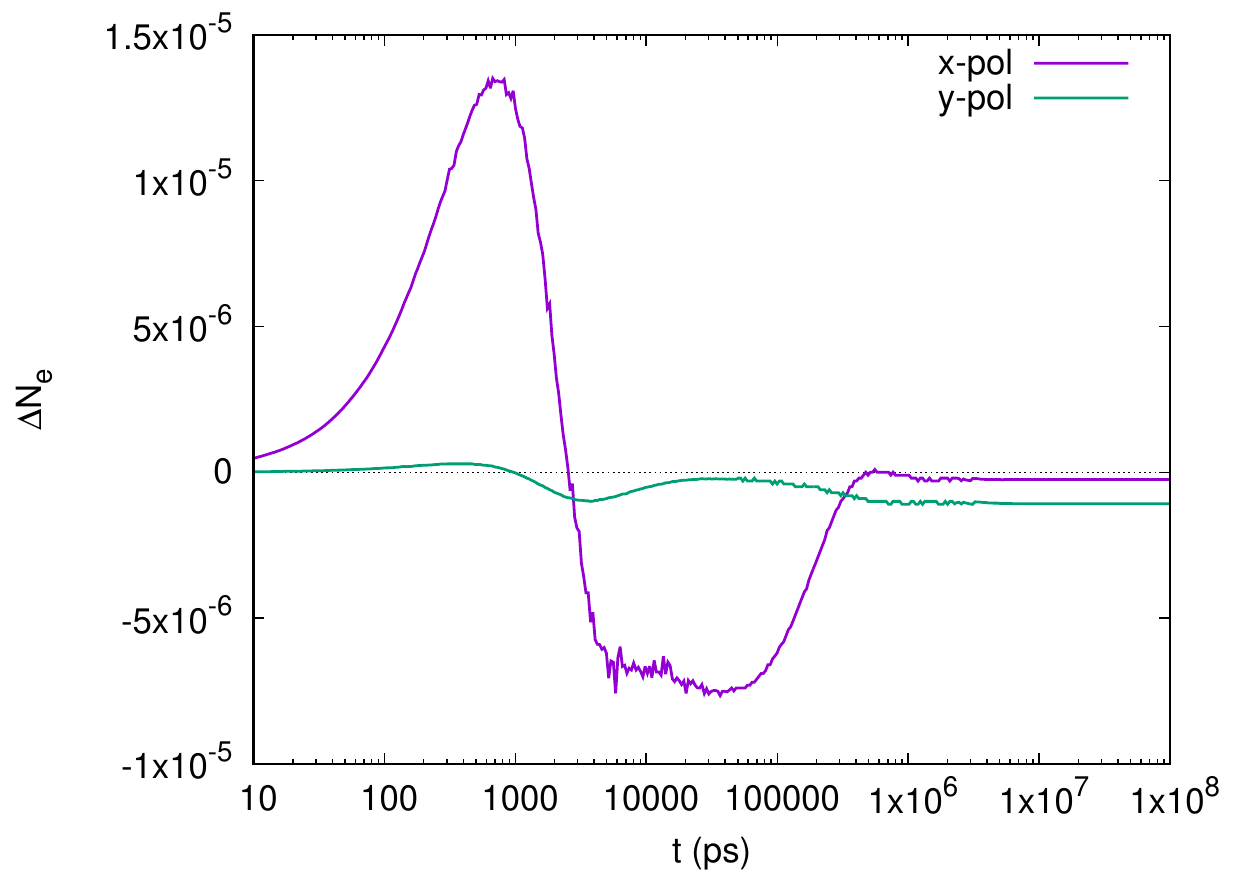}}
	{\includegraphics[width=0.48\textwidth]{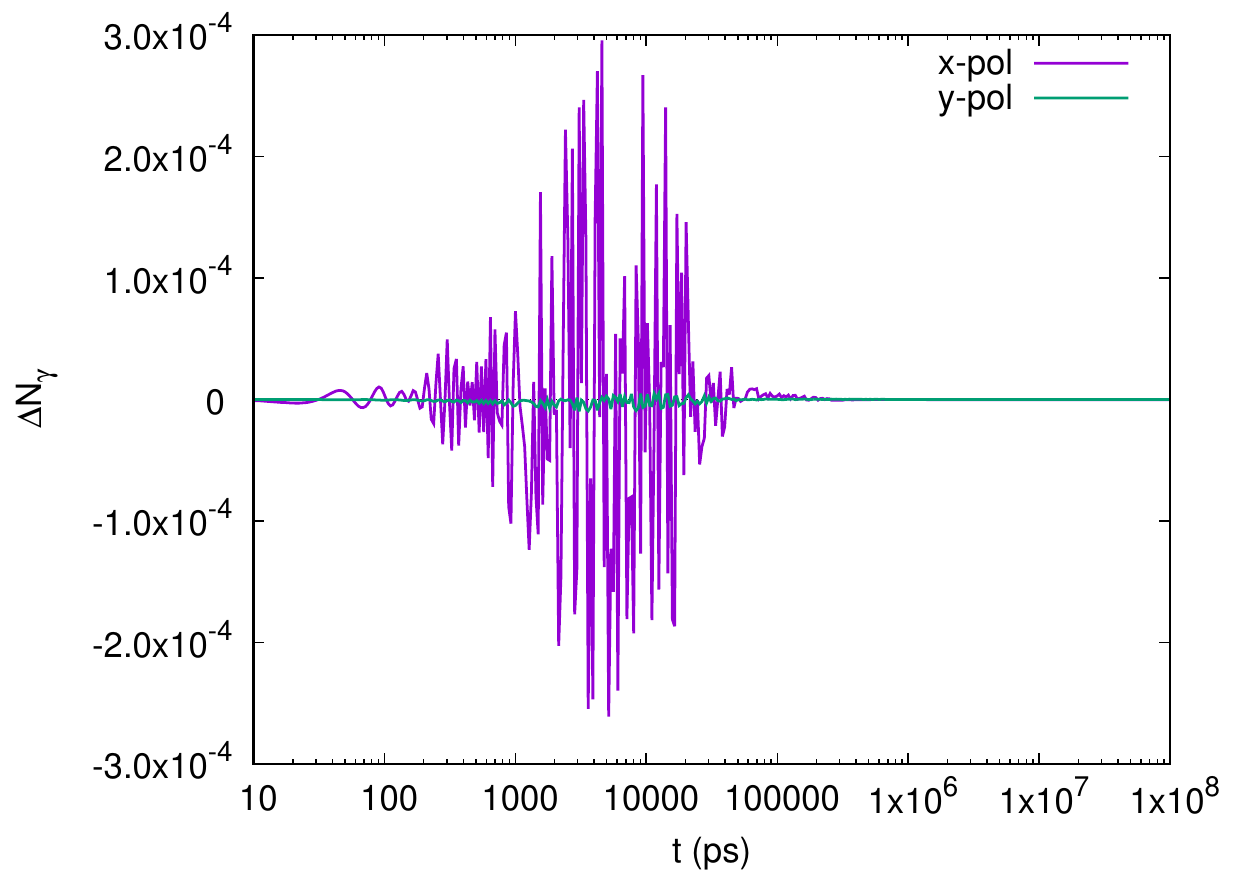}}\\
	{\includegraphics[width=0.48\textwidth]{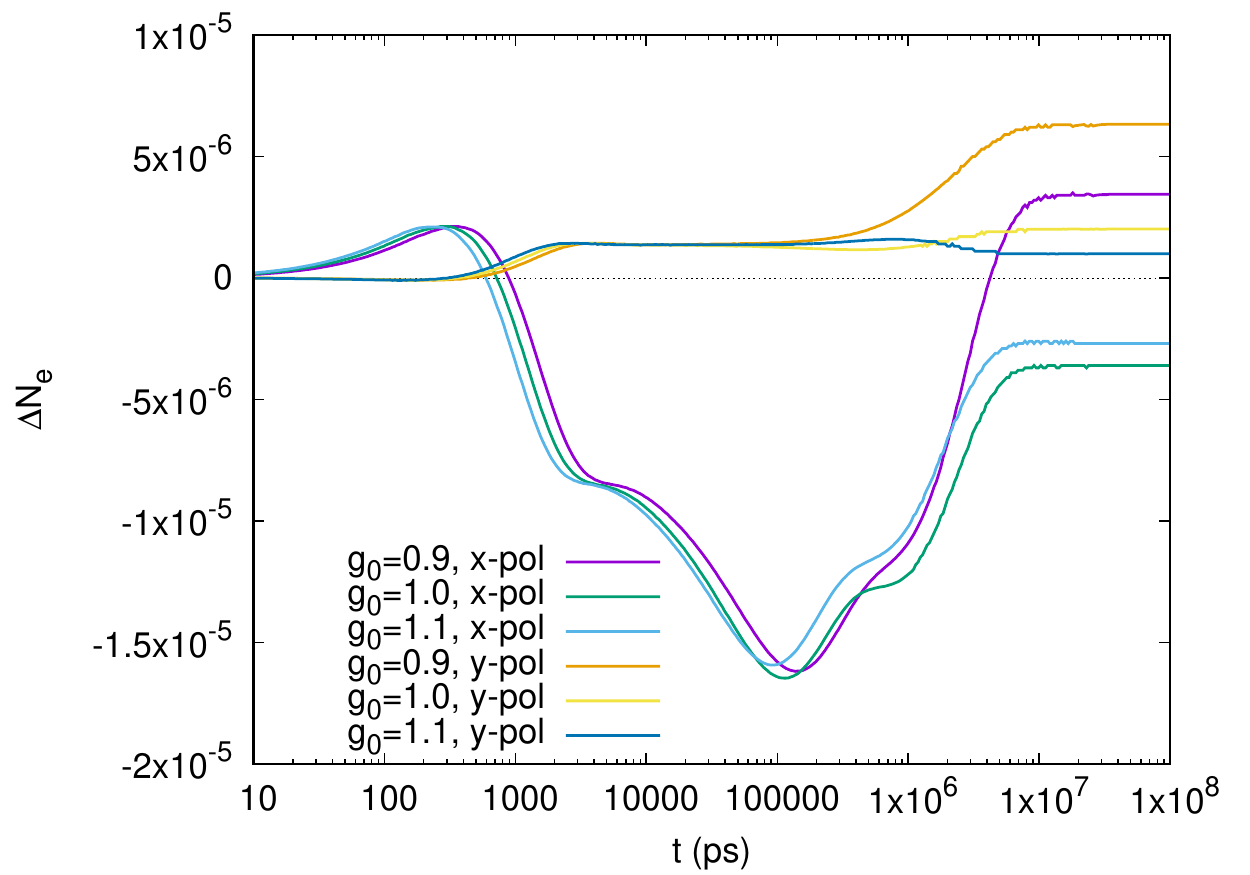}}
	{\includegraphics[width=0.48\textwidth]{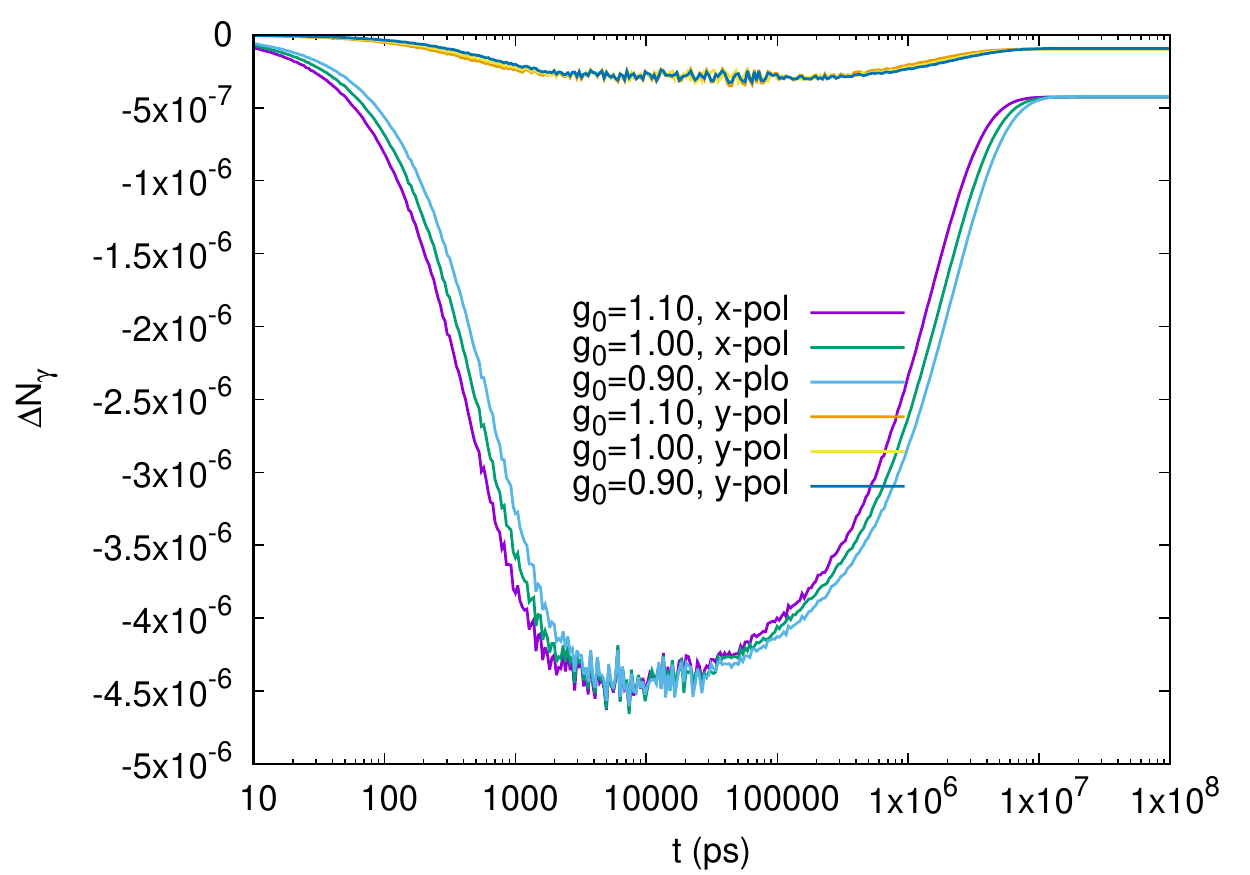}}
	\caption{The differences in the mean electron $\Delta N_e$ (left) 
	         and photon $\Delta N_\gamma$ (right) 
	         numbers for a system with an exact, and a system with a dipole 
	         approximation for the electron-photon interactions for a single 
	         cavity mode for $\hbar = 0.98$ meV (upper), and for 
	         $\hbar = 2.63$ meV (lower) with $g_0=1.0$. 
	         $\bar{n}_\mathrm{R}=0$, $g_\mathrm{EM}=0.1$ meV,
	         $B=1.0$ T, $\kappa = 1.0\times 10^{-5}$ meV, 
	         $-eV_\mathrm{g}=2.47$ meV, and $L_x=180$ nm.}
	\label{Fig07}
\end{figure*}
All the results described above have been calculated using a
numerically exact diagonalization for the single-mode electron-photon
interactions. 
We now look for dynamic effects of the higher order terms in the results. 
To this end we assume the spatial variation of the vector potential $\mathbf{A}_\gamma¸$ 
to vanish inside the electron system itself. 
One might call that a dipole approximation, but within it we also keep
the diamagnetic interaction, which is usually not done in a dipole
approximation, and we use the exact numerical diagonalization for both cases. Within this 
approximation we calculate the same quantities as in subsection III A and then 
compare the two sets of results by defining $\Delta N_e=N_e^\mathrm{exact}-N_e^\mathrm{dipole}$,
$\Delta N_\gamma =N_\gamma^\mathrm{exact}-N_\gamma^\mathrm{dipole}$, and 
$\Delta I=I^\mathrm{exact}-I^\mathrm{dipole}$.
Fig.\ \ref{Fig07} displays $\Delta N_e$ and $\Delta N_\gamma$ in the central 
system for the whole time range
from the transient regime to the steady state. The figure shows results for 
both photon energies and polarizations, and for the photon energy 
$\hbar\omega =2.63$ meV it shows results for three different values of the 
overall system-lead coupling coefficient $g_0$. We notice that especially in
the intermediate time range (ITR) the difference in the electron and photon number
becomes larger for the $x$-polarization than the $y$-polarization.

Three of the subfigures of Fig.\ \ref{Fig07} show irregular oscillations looking
like noise. This is not the case here. When a higher resolution is used for a
narrower time interval (as will be done below) the irregular oscillations are
replaced by very regular Rabi oscillations.  

In the lower panels of Fig.\ \ref{Fig07} for photon frequency $\hbar\omega = 2.63$ meV,
when no photon replicas or other electron states with a high expectation value of photons is
in the bias window we see that in the ITR both the 
expectation values for electrons and photons is reduced for the $x$-polarized
cavity field. The time variation of both $\Delta N_e$ and $\Delta N_\gamma$ can be correlated with
changes in the occupation of the states of the central system seen in the lower panel of
Fig.\ \ref{Fig06} if one considers at the same time the change in the self-energies of
the corresponding states due to the higher-order electron-photon interaction shown in 
lower panel of Fig.\ \ref{Fig15} in Appendix \ref{AppA}, and the changes in the occupation
of the states induced by the same terms seen in the lower panel of Fig.\ \ref{Fig14} in 
the same Appendix. The photon energy in the lower panels in Fig.\ \ref{Fig07} $\hbar\omega = 2.63$ meV
is just below the effective confinement energy of the short quantum wire, 
$\hbar\Omega_w=2.642$ meV favoring polarization of the electron charge density in the $x$-direction,
but at the same time allowing for a small polarization in the $y$-direction. 
In the upper panels the photon energy is $\hbar = 0.98$ meV, so polarization of the 
charge density in the $y$-direction can only be very tiny, but as there is now a multiple Rabi resonance
with the states in the bias window, the system can effectively be polarized in the $x$-direction.
Correspondingly, the time variation of both $\Delta N_e$ and $\Delta N_\gamma$ in the upper panels of
Fig.\ \ref{Fig07} is tiny for the $y$-polarized cavity field, but for the $x$-polarized field the 
effects are larger and $\Delta N_\gamma$ shows clear signs of Rabi-oscillations.

For the lower photon energy, $\hbar\omega = 0.98$ meV, we see in Appendix \ref{AppA} in the upper 
panel of Fig.\ \ref{Fig15} that, indeed, the self-energies of the relevant states due to the 
higher-order terms in the electron-photon interactions are vanishingly small for the $y$-polarized
cavity field, but for the $x$-polarized field the states in the bias window acquire no or relatively
large positive or negative self-energy contribution. The early in time photon active transitions 
seen in the differences in the occupation in the upper panel of Fig.\ \ref{Fig14} in Appendix \ref{AppA}
lead both to the strong response in $\Delta N_e$ in the upper left panel of Fig.\ \ref{Fig07}
for the early and the intermediate time.

\begin{figure}[htb]
	\centerline{\includegraphics[width=0.48\textwidth]{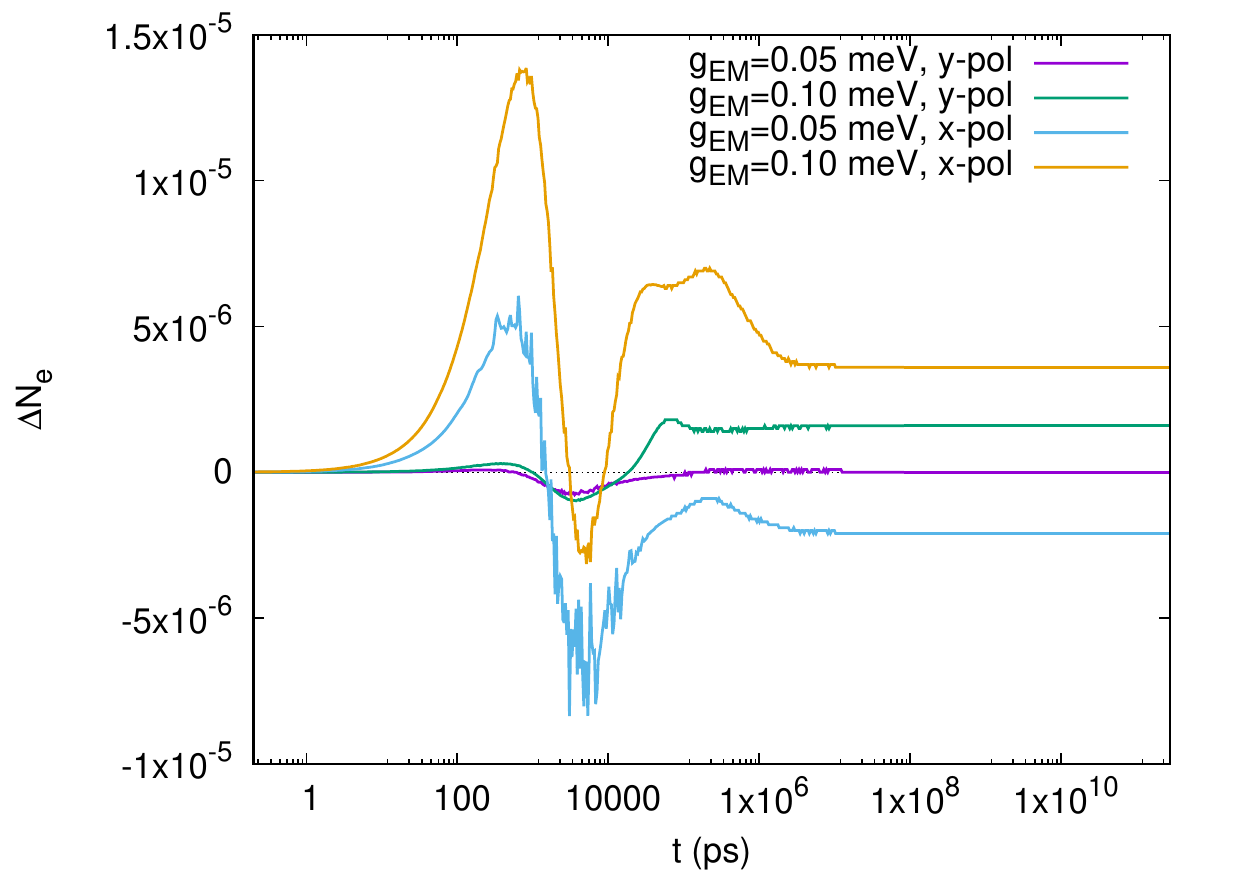}}
	\caption{The difference in the mean electron number for a system
	         with an exact, and a system with a dipole approximation for 
	         the electron-photon interactions for a single cavity mode.
	         The figure compares $\Delta N_e$ for two values of the 
	         electron-photon coupling strength $g_\mathrm{EM}$, 
	         $\bar{n}_\mathrm{R}=1$, and the polarization of the field.
	         $B=1.0$ T, $\hbar\omega = 0.98$ meV,
	         $\kappa = 1.0\times 10^{-5}$ meV, $-eV_\mathrm{g}=2.47$ meV,
	         $L_x=180$ nm, and $g_\mathrm{0}g_\mathrm{LR}a_w^{3/2}=0.101$ meV.}
	\label{Fig08}
\end{figure}

All the results in Fig.\ \ref{Fig07} are for the case when no photons flow from the photon reservoir
to the cavity, i.\ e.\ the mean value of photons in the cavity is $\bar n_\mathrm{R}=0$, and as the 
there are several one-electron states below the bias window the system ends up in a steady state of a 
Coulomb-blockade. Figure \ref{Fig08} shows how $\Delta N_e$ looks for a system with $\bar n_\mathrm{R}=1$,
that does not end up in a Coulomb-blockade, but the current through the system is maintained by 
the photons the reservoir supplies. Initially, the $\Delta N_e$ shows similar behavior as in 
the upper left panel of Fig.\ \ref{Fig07}, but in the late intermediate time there are changes
due to the nonvanishing current through the system.

\begin{figure}[htb]
	\centerline{\includegraphics[width=0.48\textwidth]{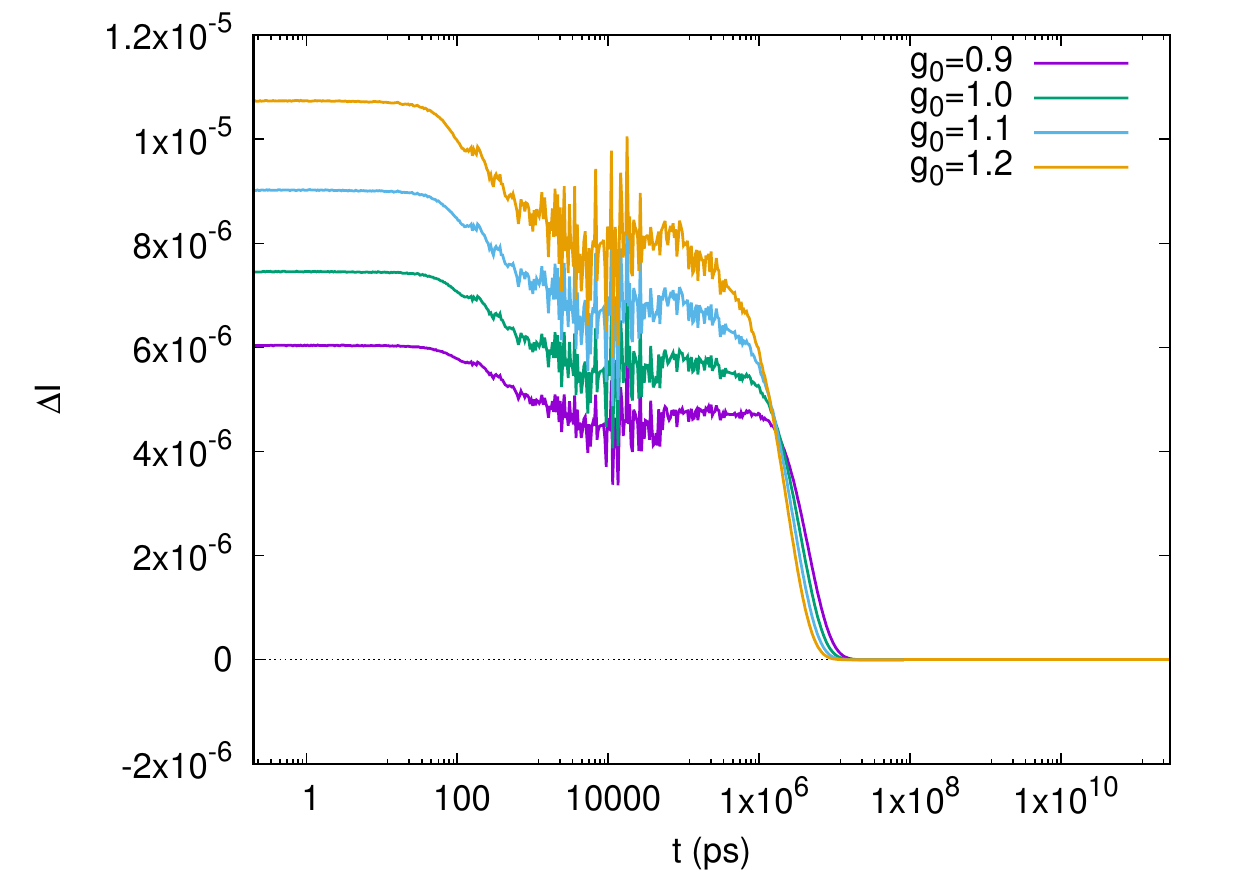}}
	\centerline{\includegraphics[width=0.48\textwidth]{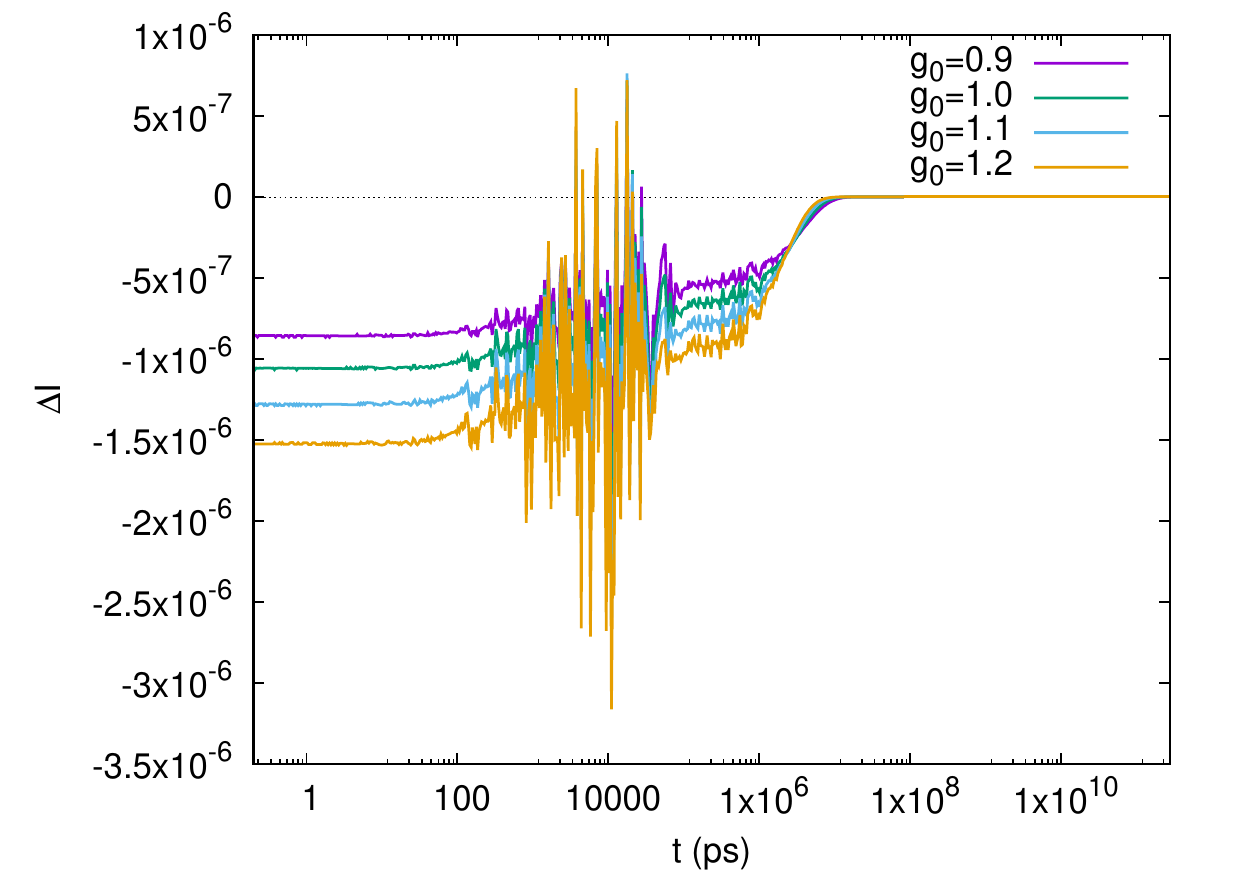}}
	\caption{The difference in the mean electron current $I=I_\mathrm{L}+I_\mathrm{R}$
	         for $x$-polarized (upper), and 
	         $y$-polarized cavity field, for a system
	         with an exact, and a system with a dipole approximation for 
	         the electron-photon interactions for a single cavity mode.
	         The figure compares $\Delta I$ for 
	         three values of the dimensionless overall lead-system coupling strength 
	         $g_\mathrm{0}$. $\bar{n}_\mathrm{R}=0$. $g_\mathrm{EM}=0.1$ meV,
	         $B=1.0$ T, $\hbar\omega = 2.63$ meV,
	         $\kappa = 1.0\times 10^{-5}$ meV, $-eV_\mathrm{g}=2.47$ meV,
	         and $L_x=180$ nm.}
	\label{Fig09}
\end{figure}

Commonly, in classical circuits the current is analyzed in order to 
find traits of self-induction. In Fig.\ \ref{Fig09} we display the change in the current
$I=I_\mathrm{L}+I_\mathrm{R}$ for the higher photon energy, $\hbar\omega = 2.63$ meV, for 
both polarizations of the cavity field, and four values of the overall coupling
to the external leads $g_0$.
For the $x$-polarized cavity field (upper panel) the difference in the current
caused by the higher-order electron-photon interactions terms increases with growing
overall coupling and is generally positive. The higher-order terms lead to more
current through the central system. On the contrary, for a $y$-polarized cavity field 
(lower panel) the current is weakly suppressed by the higher-order terms, increasing with
the overall coupling $g_0$. For $\bar{n}_\mathrm{R}=1$ (not shown here) the differences
in the current are very similar, except the higher order terms do not suppress 
the current for the $y$-polarization. It is still of the same magnitude as for 
$\bar{n}_\mathrm{R}=0$.

\begin{figure}[htb]
	\centerline{\includegraphics[width=0.48\textwidth]{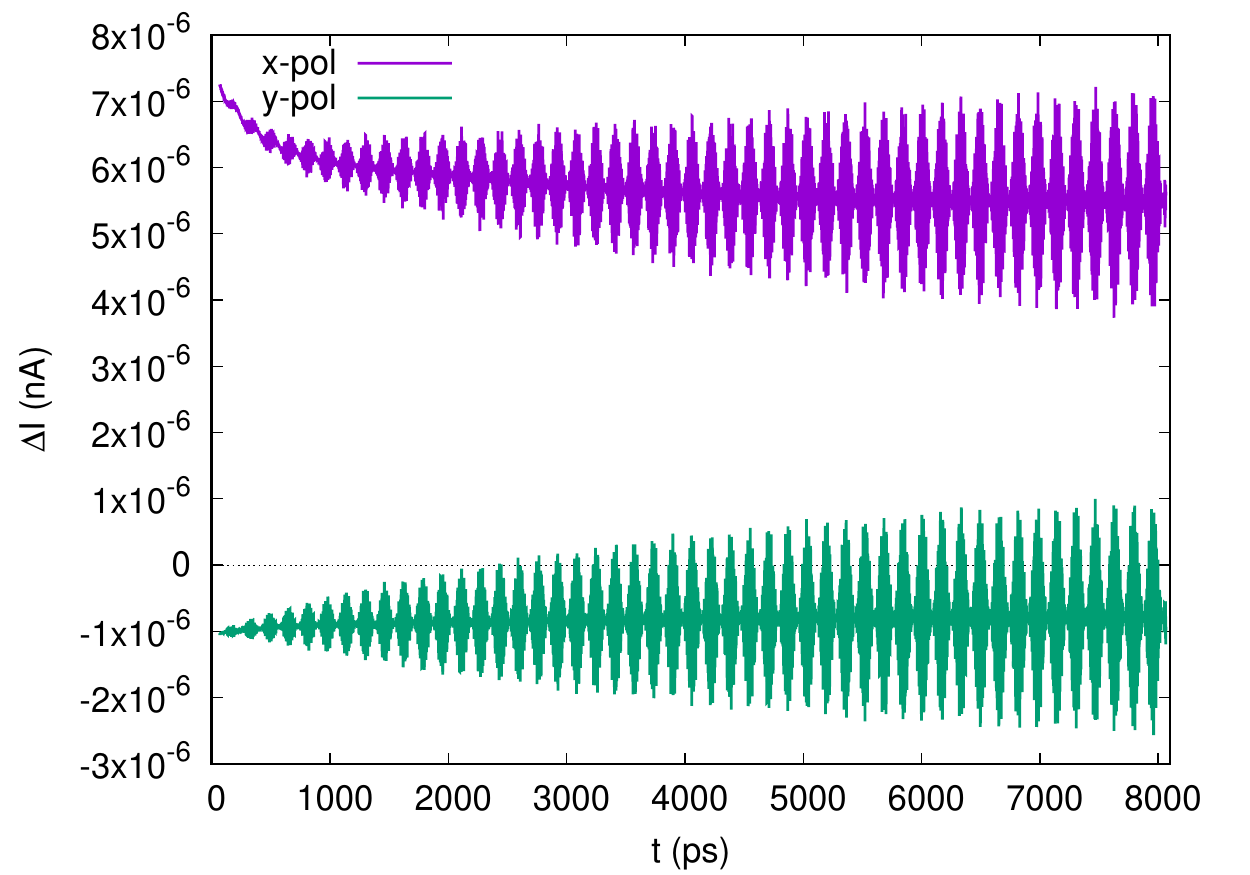}}
	\caption{The difference in the mean electron current $I=I_\mathrm{L}+I_\mathrm{R}$ for $x$- and 
	         $y$-polarized cavity fields, for a system
	         with an exact, and a system with a dipole approximation for 
	         the electron-photon interactions for a single cavity mode.
	         The figure displays partial information from Fig.\ \ref{Fig09} 
	         on a linear $t$-scale with a higher resolution.
	         $\bar{n}_\mathrm{R}=0$. $g_\mathrm{EM}=0.1$ meV,
	         $g_\mathrm{0}=1.0$, $B=1.0$ T, $\hbar\omega = 2.63$ meV,
	         $\kappa = 1.0\times 10^{-5}$ meV, $-eV_\mathrm{g}=2.47$ meV,
	         and $L_x=180$ nm.}
	\label{Fig10}
\end{figure}

In the ITR irregularly looking oscillations appear.
We focus in on these oscillations in Fig.\ \ref{Fig10} for the overall
dimensionless coupling $g_0=1.0$, in order to show that with a higher resolution
on a linear scale the oscillations are totally regular.
Below, we will show that the oscillations are to the largest extent 
a beating pattern of Rabi-oscillations with other smaller oscillations
superimposed. It is important to remember here that the states in the 
bias window are not directly participating in a Rabi-resonance for 
$\hbar\omega = 2.63$ meV.

In a simple classical serial RCL circuit we would expect the appearance
of induction always to lead to a reduction of the total current 
through it. Here, we see that the quantum counterparts, the higher order
electron-photon interaction terms, can either lead to a reduction or 
an increased current through the system.

\begin{figure}[htb]
	\centerline{\includegraphics[width=0.48\textwidth]{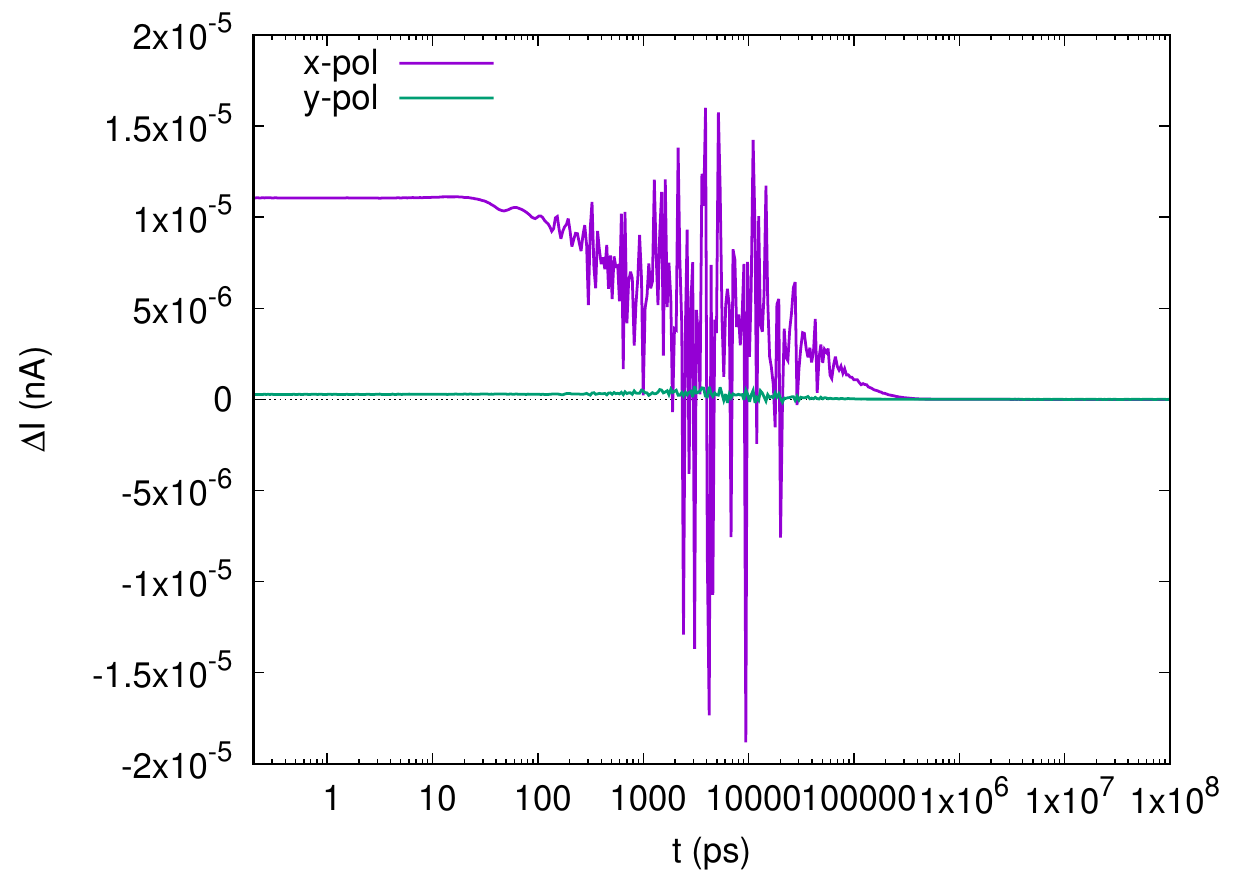}}
	\centerline{\includegraphics[width=0.48\textwidth]{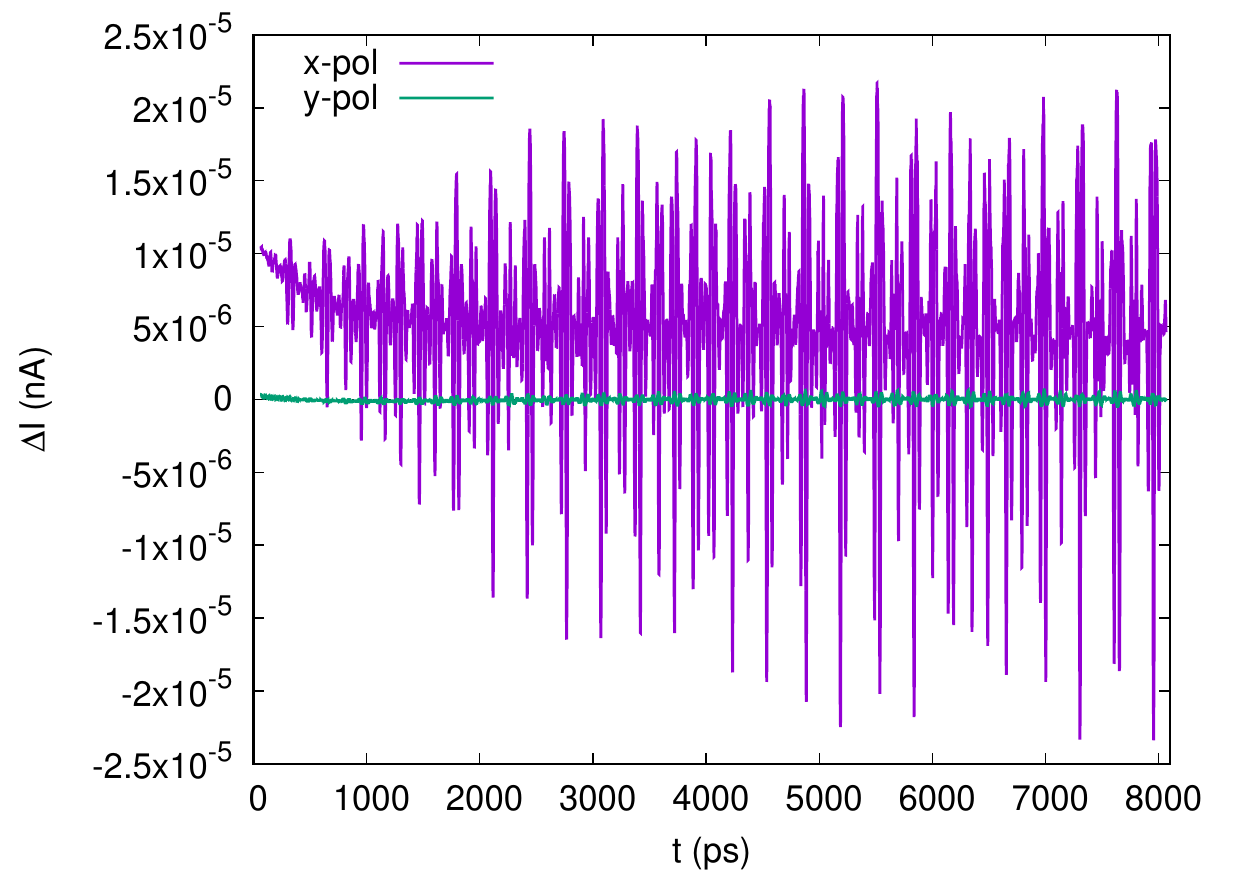}}
	\caption{The difference in the mean electron current $I=I_\mathrm{L}+I_\mathrm{R}$ for $x$- and 
	         $y$-polarized cavity fields, for a system
	         with an exact, and a system with a dipole approximation for 
	         the electron-photon interactions for a single cavity mode. 
	         $\bar{n}_\mathrm{R}=0$. $g_\mathrm{EM}=0.1$ meV,
	         $g_\mathrm{0}=1.0$, $B=1.0$ T, $\hbar\omega = 0.98$ meV,
	         $\kappa = 1.0\times 10^{-5}$ meV, $-eV_\mathrm{g}=2.47$ meV,
	         and $L_x=180$ nm.}
	\label{Fig11}
\end{figure}

The change in the current $\Delta I$ for the lower photon energy $\hbar\omega = 0.98$ meV
is displayed in Fig.\ \ref{Fig11}. On the scale appropriate for $\Delta I$ in the case
of $x$-polarization of the cavity field $\Delta I$ for the $y$-polarization is vanishingly
small.
The mean change $\Delta I$ for the $x$-polarized field is positive as before,
but a complex pattern of Rabi-oscillations emerges at the ITR, that will be analyzed below. 

\begin{figure}[htb]
	\centerline{\includegraphics[width=0.48\textwidth]{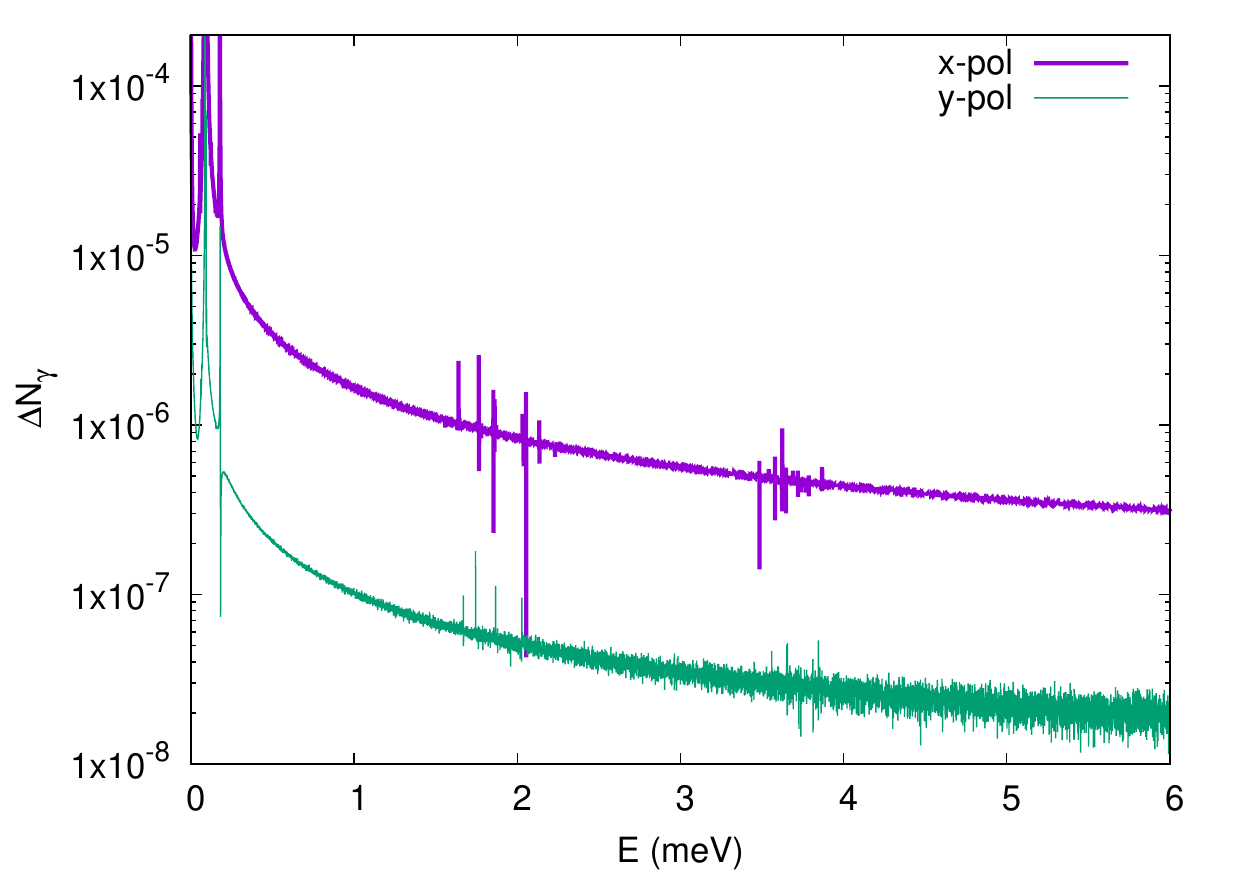}}
	\centerline{\includegraphics[width=0.48\textwidth]{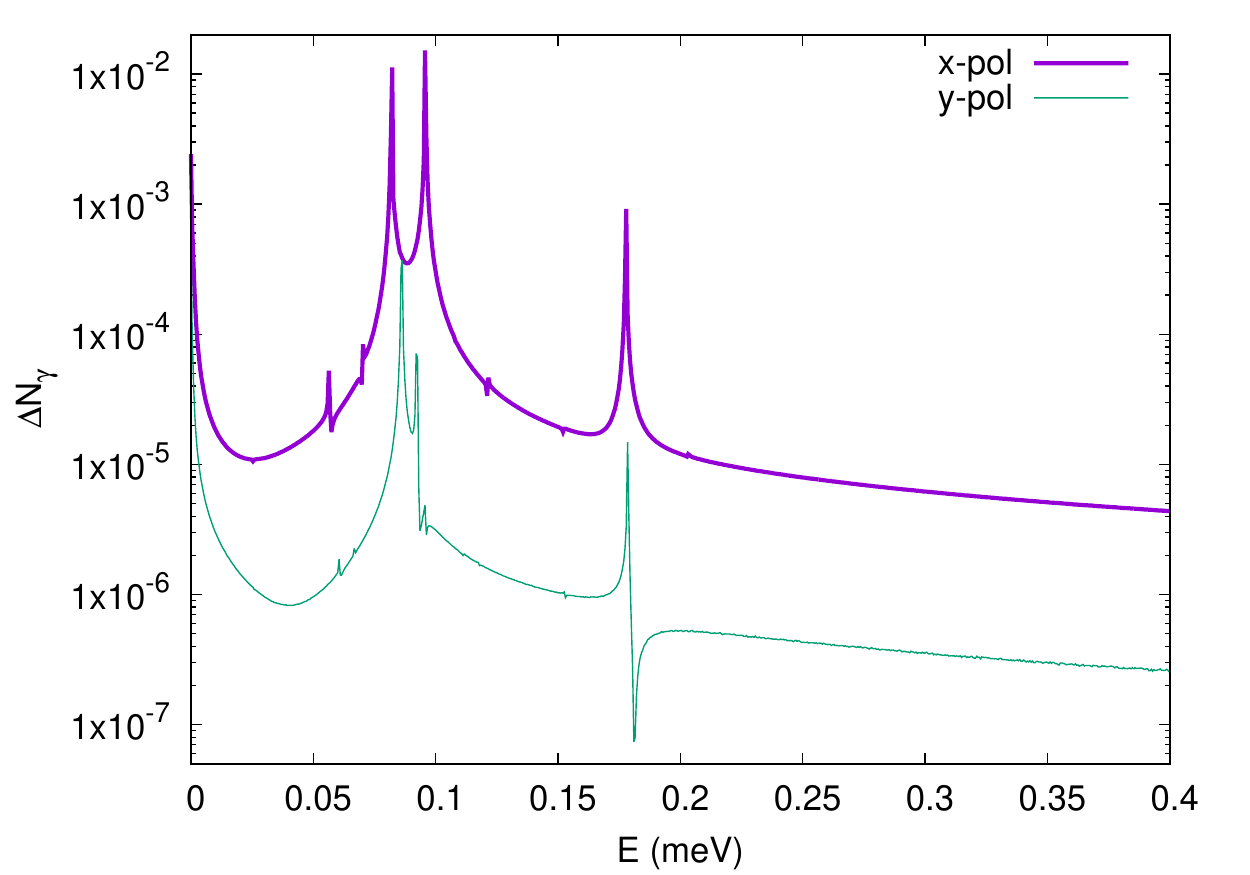}}
	\caption{The Fourier power spectrum for the difference in the mean photon
	         number for a system with an exact, and a system with a dipole 
	         approximation for the electron-photon interactions for a single 
	         cavity mode. The two panels show different energy intervals. 
	         $\hbar\omega = 0.98$ meV, $g_\mathrm{EM}=0.1$ meV, $g_\mathrm{0}=1.0$,
	         and $g_\mathrm{0}g_\mathrm{LR}a_w^{3/2}=0.101$ meV.}
	\label{Fig12}
\end{figure}
In the two remaining figures we show the Fourier power spectrum for 
the oscillations in the difference of the mean photon number, $\Delta N_\gamma$
for the two different photon energies, starting with $\hbar\omega = 0.98$ meV in
Fig.\ \ref{Fig12}. The Fourier transform is taken using 40000 points equispaced
in the time interval from 65.82 -- 8065.45 ps. In the upper panel of Fig.\ \ref{Fig12}
the whole energy range from 0 -- 6.0 meV is seen and clearly there is an order of
magnitude between the two spectra for the $x$- and the $y$-polarization of the 
cavity field. Numerical noise is visible, specially for the highest energies in the 
$y$-polarization. The lower panel of Fig.\ \ref{Fig12} presents the low energy
range of the same spectra with three main peaks caused by the Rabi-splitting of the 
states in the bias window. 

\begin{figure}[htb]
	\centerline{\includegraphics[width=0.48\textwidth]{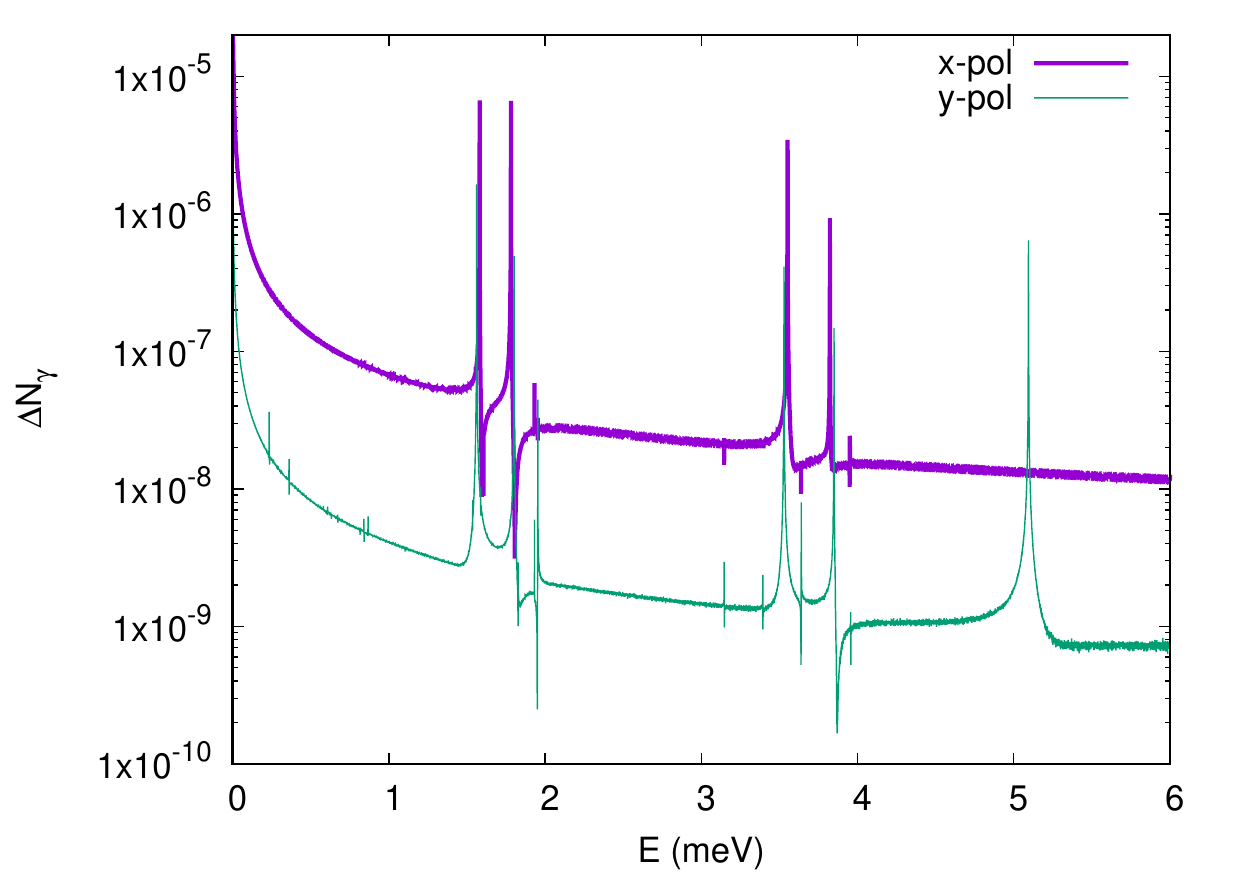}}
	\centerline{\includegraphics[width=0.48\textwidth]{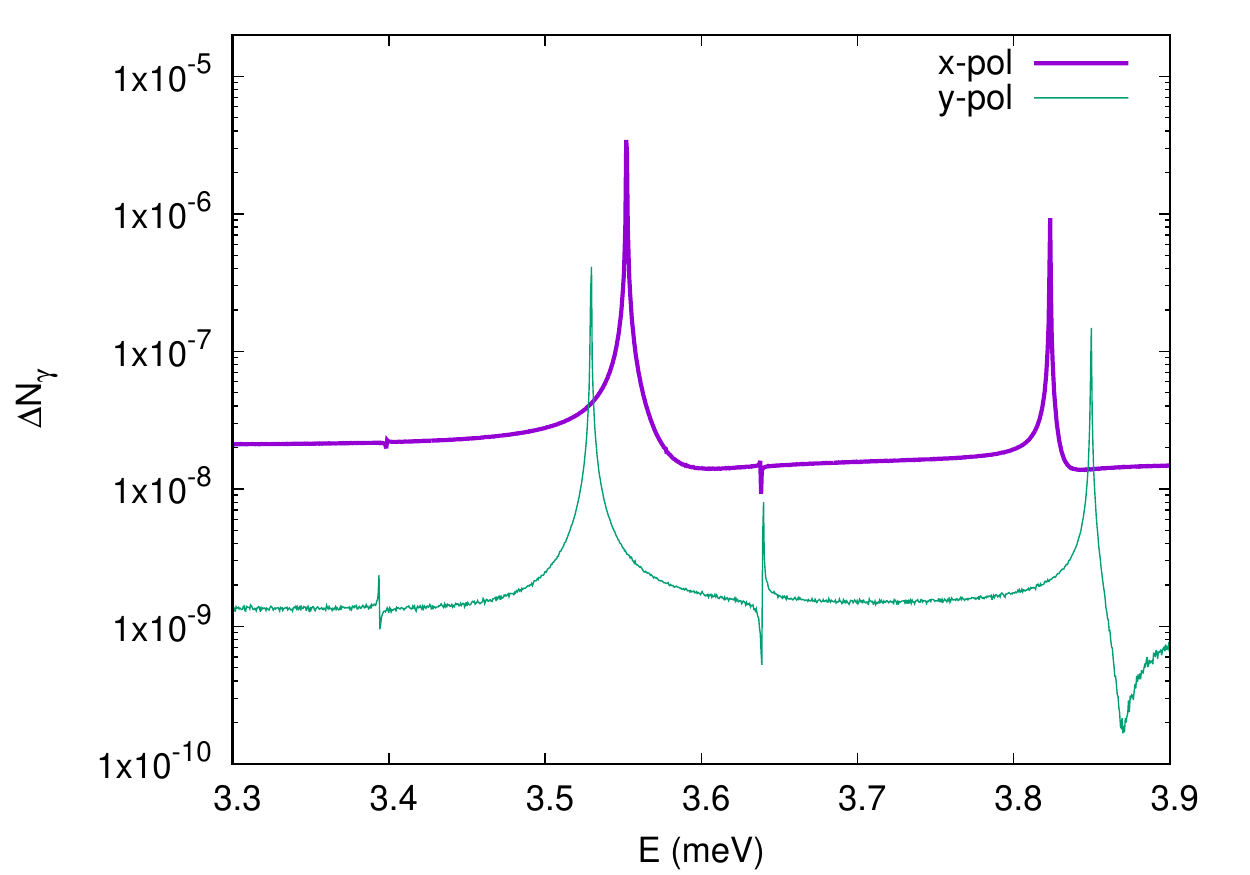}}
	\caption{The Fourier power spectrum for the difference in the mean photon
	         number for a system with an exact, and a system with a dipole 
	         approximation for the electron-photon interactions for a single 
	         cavity mode. The two panels show different energy intervals. 
	         $\hbar\omega = 2.63$ meV, $g_\mathrm{EM}=0.1$ meV, $g_\mathrm{0}=1.0$,
	         and $g_\mathrm{0}g_\mathrm{LR}a_w^{3/2}=0.101$ meV.}
	\label{Fig13}
\end{figure}
In the upper panel of Fig.\ \ref{Fig12} there are a small groups of peaks
just below 2.0 meV, and again just below 4.0 meV. These peaks can all be
correlated to active transitions between one-electron states in the system.
It is interesting to note that from the energy spectra shown in 
the left panels of Fig.\ \ref{Fig03} for $\hbar\omega = 0.98$ meV there are
clear differences in energy, and specially in the mean photon number of
many states for the two different polarizations of the cavity field.
This difference is minimal for the spectra for $\hbar\omega = 2.63$ meV seen
in the right panels of Fig.\ \ref{Fig03}, but the Fourier power spectra for 
$\Delta N_\gamma$ seen in Fig.\ \ref{Fig13} again shows the order of magnitude
between the spectra for the two polarizations.

Here are no Rabi-split states in the bias window and thus no low-energy 
Rabi-resonances are seen, instead we notice pairs of resonances emerging
corresponding to the Rabi-split states above the bias window, indicating
that these states do participate, even though very weakly, in the electron
transport through the central system. Here, we have analyzed the difference in
the mean photon number $\Delta N_\gamma$, similar analysis for the difference
in the current $\Delta I$ reveals corresponding information. The resonances
in $\Delta I$ are weaker, but in addition we then see peaks caused by transitions
between dressed electron states with only a minimal photon content.  

\section{Conclusions}
\label{Conclusions}
As expected, the dipole approximation for the electron-photon interaction 
together with the lowest order diamagnetic terms, is a good approximation
to describe the interactions of electrons and photons in nanoscale systems
embedded in a cavity.

In our central system with discrete energy spectrum and few particles far away from 
the possibility of a semi-classical interpretation
all correspondence to classical self-induction in simple circuits is lost.

Another difficulty comes from the fact that we consider a photon cavity
with a single photon mode, one fundamental frequency. Anybody using a Laplace transformation
to analyze a classical circuit driven by a step potential has noticed that the results 
always imply a spread of frequencies producing the inductance effects.
Nevertheless, the question about the self-induction is interesting in order to acquire 
a more complete understanding of the quantum transport of electrons through photon cavities, 
and a cavity leads to a dominance of their fundamental frequency in the process, as
only processes with the fundamental frequency of the cavity or integer multiples thereof
will form a standing wave (a cavity mode) with high intensity in the central system.

Our modeling of the electron transport through the central system, a nanoscale
electron system embedded in a three-dimensional photon cavity with a single FIR 
photon mode relies on a linear many-body space of photon-dressed electron 
states $\left\{|\breve{\mu})\right\}$. These states have been constructed using a
step-wise numerically exact diagonalization and truncations for all the interactions
present in the isolated central system \cite{Gudmundsson:2013.305}. By keeping this
space large enough we have used it to describe the dynamical evolution of the system
as it is coupled to the external leads with different chemical potentials. 
This evolution takes place in the many-body Fock-space $\left\{|\breve{\mu})\right\}$, or 
since we consider only the Markovian evolution of the system (with respect to the coupling
to the leads) is achieved by mapping the original non-Markovian master equation to a
larger Liouville space of transitions \cite{2016arXiv161003223J}. As the central system
has a certain geometry this approach includes higher-order wavefunction effects, i.\ e.\ the states
contain the information of polarization of their charge or probability densities, and in 
our experience, exactly this requirement puts the most difficult constrains on the size
needed for our basis used. 

The higher order effects of the electron-photon interactions stemming from the interaction 
of the electrons with their own photon field are thus included in the self-energy of the 
many-body space and their compositional changes (by admixing of other states due to the
interactions). These changes, in terms, lead to changes
in the dynamical evolution of the system, i.\ e.\ when and how the electrons transit 
between the states of the system having different self-energies.

The differences in the transport properties are caused by high-order terms in the 
electron-photon interactions. The strongest terms contributing the most are terms 
with electrical quadrupole and magnetic dipole momenta, but we can not separate
the two contributions. 

The approach using exact numerical diagonalizations, or configuration interactions, is
essential here, as this view is not easily extensible to formalisms built on mean-field theories,
in which one has to guarantee self-consistence at each time step in the calculation.
This view underlines the difference between the description of a classical circuit
with an inductance and the quantum mechanical one. If the first reaction is that the 
quantum mechanical view may seem more boring, then one has to appreciate what the 
quantum mechanical view is really offering. Our calculations indicates that if the full geometry 
of the system is taken into count, we may not be able to guess what small changes the 
higher-order effects may lead to, without a full calculation. Here, we have already seen that
even the simple question if the higher order terms will increase or decrease the current 
through the system is not simple in the quantum mechanical sense. In a direct continuation 
we are not able to predict what may happen in systems with higher electron-photon coupling
and different geometries, even though the effects seen here are by no means large.

Our calculation have been performed for a relatively small system in order to conserve
the RAM-memory size needed, but the inductance of the electronic system can be made larger
by increasing its size, as there is space enough in the cavity. Increased system size
would lead to the need to include more electrons in the system, and the evolution
of the results with the system size should point out the path to the porperties of the 
classical system.

\begin{acknowledgments}
This work was financially supported by the Research Fund of the University
of Iceland, and the Icelandic Infrastructure Fund. The computations were performed on resources
provided by the Icelandic High Performance Computing Centre at the University of Iceland. 
V.M.\ acknowledges financial support from the Romanian Core Program PN19-03
(contract No.\ 21 N/08.02.2019).
\end{acknowledgments}

%
\appendix

\section{Information about the numerical computations}
\label{AppNumDetails}
We use a scheme of step-wise introduction of model complexities and a step-wise 
truncation of the ensuing many-body spaces to guarantee the accuracy of the 
calculations and to contain the RAM memory needed \cite{Gudmundsson:2013.305}.
Initially, a large single-electron basis \VG{$\{|a\rangle\}$} with 2688 elements is made for the central
system, out of which the lowest 53 in energy are used to build a Fock space with
1 0-electron, 52 1-electron, 1326 2-electron, and 16 3-electron states. All together
1395 Fock many-body eigenstates for the noninteracting central system in an external
magnetic field. This eigenbasis is used to diagonalize the many-body Hamiltonian
of the electrons with their mutual Coulomb interaction. The lowest in energy 512 
eigenstates in this basis are then tensor multiplied by the 17 lowest eigenstates
of the photon number operator to form a basis with 8704 elements that is used to
diagonalize the Hamilton matrix including the electron-photon interactions,
Eq.\ \ref{H-e-EM-q}.
Of the resulting eigenstates $|\breve{\mu})$ for the fully interacting central system, 
the lowest in energy 128 states are used to build the 16384-dimensional Liouville 
basis of transitions in which the Markovian version of the master equation is solved. 
The rotating wave 
approximation is not used for the electron-photon interaction in the central system,
but it is used when deriving the dissipation terms for the coupling of the cavity
to the external photon reservoir. When constructing these therms special care has been
taken as they have to be transformed from the basis of the non-interacting photons to 
the basis of interacting electrons and 
photons \cite{PhysRev.129.2342,PhysRevA.31.3761,PhysRevA.80.053810,PhysRevA.75.013811,PhysRevA.84.043832}. 
In the Schr{\"o}dinger picture used here this is implemented by dismissing all
creation terms in the transformed annihilation operators and all
annihilation terms in the transformed creation operators. This
guarantees that an open system will evolve into the correct
physical steady state with respect to the photon decay \cite{Gudmundsson19:10}.
The processing time is reduced by a general use of parallelism and a heavy use of off-loading 
large linear algebra tasks to fast GPU cards \cite{2016arXiv161003223J}.

\section{Changes in the self-energy of states and their occupation}
\label{AppA}
\begin{figure}[htb]
	\centerline{\includegraphics[width=0.48\textwidth]{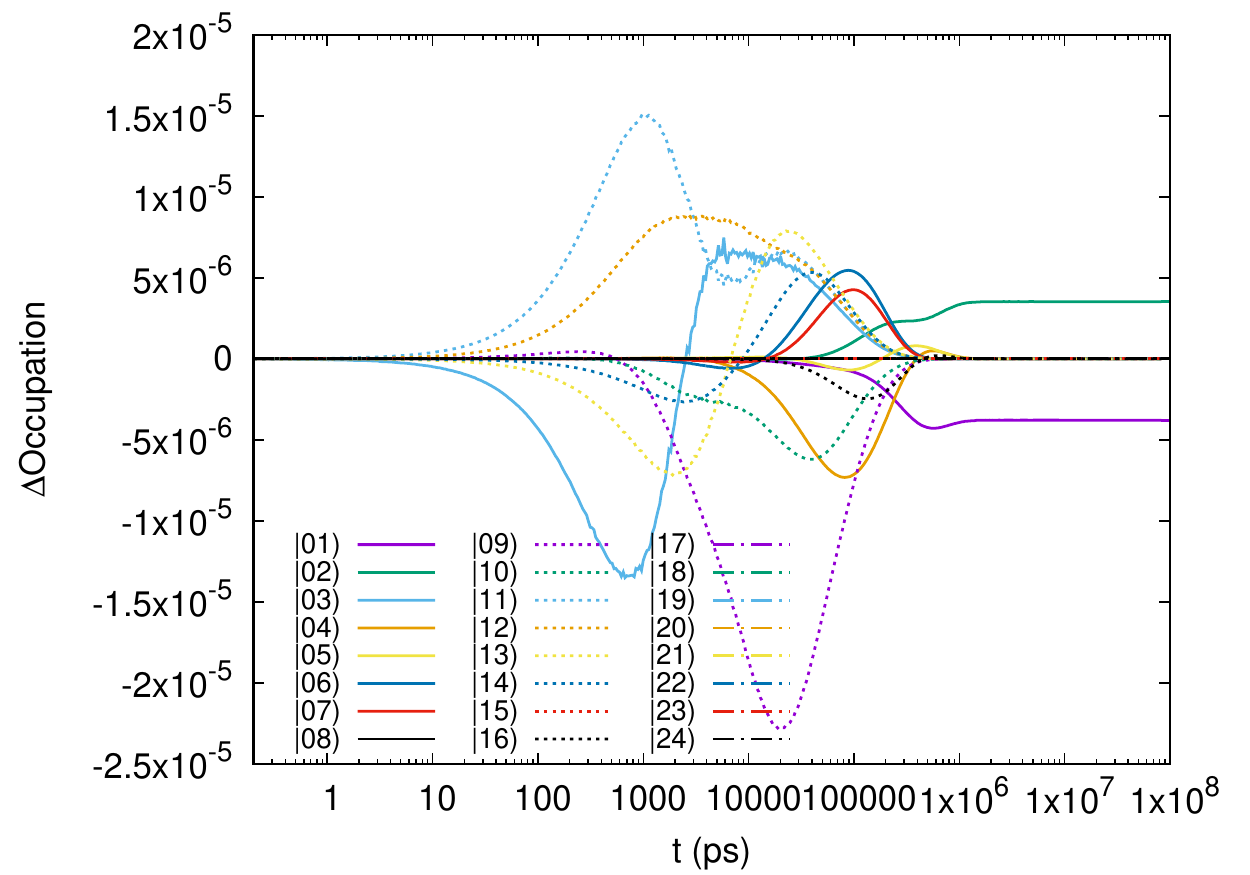}}
	\centerline{\includegraphics[width=0.48\textwidth]{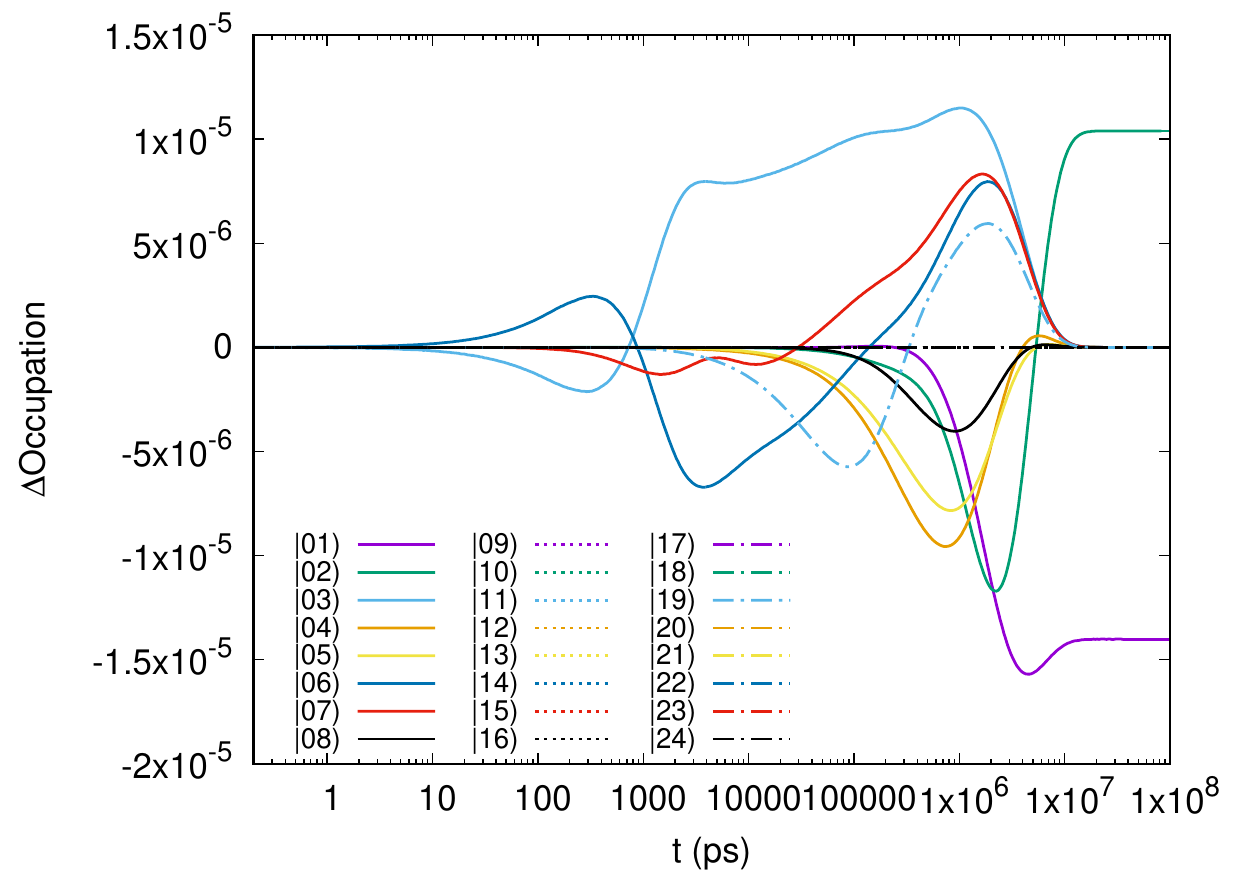}}
	\caption{The difference in the time-dependent occupation of selected many-body 
	         states $|\breve{\mu})$ for a system
	         with an exact, and a system with a dipole approximation for 
	         the electron-photon interactions for a single cavity mode,
	         for $\hbar\omega = 0.98$ meV (upper), and $\hbar\omega = 2.63$ meV (lower).
             $g_\mathrm{EM}=0.1$ meV, $B=1.0$ T, $\bar{n}_\mathrm{R}=0$,
             $-eV_\mathrm{g}=2.47$ meV, and $L_x=180$ nm, and 
             $g_\mathrm{0}g_\mathrm{LR}a_w^{3/2}=0.101$ meV.}
	\label{Fig14}
\end{figure}
The electron transport through the central system is modeled using a 
quantum many-body formalism, for the electrons in the central system,
the single-photon mode of the cavity electromagnetic field, and the 
electrons in the external leads. The search for self-induction and 
similar higher-order effects where an electron interacts with its own
electromagnetic field is carried out in a model based on linear many-body spaces, 
instead of grids, we thus always have to consider the self-energy of
the cavity photon dressed electron states. To aid this exploration 
we present here Fig.\ \ref{Fig14} showing the changes in the dynamical
occupation of the lowest many-body states of the central system.

\begin{figure}[htb]
	\centerline{\includegraphics[width=0.48\textwidth]{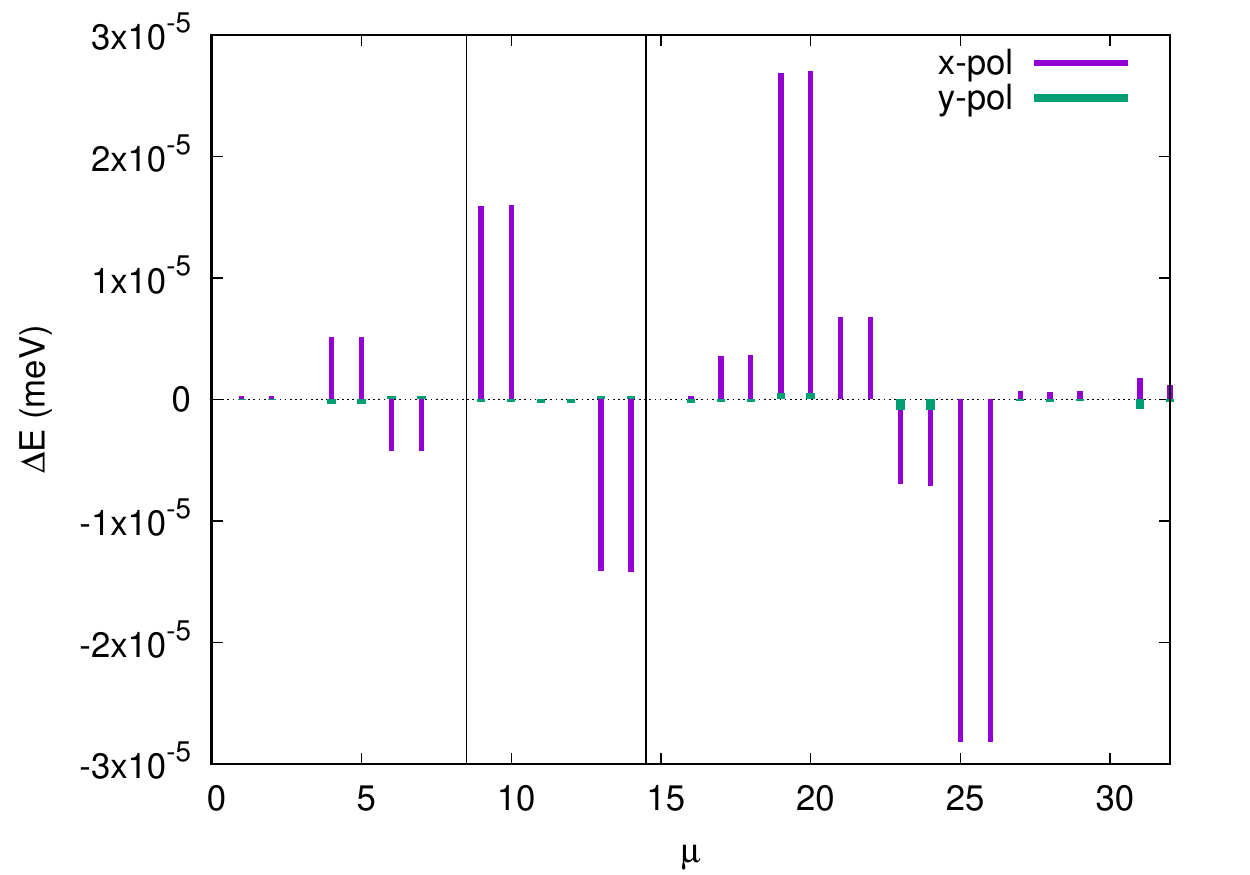}}
	\centerline{\includegraphics[width=0.48\textwidth]{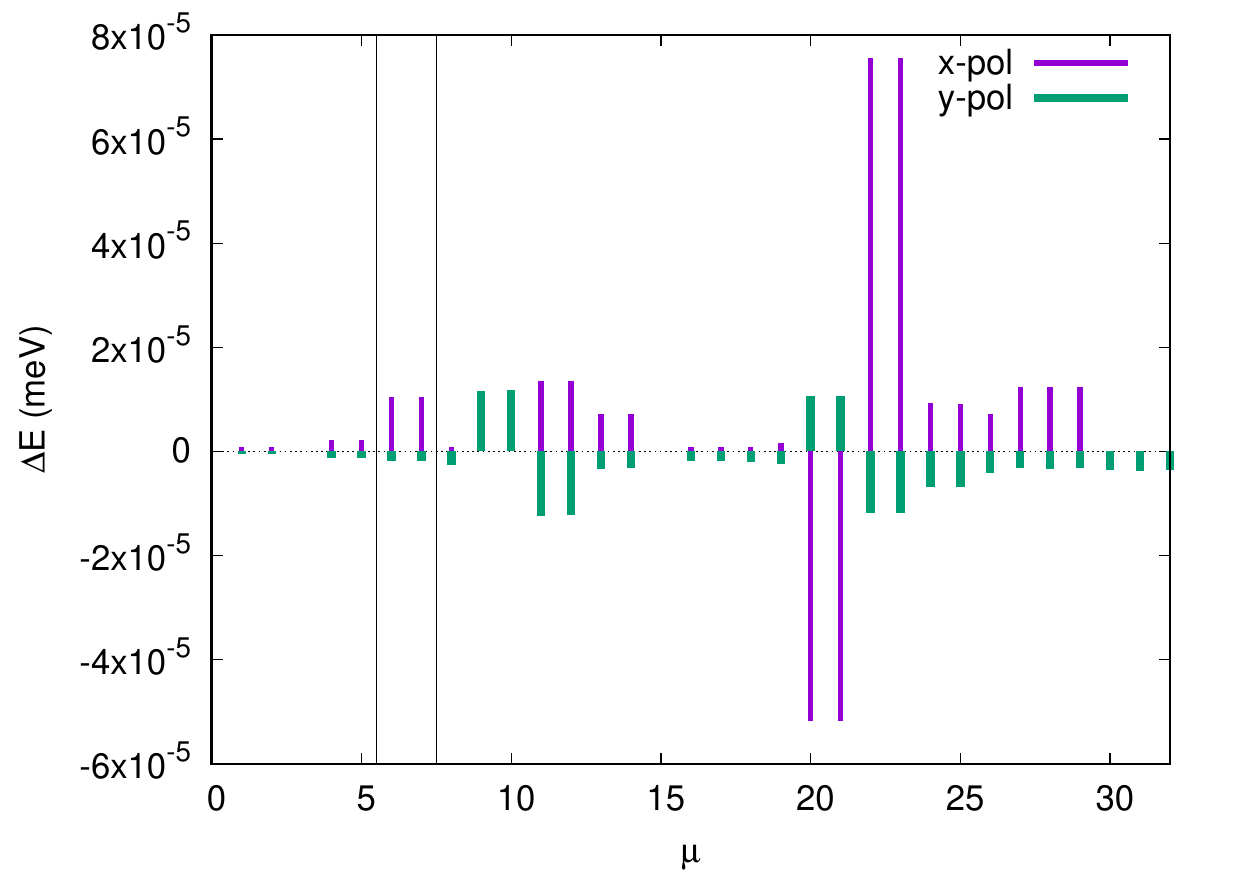}}
	\caption{The difference in the self-energy of the 32 lowest many-body 
	         states $|\breve{\mu})$ for a system
	         with an exact, and a system with a dipole approximation for 
	         the electron-photon interactions for a single cavity mode,
	         for $\hbar\omega = 0.98$ meV (upper), and $\hbar\omega = 2.63$ meV (lower).
	         The states within the two vertical black lines are the 
	         states in the bias window.
             $g_\mathrm{EM}=0.1$ meV, $B=1.0$ T, $\bar{n}_\mathrm{R}=0$,
             $-eV_\mathrm{g}=2.47$ meV, and $L_x=180$ nm, and 
             $g_\mathrm{0}g_\mathrm{LR}a_w^{3/2}=0.101$ meV.}
	\label{Fig15}
\end{figure}
In addition, we find it necessary to display in Fig.\ \ref{Fig15} the underlying
changes in the self-energy of each many-body state $|\breve{\mu})$ caused by all 
higher-order terms of the electron-photon interactions.
In Fig.\ \ref{Fig15} we show which states are within the bias window for both 
energies of the cavity photons we have chosen. Note the differences in the energy
scales in Fig.\ \ref{Fig15}.

%
\frenchspacing

\begin{thebibliography}{53}%
\makeatletter
\providecommand \@ifxundefined [1]{%
 \@ifx{#1\undefined}
}%
\providecommand \@ifnum [1]{%
 \ifnum #1\expandafter \@firstoftwo
 \else \expandafter \@secondoftwo
 \fi
}%
\providecommand \@ifx [1]{%
 \ifx #1\expandafter \@firstoftwo
 \else \expandafter \@secondoftwo
 \fi
}%
\providecommand \natexlab [1]{#1}%
\providecommand \enquote  [1]{``#1''}%
\providecommand \bibnamefont  [1]{#1}%
\providecommand \bibfnamefont [1]{#1}%
\providecommand \citenamefont [1]{#1}%
\providecommand \href@noop [0]{\@secondoftwo}%
\providecommand \href [0]{\begingroup \@sanitize@url \@href}%
\providecommand \@href[1]{\@@startlink{#1}\@@href}%
\providecommand \@@href[1]{\endgroup#1\@@endlink}%
\providecommand \@sanitize@url [0]{\catcode `\\12\catcode `\$12\catcode
  `\&12\catcode `\#12\catcode `\^12\catcode `\_12\catcode `\%12\relax}%
\providecommand \@@startlink[1]{}%
\providecommand \@@endlink[0]{}%
\providecommand \url  [0]{\begingroup\@sanitize@url \@url }%
\providecommand \@url [1]{\endgroup\@href {#1}{\urlprefix }}%
\providecommand \urlprefix  [0]{URL }%
\providecommand \Eprint [0]{\href }%
\providecommand \doibase [0]{https://doi.org/}%
\providecommand \selectlanguage [0]{\@gobble}%
\providecommand \bibinfo  [0]{\@secondoftwo}%
\providecommand \bibfield  [0]{\@secondoftwo}%
\providecommand \translation [1]{[#1]}%
\providecommand \BibitemOpen [0]{}%
\providecommand \bibitemStop [0]{}%
\providecommand \bibitemNoStop [0]{.\EOS\space}%
\providecommand \EOS [0]{\spacefactor3000\relax}%
\providecommand \BibitemShut  [1]{\csname bibitem#1\endcsname}%
\let\auto@bib@innerbib\@empty
\bibitem [{\citenamefont {Zhang}\ \emph {et~al.}(2016)\citenamefont {Zhang},
  \citenamefont {Lou}, \citenamefont {Li}, \citenamefont {Reno}, \citenamefont
  {Pan}, \citenamefont {Watson}, \citenamefont {Manfra},\ and\ \citenamefont
  {Kono}}]{Zhang1005:2016}%
  \BibitemOpen
  \bibfield  {author} {\bibinfo {author} {\bibfnamefont {Q.}~\bibnamefont
  {Zhang}}, \bibinfo {author} {\bibfnamefont {M.}~\bibnamefont {Lou}}, \bibinfo
  {author} {\bibfnamefont {X.}~\bibnamefont {Li}}, \bibinfo {author}
  {\bibfnamefont {J.~L.}\ \bibnamefont {Reno}}, \bibinfo {author}
  {\bibfnamefont {W.}~\bibnamefont {Pan}}, \bibinfo {author} {\bibfnamefont
  {J.~D.}\ \bibnamefont {Watson}}, \bibinfo {author} {\bibfnamefont {M.~J.}\
  \bibnamefont {Manfra}},\ and\ \bibinfo {author} {\bibfnamefont
  {J.}~\bibnamefont {Kono}},\ }\bibfield  {title} {\bibinfo {title}
  {{Collective non-perturbative coupling of 2D electrons with
  high-quality-factor terahertz cavity photons}},\ }\href
  {https://doi.org/10.1038/nphys3850; 10.1038/nphys3850} {\bibfield  {journal}
  {\bibinfo  {journal} {Nature Physics}\ }\textbf {\bibinfo {volume} {12}},\
  \bibinfo {pages} {1005} (\bibinfo {year} {2016})}\BibitemShut {NoStop}%
\bibitem [{\citenamefont {Bruhat}\ \emph {et~al.}(2016)\citenamefont {Bruhat},
  \citenamefont {Viennot}, \citenamefont {Dartiailh}, \citenamefont
  {Desjardins}, \citenamefont {Kontos},\ and\ \citenamefont
  {Cottet}}]{PhysRevX.6.021014}%
  \BibitemOpen
  \bibfield  {author} {\bibinfo {author} {\bibfnamefont {L.~E.}\ \bibnamefont
  {Bruhat}}, \bibinfo {author} {\bibfnamefont {J.~J.}\ \bibnamefont {Viennot}},
  \bibinfo {author} {\bibfnamefont {M.~C.}\ \bibnamefont {Dartiailh}}, \bibinfo
  {author} {\bibfnamefont {M.~M.}\ \bibnamefont {Desjardins}}, \bibinfo
  {author} {\bibfnamefont {T.}~\bibnamefont {Kontos}},\ and\ \bibinfo {author}
  {\bibfnamefont {A.}~\bibnamefont {Cottet}},\ }\bibfield  {title} {\bibinfo
  {title} {{Cavity Photons as a Probe for Charge Relaxation Resistance and
  Photon Emission in a Quantum Dot Coupled to Normal and Superconducting
  Continua}},\ }\href {https://doi.org/10.1103/PhysRevX.6.021014} {\bibfield
  {journal} {\bibinfo  {journal} {Phys. Rev. X}\ }\textbf {\bibinfo {volume}
  {6}},\ \bibinfo {pages} {021014} (\bibinfo {year} {2016})}\BibitemShut
  {NoStop}%
\bibitem [{\citenamefont {Cottet}\ \emph {et~al.}(2017)\citenamefont {Cottet},
  \citenamefont {Dartiailh}, \citenamefont {Desjardins}, \citenamefont
  {Cubaynes}, \citenamefont {Contamin}, \citenamefont {Delbecq}, \citenamefont
  {Viennot}, \citenamefont {Bruhat}, \citenamefont {Dou{\c c}ot},\ and\
  \citenamefont {Kontos}}]{0953-8984-29-43-433002}%
  \BibitemOpen
  \bibfield  {author} {\bibinfo {author} {\bibfnamefont {A.}~\bibnamefont
  {Cottet}}, \bibinfo {author} {\bibfnamefont {M.~C.}\ \bibnamefont
  {Dartiailh}}, \bibinfo {author} {\bibfnamefont {M.~M.}\ \bibnamefont
  {Desjardins}}, \bibinfo {author} {\bibfnamefont {T.}~\bibnamefont
  {Cubaynes}}, \bibinfo {author} {\bibfnamefont {L.~C.}\ \bibnamefont
  {Contamin}}, \bibinfo {author} {\bibfnamefont {M.}~\bibnamefont {Delbecq}},
  \bibinfo {author} {\bibfnamefont {J.~J.}\ \bibnamefont {Viennot}}, \bibinfo
  {author} {\bibfnamefont {L.~E.}\ \bibnamefont {Bruhat}}, \bibinfo {author}
  {\bibfnamefont {B.}~\bibnamefont {Dou{\c c}ot}},\ and\ \bibinfo {author}
  {\bibfnamefont {T.}~\bibnamefont {Kontos}},\ }\bibfield  {title} {\bibinfo
  {title} {{Cavity QED with hybrid nanocircuits: from atomic-like physics to
  condensed matter phenomena}},\ }\href
  {http://stacks.iop.org/0953-8984/29/i=43/a=433002} {\bibfield  {journal}
  {\bibinfo  {journal} {Journal of Physics: Condensed Matter}\ }\textbf
  {\bibinfo {volume} {29}},\ \bibinfo {pages} {433002} (\bibinfo {year}
  {2017})}\BibitemShut {NoStop}%
\bibitem [{\citenamefont {Delbecq}\ \emph {et~al.}(2011)\citenamefont
  {Delbecq}, \citenamefont {Schmitt}, \citenamefont {Parmentier}, \citenamefont
  {Roch}, \citenamefont {Viennot}, \citenamefont {F{\`e}ve}, \citenamefont
  {Huard}, \citenamefont {Mora}, \citenamefont {Cottet},\ and\ \citenamefont
  {Kontos}}]{Delbecq11:01}%
  \BibitemOpen
  \bibfield  {author} {\bibinfo {author} {\bibfnamefont {M.}~\bibnamefont
  {Delbecq}}, \bibinfo {author} {\bibfnamefont {V.}~\bibnamefont {Schmitt}},
  \bibinfo {author} {\bibfnamefont {F.}~\bibnamefont {Parmentier}}, \bibinfo
  {author} {\bibfnamefont {N.}~\bibnamefont {Roch}}, \bibinfo {author}
  {\bibfnamefont {J.}~\bibnamefont {Viennot}}, \bibinfo {author} {\bibfnamefont
  {G.}~\bibnamefont {F{\`e}ve}}, \bibinfo {author} {\bibfnamefont
  {B.}~\bibnamefont {Huard}}, \bibinfo {author} {\bibfnamefont
  {C.}~\bibnamefont {Mora}}, \bibinfo {author} {\bibfnamefont {A.}~\bibnamefont
  {Cottet}},\ and\ \bibinfo {author} {\bibfnamefont {T.}~\bibnamefont
  {Kontos}},\ }\bibfield  {title} {\bibinfo {title} {{Coupling a quantum dot,
  fermionic leads and a microwave cavity on-chip}},\ }\href
  {https://doi.org/10.1103/PhysRevLett.107.256804} {\bibfield  {journal}
  {\bibinfo  {journal} {Phys. Rev. Lett.}\ }\textbf {\bibinfo {volume} {107}},\
  \bibinfo {pages} {256804} (\bibinfo {year} {2011})}\BibitemShut {NoStop}%
\bibitem [{\citenamefont {{Liu}}\ \emph {et~al.}(2017)\citenamefont {{Liu}},
  \citenamefont {{Stehlik}}, \citenamefont {{Eichler}}, \citenamefont {{Mi}},
  \citenamefont {{Hartke}}, \citenamefont {{Gullans}}, \citenamefont
  {{Taylor}},\ and\ \citenamefont {{Petta}}}]{2017arXiv170401961L}%
  \BibitemOpen
  \bibfield  {author} {\bibinfo {author} {\bibfnamefont {Y.-Y.}\ \bibnamefont
  {{Liu}}}, \bibinfo {author} {\bibfnamefont {J.}~\bibnamefont {{Stehlik}}},
  \bibinfo {author} {\bibfnamefont {C.}~\bibnamefont {{Eichler}}}, \bibinfo
  {author} {\bibfnamefont {X.}~\bibnamefont {{Mi}}}, \bibinfo {author}
  {\bibfnamefont {T.}~\bibnamefont {{Hartke}}}, \bibinfo {author}
  {\bibfnamefont {M.~J.}\ \bibnamefont {{Gullans}}}, \bibinfo {author}
  {\bibfnamefont {J.~M.}\ \bibnamefont {{Taylor}}},\ and\ \bibinfo {author}
  {\bibfnamefont {J.~R.}\ \bibnamefont {{Petta}}},\ }\bibfield  {title}
  {\bibinfo {title} {{Threshold Dynamics of a Semiconductor Single Atom
  Maser}},\ }\href@noop {} {\bibfield  {journal} {\bibinfo  {journal} {Phys.
  Rev. Lett.}\ }\textbf {\bibinfo {volume} {119}},\ \bibinfo {pages} {097702}
  (\bibinfo {year} {2017})},\ \Eprint {https://arxiv.org/abs/1704.01961}
  {arXiv:1704.01961 [cond-mat.mes-hall]} \BibitemShut {NoStop}%
\bibitem [{\citenamefont {Stockklauser}\ \emph {et~al.}(2017)\citenamefont
  {Stockklauser}, \citenamefont {Scarlino}, \citenamefont {Koski},
  \citenamefont {Gasparinetti}, \citenamefont {Andersen}, \citenamefont
  {Reichl}, \citenamefont {Wegscheider}, \citenamefont {Ihn}, \citenamefont
  {Ensslin},\ and\ \citenamefont {Wallraff}}]{PhysRevX.7.011030}%
  \BibitemOpen
  \bibfield  {author} {\bibinfo {author} {\bibfnamefont {A.}~\bibnamefont
  {Stockklauser}}, \bibinfo {author} {\bibfnamefont {P.}~\bibnamefont
  {Scarlino}}, \bibinfo {author} {\bibfnamefont {J.~V.}\ \bibnamefont {Koski}},
  \bibinfo {author} {\bibfnamefont {S.}~\bibnamefont {Gasparinetti}}, \bibinfo
  {author} {\bibfnamefont {C.~K.}\ \bibnamefont {Andersen}}, \bibinfo {author}
  {\bibfnamefont {C.}~\bibnamefont {Reichl}}, \bibinfo {author} {\bibfnamefont
  {W.}~\bibnamefont {Wegscheider}}, \bibinfo {author} {\bibfnamefont
  {T.}~\bibnamefont {Ihn}}, \bibinfo {author} {\bibfnamefont {K.}~\bibnamefont
  {Ensslin}},\ and\ \bibinfo {author} {\bibfnamefont {A.}~\bibnamefont
  {Wallraff}},\ }\bibfield  {title} {\bibinfo {title} {{Strong Coupling Cavity
  QED with Gate-Defined Double Quantum Dots Enabled by a High Impedance
  Resonator}},\ }\href {https://doi.org/10.1103/PhysRevX.7.011030} {\bibfield
  {journal} {\bibinfo  {journal} {Phys. Rev. X}\ }\textbf {\bibinfo {volume}
  {7}},\ \bibinfo {pages} {011030} (\bibinfo {year} {2017})}\BibitemShut
  {NoStop}%
\bibitem [{\citenamefont {Cirio}\ \emph {et~al.}(2016)\citenamefont {Cirio},
  \citenamefont {{De Liberato}}, \citenamefont {Lambert},\ and\ \citenamefont
  {Nori}}]{PhysRevLett.116.113601}%
  \BibitemOpen
  \bibfield  {author} {\bibinfo {author} {\bibfnamefont {M.}~\bibnamefont
  {Cirio}}, \bibinfo {author} {\bibfnamefont {S.}~\bibnamefont {{De
  Liberato}}}, \bibinfo {author} {\bibfnamefont {N.}~\bibnamefont {Lambert}},\
  and\ \bibinfo {author} {\bibfnamefont {F.}~\bibnamefont {Nori}},\ }\bibfield
  {title} {\bibinfo {title} {{Ground State Electroluminescence}},\ }\href
  {https://doi.org/10.1103/PhysRevLett.116.113601} {\bibfield  {journal}
  {\bibinfo  {journal} {Phys. Rev. Lett.}\ }\textbf {\bibinfo {volume} {116}},\
  \bibinfo {pages} {113601} (\bibinfo {year} {2016})}\BibitemShut {NoStop}%
\bibitem [{\citenamefont {Yang}\ \emph {et~al.}(2015)\citenamefont {Yang},
  \citenamefont {Lin},\ and\ \citenamefont {Zhang}}]{PhysRevB.92.165403}%
  \BibitemOpen
  \bibfield  {author} {\bibinfo {author} {\bibfnamefont {P.-Y.}\ \bibnamefont
  {Yang}}, \bibinfo {author} {\bibfnamefont {C.-Y.}\ \bibnamefont {Lin}},\ and\
  \bibinfo {author} {\bibfnamefont {W.-M.}\ \bibnamefont {Zhang}},\ }\bibfield
  {title} {\bibinfo {title} {{Master equation approach to transient quantum
  transport in nanostructures incorporating initial correlations}},\ }\href
  {https://doi.org/10.1103/PhysRevB.92.165403} {\bibfield  {journal} {\bibinfo
  {journal} {Phys. Rev. B}\ }\textbf {\bibinfo {volume} {92}},\ \bibinfo
  {pages} {165403} (\bibinfo {year} {2015})}\BibitemShut {NoStop}%
\bibitem [{\citenamefont {{Gudmundsson}}\ \emph {et~al.}(2016)\citenamefont
  {{Gudmundsson}}, \citenamefont {{Jonsson}}, \citenamefont {{Bernodusson}},
  \citenamefont {{Abdullah}}, \citenamefont {{Sitek}}, \citenamefont {{Goan}},
  \citenamefont {{Tang}},\ and\ \citenamefont
  {{Manolescu}}}]{Gudmundsson16:AdP_10}%
  \BibitemOpen
  \bibfield  {author} {\bibinfo {author} {\bibfnamefont {V.}~\bibnamefont
  {{Gudmundsson}}}, \bibinfo {author} {\bibfnamefont {T.~H.}\ \bibnamefont
  {{Jonsson}}}, \bibinfo {author} {\bibfnamefont {M.~L.}\ \bibnamefont
  {{Bernodusson}}}, \bibinfo {author} {\bibfnamefont {N.~R.}\ \bibnamefont
  {{Abdullah}}}, \bibinfo {author} {\bibfnamefont {A.}~\bibnamefont {{Sitek}}},
  \bibinfo {author} {\bibfnamefont {H.-S.}\ \bibnamefont {{Goan}}}, \bibinfo
  {author} {\bibfnamefont {C.-S.}\ \bibnamefont {{Tang}}},\ and\ \bibinfo
  {author} {\bibfnamefont {A.}~\bibnamefont {{Manolescu}}},\ }\bibfield
  {title} {\bibinfo {title} {{Regimes of radiative and nonradiative transitions
  in transport through an electronic system in a photon cavity reaching a
  steady state}},\ }\href@noop {} {\bibfield  {journal} {\bibinfo  {journal}
  {Ann. Phys.}\ }\textbf {\bibinfo {volume} {529}},\ \bibinfo {pages} {1600177}
  (\bibinfo {year} {2016})}\BibitemShut {NoStop}%
\bibitem [{\citenamefont {{Hagenm{\"u}ller}}\ \emph {et~al.}(2017)\citenamefont
  {{Hagenm{\"u}ller}}, \citenamefont {{Schachenmayer}}, \citenamefont
  {{Sch{\"u}tz}}, \citenamefont {{Genes}},\ and\ \citenamefont
  {{Pupillo}}}]{2017arXiv170300803H}%
  \BibitemOpen
  \bibfield  {author} {\bibinfo {author} {\bibfnamefont {D.}~\bibnamefont
  {{Hagenm{\"u}ller}}}, \bibinfo {author} {\bibfnamefont {J.}~\bibnamefont
  {{Schachenmayer}}}, \bibinfo {author} {\bibfnamefont {S.}~\bibnamefont
  {{Sch{\"u}tz}}}, \bibinfo {author} {\bibfnamefont {C.}~\bibnamefont
  {{Genes}}},\ and\ \bibinfo {author} {\bibfnamefont {G.}~\bibnamefont
  {{Pupillo}}},\ }\bibfield  {title} {\bibinfo {title} {{Cavity-enhanced
  transport of charge}},\ }\href@noop {} {\bibfield  {journal} {\bibinfo
  {journal} {Phys. Rev. Lett.}\ }\textbf {\bibinfo {volume} {119}},\ \bibinfo
  {pages} {043502} (\bibinfo {year} {2017})},\ \Eprint
  {https://arxiv.org/abs/1703.00803} {arXiv:1703.00803 [quant-ph]} \BibitemShut
  {NoStop}%
\bibitem [{\citenamefont {Dinu}\ \emph {et~al.}(2018)\citenamefont {Dinu},
  \citenamefont {Moldoveanu},\ and\ \citenamefont
  {Gartner}}]{PhysRevB.97.195442}%
  \BibitemOpen
  \bibfield  {author} {\bibinfo {author} {\bibfnamefont {I.~V.}\ \bibnamefont
  {Dinu}}, \bibinfo {author} {\bibfnamefont {V.}~\bibnamefont {Moldoveanu}},\
  and\ \bibinfo {author} {\bibfnamefont {P.}~\bibnamefont {Gartner}},\
  }\bibfield  {title} {\bibinfo {title} {{Many-body effects in transport
  through a quantum-dot cavity system}},\ }\href
  {https://doi.org/10.1103/PhysRevB.97.195442} {\bibfield  {journal} {\bibinfo
  {journal} {Phys. Rev. B}\ }\textbf {\bibinfo {volume} {97}},\ \bibinfo
  {pages} {195442} (\bibinfo {year} {2018})}\BibitemShut {NoStop}%
\bibitem [{\citenamefont {Agarwalla}\ \emph {et~al.}(2016)\citenamefont
  {Agarwalla}, \citenamefont {Kulkarni}, \citenamefont {Mukamel},\ and\
  \citenamefont {Segal}}]{PhysRevB.94.035434}%
  \BibitemOpen
  \bibfield  {author} {\bibinfo {author} {\bibfnamefont {B.~K.}\ \bibnamefont
  {Agarwalla}}, \bibinfo {author} {\bibfnamefont {M.}~\bibnamefont {Kulkarni}},
  \bibinfo {author} {\bibfnamefont {S.}~\bibnamefont {Mukamel}},\ and\ \bibinfo
  {author} {\bibfnamefont {D.}~\bibnamefont {Segal}},\ }\bibfield  {title}
  {\bibinfo {title} {{Tunable photonic cavity coupled to a voltage-biased
  double quantum dot system: Diagrammatic nonequilibrium Green's function
  approach}},\ }\href {https://doi.org/10.1103/PhysRevB.94.035434} {\bibfield
  {journal} {\bibinfo  {journal} {Phys. Rev. B}\ }\textbf {\bibinfo {volume}
  {94}},\ \bibinfo {pages} {035434} (\bibinfo {year} {2016})}\BibitemShut
  {NoStop}%
\bibitem [{\citenamefont {Hagenm{\"u}ller}\ \emph {et~al.}(2017)\citenamefont
  {Hagenm{\"u}ller}, \citenamefont {Schachenmayer}, \citenamefont {Sch{\"u}tz},
  \citenamefont {Genes},\ and\ \citenamefont
  {Pupillo}}]{PhysRevLett.119.223601}%
  \BibitemOpen
  \bibfield  {author} {\bibinfo {author} {\bibfnamefont {D.}~\bibnamefont
  {Hagenm{\"u}ller}}, \bibinfo {author} {\bibfnamefont {J.}~\bibnamefont
  {Schachenmayer}}, \bibinfo {author} {\bibfnamefont {S.}~\bibnamefont
  {Sch{\"u}tz}}, \bibinfo {author} {\bibfnamefont {C.}~\bibnamefont {Genes}},\
  and\ \bibinfo {author} {\bibfnamefont {G.}~\bibnamefont {Pupillo}},\
  }\bibfield  {title} {\bibinfo {title} {{Cavity-Enhanced Transport of
  Charge}},\ }\href {https://doi.org/10.1103/PhysRevLett.119.223601} {\bibfield
   {journal} {\bibinfo  {journal} {Phys. Rev. Lett.}\ }\textbf {\bibinfo
  {volume} {119}},\ \bibinfo {pages} {223601} (\bibinfo {year}
  {2017})}\BibitemShut {NoStop}%
\bibitem [{\citenamefont {Lu}\ \emph {et~al.}(2019)\citenamefont {Lu},
  \citenamefont {Wang}, \citenamefont {Ren}, \citenamefont {Kulkarni},\ and\
  \citenamefont {Jiang}}]{PhysRevB.99.035129}%
  \BibitemOpen
  \bibfield  {author} {\bibinfo {author} {\bibfnamefont {J.}~\bibnamefont
  {Lu}}, \bibinfo {author} {\bibfnamefont {R.}~\bibnamefont {Wang}}, \bibinfo
  {author} {\bibfnamefont {J.}~\bibnamefont {Ren}}, \bibinfo {author}
  {\bibfnamefont {M.}~\bibnamefont {Kulkarni}},\ and\ \bibinfo {author}
  {\bibfnamefont {J.-H.}\ \bibnamefont {Jiang}},\ }\bibfield  {title} {\bibinfo
  {title} {{Quantum-dot circuit-QED thermoelectric diodes and transistors}},\
  }\href {https://doi.org/10.1103/PhysRevB.99.035129} {\bibfield  {journal}
  {\bibinfo  {journal} {Phys. Rev. B}\ }\textbf {\bibinfo {volume} {99}},\
  \bibinfo {pages} {035129} (\bibinfo {year} {2019})}\BibitemShut {NoStop}%
\bibitem [{\citenamefont {Kulkarni}\ \emph {et~al.}(2014)\citenamefont
  {Kulkarni}, \citenamefont {Cotlet},\ and\ \citenamefont
  {T{\"u}reci}}]{PhysRevB.90.125402}%
  \BibitemOpen
  \bibfield  {author} {\bibinfo {author} {\bibfnamefont {M.}~\bibnamefont
  {Kulkarni}}, \bibinfo {author} {\bibfnamefont {O.}~\bibnamefont {Cotlet}},\
  and\ \bibinfo {author} {\bibfnamefont {H.~E.}\ \bibnamefont {T{\"u}reci}},\
  }\bibfield  {title} {\bibinfo {title} {{Cavity-coupled double-quantum dot at
  finite bias: Analogy with lasers and beyond}},\ }\href
  {https://doi.org/10.1103/PhysRevB.90.125402} {\bibfield  {journal} {\bibinfo
  {journal} {Phys. Rev. B}\ }\textbf {\bibinfo {volume} {90}},\ \bibinfo
  {pages} {125402} (\bibinfo {year} {2014})}\BibitemShut {NoStop}%
\bibitem [{\citenamefont {van~den Berg}\ \emph {et~al.}(2014)\citenamefont
  {van~den Berg}, \citenamefont {Bergenfeldt},\ and\ \citenamefont
  {Samuelsson}}]{PhysRevB.90.085416}%
  \BibitemOpen
  \bibfield  {author} {\bibinfo {author} {\bibfnamefont {T.~L.}\ \bibnamefont
  {van~den Berg}}, \bibinfo {author} {\bibfnamefont {C.}~\bibnamefont
  {Bergenfeldt}},\ and\ \bibinfo {author} {\bibfnamefont {P.}~\bibnamefont
  {Samuelsson}},\ }\bibfield  {title} {\bibinfo {title} {{Pump-probe scheme for
  electron-photon dynamics in hybrid conductor-cavity systems}},\ }\href
  {https://doi.org/10.1103/PhysRevB.90.085416} {\bibfield  {journal} {\bibinfo
  {journal} {Phys. Rev. B}\ }\textbf {\bibinfo {volume} {90}},\ \bibinfo
  {pages} {085416} (\bibinfo {year} {2014})}\BibitemShut {NoStop}%
\bibitem [{\citenamefont {Moldoveanu}\ \emph {et~al.}(2019)\citenamefont
  {Moldoveanu}, \citenamefont {Manolescu},\ and\ \citenamefont
  {Gudmundsson}}]{Entropy19:731}%
  \BibitemOpen
  \bibfield  {author} {\bibinfo {author} {\bibfnamefont {V.}~\bibnamefont
  {Moldoveanu}}, \bibinfo {author} {\bibfnamefont {A.}~\bibnamefont
  {Manolescu}},\ and\ \bibinfo {author} {\bibfnamefont {V.}~\bibnamefont
  {Gudmundsson}},\ }\bibfield  {title} {\bibinfo {title} {{Generalized Master
  Equation Approach to Time-Dependent Many-Body Transport}},\ }\href
  {https://doi.org/10.3390/e21080731} {\bibfield  {journal} {\bibinfo
  {journal} {Entropy}\ }\textbf {\bibinfo {volume} {21}},\ \bibinfo {pages}
  {731} (\bibinfo {year} {2019})}\BibitemShut {NoStop}%
\bibitem [{\citenamefont {{S{\'a}nchez Mu{\~n}oz}}\ \emph
  {et~al.}(2018)\citenamefont {{S{\'a}nchez Mu{\~n}oz}}, \citenamefont {Nori},\
  and\ \citenamefont {{De Liberato}}}]{SanchezMunoz2018}%
  \BibitemOpen
  \bibfield  {author} {\bibinfo {author} {\bibfnamefont {C.}~\bibnamefont
  {{S{\'a}nchez Mu{\~n}oz}}}, \bibinfo {author} {\bibfnamefont
  {F.}~\bibnamefont {Nori}},\ and\ \bibinfo {author} {\bibfnamefont
  {S.}~\bibnamefont {{De Liberato}}},\ }\bibfield  {title} {\bibinfo {title}
  {{Resolution of superluminal signalling in non-perturbative cavity quantum
  electrodynamics}},\ }\href {https://doi.org/10.1038/s41467-018-04339-w}
  {\bibfield  {journal} {\bibinfo  {journal} {Nature Communications}\ }\textbf
  {\bibinfo {volume} {9}},\ \bibinfo {pages} {1924} (\bibinfo {year}
  {2018})}\BibitemShut {NoStop}%
\bibitem [{\citenamefont {Lopp}\ \emph {et~al.}(2018)\citenamefont {Lopp},
  \citenamefont {Mart{\'\i}n-Mart{\'\i}nez},\ and\ \citenamefont
  {Page}}]{Lopp_2018}%
  \BibitemOpen
  \bibfield  {author} {\bibinfo {author} {\bibfnamefont {R.}~\bibnamefont
  {Lopp}}, \bibinfo {author} {\bibfnamefont {E.}~\bibnamefont
  {Mart{\'\i}n-Mart{\'\i}nez}},\ and\ \bibinfo {author} {\bibfnamefont {D.~N.}\
  \bibnamefont {Page}},\ }\bibfield  {title} {\bibinfo {title} {{Relativity and
  quantum optics: accelerated atoms in optical cavities}},\ }\href
  {https://doi.org/10.1088/1361-6382/aae750} {\bibfield  {journal} {\bibinfo
  {journal} {Classical and Quantum Gravity}\ }\textbf {\bibinfo {volume}
  {35}},\ \bibinfo {pages} {224001} (\bibinfo {year} {2018})}\BibitemShut
  {NoStop}%
\bibitem [{\citenamefont {Lax}(1963)}]{PhysRev.129.2342}%
  \BibitemOpen
  \bibfield  {author} {\bibinfo {author} {\bibfnamefont {M.}~\bibnamefont
  {Lax}},\ }\bibfield  {title} {\bibinfo {title} {{Formal Theory of Quantum
  Fluctuations from a Driven State}},\ }\href
  {https://doi.org/10.1103/PhysRev.129.2342} {\bibfield  {journal} {\bibinfo
  {journal} {Phys. Rev.}\ }\textbf {\bibinfo {volume} {129}},\ \bibinfo {pages}
  {2342} (\bibinfo {year} {1963})}\BibitemShut {NoStop}%
\bibitem [{\citenamefont {Beaudoin}\ \emph {et~al.}(2011)\citenamefont
  {Beaudoin}, \citenamefont {Gambetta},\ and\ \citenamefont
  {Blais}}]{PhysRevA.84.043832}%
  \BibitemOpen
  \bibfield  {author} {\bibinfo {author} {\bibfnamefont {F.}~\bibnamefont
  {Beaudoin}}, \bibinfo {author} {\bibfnamefont {J.~M.}\ \bibnamefont
  {Gambetta}},\ and\ \bibinfo {author} {\bibfnamefont {A.}~\bibnamefont
  {Blais}},\ }\bibfield  {title} {\bibinfo {title} {{Dissipation and
  ultrastrong coupling in circuit QED}},\ }\href
  {https://doi.org/10.1103/PhysRevA.84.043832} {\bibfield  {journal} {\bibinfo
  {journal} {Phys. Rev. A}\ }\textbf {\bibinfo {volume} {84}},\ \bibinfo
  {pages} {043832} (\bibinfo {year} {2011})}\BibitemShut {NoStop}%
\bibitem [{\citenamefont {Scully}\ and\ \citenamefont
  {Zubairy}(1997)}]{Scully97}%
  \BibitemOpen
  \bibfield  {author} {\bibinfo {author} {\bibfnamefont {M.~O.}\ \bibnamefont
  {Scully}}\ and\ \bibinfo {author} {\bibfnamefont {M.~S.}\ \bibnamefont
  {Zubairy}},\ }\href@noop {} {\emph {\bibinfo {title} {{Quantum optics}}}}\
  (\bibinfo  {publisher} {Cambridge University Press},\ \bibinfo {year}
  {1997})\BibitemShut {NoStop}%
\bibitem [{\citenamefont {G.W.Ford}\ and\ \citenamefont
  {R.F.O'Connell}(1997)}]{Ford97:377}%
  \BibitemOpen
  \bibfield  {author} {\bibinfo {author} {\bibnamefont {G.W.Ford}}\ and\
  \bibinfo {author} {\bibnamefont {R.F.O'Connell}},\ }\bibfield  {title}
  {\bibinfo {title} {{The rotating wave approximation (RWA) of quantum optics:
  serious defect}},\ }\href {https://doi.org/10.1016/S0378-4371(97)00265-3}
  {\bibfield  {journal} {\bibinfo  {journal} {Physica A}\ }\textbf {\bibinfo
  {volume} {243}},\ \bibinfo {pages} {377} (\bibinfo {year}
  {1997})}\BibitemShut {NoStop}%
\bibitem [{\citenamefont {{Rza\ifmmode \dot{z}\else {\.z}\fi{}ewski}}\ \emph
  {et~al.}(1975)\citenamefont {{Rza\ifmmode \dot{z}\else {\.z}\fi{}ewski}},
  \citenamefont {W{\'o}dkiewicz},\ and\ \citenamefont {{\ifmmode \dot{Z}\else
  {\.Z}\fi{}akowicz}}}]{PhysRevLett.35.432}%
  \BibitemOpen
  \bibfield  {author} {\bibinfo {author} {\bibfnamefont {K.}~\bibnamefont
  {{Rza\ifmmode \dot{z}\else {\.z}\fi{}ewski}}}, \bibinfo {author}
  {\bibfnamefont {K.}~\bibnamefont {W{\'o}dkiewicz}},\ and\ \bibinfo {author}
  {\bibfnamefont {W.}~\bibnamefont {{\ifmmode \dot{Z}\else
  {\.Z}\fi{}akowicz}}},\ }\bibfield  {title} {\bibinfo {title} {{Phase
  Transitions, Two-Level Atoms, and the ${A}^{2}$ Term}},\ }\href
  {https://doi.org/10.1103/PhysRevLett.35.432} {\bibfield  {journal} {\bibinfo
  {journal} {Phys. Rev. Lett.}\ }\textbf {\bibinfo {volume} {35}},\ \bibinfo
  {pages} {432} (\bibinfo {year} {1975})}\BibitemShut {NoStop}%
\bibitem [{\citenamefont {{Gudmundsson}}\ \emph {et~al.}()\citenamefont
  {{Gudmundsson}}, \citenamefont {{Abdullah}}, \citenamefont {{Sitek}},
  \citenamefont {{Goan}}, \citenamefont {{Tang}},\ and\ \citenamefont
  {{Manolescu}}}]{doi:10.1002/andp.201700334}%
  \BibitemOpen
  \bibfield  {author} {\bibinfo {author} {\bibfnamefont {V.}~\bibnamefont
  {{Gudmundsson}}}, \bibinfo {author} {\bibfnamefont {N.~R.}\ \bibnamefont
  {{Abdullah}}}, \bibinfo {author} {\bibfnamefont {A.}~\bibnamefont {{Sitek}}},
  \bibinfo {author} {\bibfnamefont {H.-S.}\ \bibnamefont {{Goan}}}, \bibinfo
  {author} {\bibfnamefont {C.-S.}\ \bibnamefont {{Tang}}},\ and\ \bibinfo
  {author} {\bibfnamefont {A.}~\bibnamefont {{Manolescu}}},\ }\bibfield
  {title} {\bibinfo {title} {{Electroluminescence Caused by the Transport of
  Interacting Electrons through Parallel Quantum Dots in a Photon Cavity}},\
  }\href {https://doi.org/10.1002/andp.201700334} {\bibfield  {journal}
  {\bibinfo  {journal} {Annalen der Physik}\ }\textbf {\bibinfo {volume}
  {530}},\ \bibinfo {pages} {1700334}}\BibitemShut {NoStop}%
\bibitem [{\citenamefont {Nataf}\ and\ \citenamefont
  {Ciuti}(2010)}]{Nataf2010}%
  \BibitemOpen
  \bibfield  {author} {\bibinfo {author} {\bibfnamefont {P.}~\bibnamefont
  {Nataf}}\ and\ \bibinfo {author} {\bibfnamefont {C.}~\bibnamefont {Ciuti}},\
  }\bibfield  {title} {\bibinfo {title} {{No-go theorem for superradiant
  quantum phase transitions in cavity QED and counter-example in circuit
  QED}},\ }\href {https://doi.org/10.1038/ncomms1069} {\bibfield  {journal}
  {\bibinfo  {journal} {Nature Communications}\ }\textbf {\bibinfo {volume}
  {1}},\ \bibinfo {pages} {72} (\bibinfo {year} {2010})}\BibitemShut {NoStop}%
\bibitem [{\citenamefont {Shiga}\ \emph {et~al.}(2011)\citenamefont {Shiga},
  \citenamefont {Itano},\ and\ \citenamefont {Bollinger}}]{PhysRevA.84.012510}%
  \BibitemOpen
  \bibfield  {author} {\bibinfo {author} {\bibfnamefont {N.}~\bibnamefont
  {Shiga}}, \bibinfo {author} {\bibfnamefont {W.~M.}\ \bibnamefont {Itano}},\
  and\ \bibinfo {author} {\bibfnamefont {J.~J.}\ \bibnamefont {Bollinger}},\
  }\bibfield  {title} {\bibinfo {title} {{Diamagnetic correction to the
  ${}^{9}$Be${}^{+}$ ground-state hyperfine constant}},\ }\href
  {https://doi.org/10.1103/PhysRevA.84.012510} {\bibfield  {journal} {\bibinfo
  {journal} {Phys. Rev. A}\ }\textbf {\bibinfo {volume} {84}},\ \bibinfo
  {pages} {012510} (\bibinfo {year} {2011})}\BibitemShut {NoStop}%
\bibitem [{\citenamefont {Feranchuk}\ \emph {et~al.}(1996)\citenamefont
  {Feranchuk}, \citenamefont {Komarov},\ and\ \citenamefont
  {Ulyanenkov}}]{Feranchuk96:4035}%
  \BibitemOpen
  \bibfield  {author} {\bibinfo {author} {\bibfnamefont {I.~D.}\ \bibnamefont
  {Feranchuk}}, \bibinfo {author} {\bibfnamefont {L.~I.}\ \bibnamefont
  {Komarov}},\ and\ \bibinfo {author} {\bibfnamefont {A.~P.}\ \bibnamefont
  {Ulyanenkov}},\ }\bibfield  {title} {\bibinfo {title} {{Two-level system in a
  one-mode quantum field: numerical solution on the basis of the operator
  method}},\ }\href {http://iopscience.iop.org/0305-4470/29/14/026} {\bibfield
  {journal} {\bibinfo  {journal} {J. Phys. A: Math. Gen.}\ }\textbf {\bibinfo
  {volume} {29}},\ \bibinfo {pages} {4035} (\bibinfo {year}
  {1996})}\BibitemShut {NoStop}%
\bibitem [{\citenamefont {Wu}\ and\ \citenamefont
  {Yang}(2007)}]{PhysRevLett.98.013601}%
  \BibitemOpen
  \bibfield  {author} {\bibinfo {author} {\bibfnamefont {Y.}~\bibnamefont
  {Wu}}\ and\ \bibinfo {author} {\bibfnamefont {X.}~\bibnamefont {Yang}},\
  }\bibfield  {title} {\bibinfo {title} {{Strong-Coupling Theory of
  Periodically Driven Two-Level Systems}},\ }\href
  {https://doi.org/10.1103/PhysRevLett.98.013601} {\bibfield  {journal}
  {\bibinfo  {journal} {Phys. Rev. Lett.}\ }\textbf {\bibinfo {volume} {98}},\
  \bibinfo {pages} {013601} (\bibinfo {year} {2007})}\BibitemShut {NoStop}%
\bibitem [{\citenamefont {Gudmundsson}\ \emph
  {et~al.}(2019{\natexlab{a}})\citenamefont {Gudmundsson}, \citenamefont
  {Abdullah}, \citenamefont {Tang}, \citenamefont {Manolescu},\ and\
  \citenamefont {Moldoveanu}}]{doi:10.1002/andp.201900306}%
  \BibitemOpen
  \bibfield  {author} {\bibinfo {author} {\bibfnamefont {V.}~\bibnamefont
  {Gudmundsson}}, \bibinfo {author} {\bibfnamefont {N.~R.}\ \bibnamefont
  {Abdullah}}, \bibinfo {author} {\bibfnamefont {C.-S.}\ \bibnamefont {Tang}},
  \bibinfo {author} {\bibfnamefont {A.}~\bibnamefont {Manolescu}},\ and\
  \bibinfo {author} {\bibfnamefont {V.}~\bibnamefont {Moldoveanu}},\ }\bibfield
   {title} {\bibinfo {title} {{Cavity-Photon-Induced High-Order Transitions
  between Ground States of Quantum Dots}},\ }\href
  {https://doi.org/10.1002/andp.201900306} {\bibfield  {journal} {\bibinfo
  {journal} {Annalen der Physik}\ }\textbf {\bibinfo {volume} {531}},\ \bibinfo
  {pages} {1900306} (\bibinfo {year} {2019}{\natexlab{a}})}\BibitemShut
  {NoStop}%
\bibitem [{\citenamefont {Gudmundsson}\ \emph
  {et~al.}(2019{\natexlab{b}})\citenamefont {Gudmundsson}, \citenamefont
  {Gestsson}, \citenamefont {Abdullah}, \citenamefont {Tang}, \citenamefont
  {Manolescu},\ and\ \citenamefont {Moldoveanu}}]{Gudmundsson19:10}%
  \BibitemOpen
  \bibfield  {author} {\bibinfo {author} {\bibfnamefont {V.}~\bibnamefont
  {Gudmundsson}}, \bibinfo {author} {\bibfnamefont {H.}~\bibnamefont
  {Gestsson}}, \bibinfo {author} {\bibfnamefont {N.~R.}\ \bibnamefont
  {Abdullah}}, \bibinfo {author} {\bibfnamefont {C.-S.}\ \bibnamefont {Tang}},
  \bibinfo {author} {\bibfnamefont {A.}~\bibnamefont {Manolescu}},\ and\
  \bibinfo {author} {\bibfnamefont {V.}~\bibnamefont {Moldoveanu}},\ }\bibfield
   {title} {\bibinfo {title} {{Coexisting spin and Rabi oscillations at
  intermediate time regimes in electron transport through a photon cavity}},\
  }\href {https://doi.org/10.3762/bjnano.10.61} {\bibfield  {journal} {\bibinfo
   {journal} {Beilstein J. Nanotechnol.}\ }\textbf {\bibinfo {volume} {10}},\
  \bibinfo {pages} {606} (\bibinfo {year} {2019}{\natexlab{b}})}\BibitemShut
  {NoStop}%
\bibitem [{\citenamefont {Yan}\ \emph {et~al.}(2017)\citenamefont {Yan},
  \citenamefont {L{\"u}}, \citenamefont {Luo},\ and\ \citenamefont
  {Zheng}}]{PhysRevA.96.033802}%
  \BibitemOpen
  \bibfield  {author} {\bibinfo {author} {\bibfnamefont {Y.}~\bibnamefont
  {Yan}}, \bibinfo {author} {\bibfnamefont {Z.}~\bibnamefont {L{\"u}}},
  \bibinfo {author} {\bibfnamefont {J.}~\bibnamefont {Luo}},\ and\ \bibinfo
  {author} {\bibfnamefont {H.}~\bibnamefont {Zheng}},\ }\bibfield  {title}
  {\bibinfo {title} {{Effects of counter-rotating couplings of the Rabi model
  with frequency modulation}},\ }\href
  {https://doi.org/10.1103/PhysRevA.96.033802} {\bibfield  {journal} {\bibinfo
  {journal} {Phys. Rev. A}\ }\textbf {\bibinfo {volume} {96}},\ \bibinfo
  {pages} {033802} (\bibinfo {year} {2017})}\BibitemShut {NoStop}%
\bibitem [{\citenamefont {Wang}\ \emph {et~al.}(2017)\citenamefont {Wang},
  \citenamefont {Miranowicz}, \citenamefont {Li},\ and\ \citenamefont
  {Nori}}]{PhysRevA.96.063820}%
  \BibitemOpen
  \bibfield  {author} {\bibinfo {author} {\bibfnamefont {X.}~\bibnamefont
  {Wang}}, \bibinfo {author} {\bibfnamefont {A.}~\bibnamefont {Miranowicz}},
  \bibinfo {author} {\bibfnamefont {H.-R.}\ \bibnamefont {Li}},\ and\ \bibinfo
  {author} {\bibfnamefont {F.}~\bibnamefont {Nori}},\ }\bibfield  {title}
  {\bibinfo {title} {{Observing pure effects of counter-rotating terms without
  ultrastrong coupling: A single photon can simultaneously excite two
  qubits}},\ }\href {https://doi.org/10.1103/PhysRevA.96.063820} {\bibfield
  {journal} {\bibinfo  {journal} {Phys. Rev. A}\ }\textbf {\bibinfo {volume}
  {96}},\ \bibinfo {pages} {063820} (\bibinfo {year} {2017})}\BibitemShut
  {NoStop}%
\bibitem [{\citenamefont {Forn-D{\'\i}az}\ \emph {et~al.}(2019)\citenamefont
  {Forn-D{\'\i}az}, \citenamefont {Lamata}, \citenamefont {Rico}, \citenamefont
  {Kono},\ and\ \citenamefont {Solano}}]{RevModPhys.91.025005}%
  \BibitemOpen
  \bibfield  {author} {\bibinfo {author} {\bibfnamefont {P.}~\bibnamefont
  {Forn-D{\'\i}az}}, \bibinfo {author} {\bibfnamefont {L.}~\bibnamefont
  {Lamata}}, \bibinfo {author} {\bibfnamefont {E.}~\bibnamefont {Rico}},
  \bibinfo {author} {\bibfnamefont {J.}~\bibnamefont {Kono}},\ and\ \bibinfo
  {author} {\bibfnamefont {E.}~\bibnamefont {Solano}},\ }\bibfield  {title}
  {\bibinfo {title} {{Ultrastrong coupling regimes of light-matter
  interaction}},\ }\href {https://doi.org/10.1103/RevModPhys.91.025005}
  {\bibfield  {journal} {\bibinfo  {journal} {Rev. Mod. Phys.}\ }\textbf
  {\bibinfo {volume} {91}},\ \bibinfo {pages} {025005} (\bibinfo {year}
  {2019})}\BibitemShut {NoStop}%
\bibitem [{\citenamefont {Gudmundsson}\ \emph {et~al.}(2009)\citenamefont
  {Gudmundsson}, \citenamefont {Gainar}, \citenamefont {Tang}, \citenamefont
  {Moldoveanu},\ and\ \citenamefont {Manolescu}}]{Gudmundsson09:113007}%
  \BibitemOpen
  \bibfield  {author} {\bibinfo {author} {\bibfnamefont {V.}~\bibnamefont
  {Gudmundsson}}, \bibinfo {author} {\bibfnamefont {C.}~\bibnamefont {Gainar}},
  \bibinfo {author} {\bibfnamefont {C.-S.}\ \bibnamefont {Tang}}, \bibinfo
  {author} {\bibfnamefont {V.}~\bibnamefont {Moldoveanu}},\ and\ \bibinfo
  {author} {\bibfnamefont {A.}~\bibnamefont {Manolescu}},\ }\bibfield  {title}
  {\bibinfo {title} {{Time-dependent transport via the generalized master
  equation through a finite quantum wire with an embedded subsystem}},\ }\href
  {http://stacks.iop.org/1367-2630/11/113007} {\bibfield  {journal} {\bibinfo
  {journal} {New Journal of Physics}\ }\textbf {\bibinfo {volume} {11}},\
  \bibinfo {pages} {113007} (\bibinfo {year} {2009})}\BibitemShut {NoStop}%
\bibitem [{\citenamefont {Moldoveanu}\ \emph {et~al.}(2009)\citenamefont
  {Moldoveanu}, \citenamefont {Manolescu},\ and\ \citenamefont
  {Gudmundsson}}]{Moldoveanu09:073019}%
  \BibitemOpen
  \bibfield  {author} {\bibinfo {author} {\bibfnamefont {V.}~\bibnamefont
  {Moldoveanu}}, \bibinfo {author} {\bibfnamefont {A.}~\bibnamefont
  {Manolescu}},\ and\ \bibinfo {author} {\bibfnamefont {V.}~\bibnamefont
  {Gudmundsson}},\ }\bibfield  {title} {\bibinfo {title} {{Geometrical effects
  and signal delay in time-dependent transport at the nanoscale}},\ }\href
  {http://stacks.iop.org/1367-2630/11/073019} {\bibfield  {journal} {\bibinfo
  {journal} {New Journal of Physics}\ }\textbf {\bibinfo {volume} {11}},\
  \bibinfo {pages} {073019} (\bibinfo {year} {2009})}\BibitemShut {NoStop}%
\bibitem [{\citenamefont {Gudmundsson}\ \emph {et~al.}(2012)\citenamefont
  {Gudmundsson}, \citenamefont {Jonasson}, \citenamefont {Tang}, \citenamefont
  {Goan},\ and\ \citenamefont {Manolescu}}]{Gudmundsson12:1109.4728}%
  \BibitemOpen
  \bibfield  {author} {\bibinfo {author} {\bibfnamefont {V.}~\bibnamefont
  {Gudmundsson}}, \bibinfo {author} {\bibfnamefont {O.}~\bibnamefont
  {Jonasson}}, \bibinfo {author} {\bibfnamefont {C.-S.}\ \bibnamefont {Tang}},
  \bibinfo {author} {\bibfnamefont {H.-S.}\ \bibnamefont {Goan}},\ and\
  \bibinfo {author} {\bibfnamefont {A.}~\bibnamefont {Manolescu}},\ }\bibfield
  {title} {\bibinfo {title} {{Time-dependent transport of electrons through a
  photon cavity}},\ }\href {https://doi.org/10.1103/PhysRevB.85.075306}
  {\bibfield  {journal} {\bibinfo  {journal} {Phys. Rev. B}\ }\textbf {\bibinfo
  {volume} {85}},\ \bibinfo {pages} {075306} (\bibinfo {year}
  {2012})}\BibitemShut {NoStop}%
\bibitem [{\citenamefont {Weidlich}(1971)}]{Weidlich71:325}%
  \BibitemOpen
  \bibfield  {author} {\bibinfo {author} {\bibfnamefont {W.}~\bibnamefont
  {Weidlich}},\ }\bibfield  {title} {\bibinfo {title} {{Liouville-space
  formalism for quantum systems in contact with reservoirs}},\ }\href
  {http://link.springer.com/article/10.1007%2FBF01395429} {\bibfield  {journal}
  {\bibinfo  {journal} {Zeitschrift f{\"u}r Physik}\ }\textbf {\bibinfo
  {volume} {241}},\ \bibinfo {pages} {325} (\bibinfo {year}
  {1971})}\BibitemShut {NoStop}%
\bibitem [{\citenamefont {Nakano}\ \emph {et~al.}(2010)\citenamefont {Nakano},
  \citenamefont {Hatano},\ and\ \citenamefont {Petrosky}}]{Nakano2010}%
  \BibitemOpen
  \bibfield  {author} {\bibinfo {author} {\bibfnamefont {R.}~\bibnamefont
  {Nakano}}, \bibinfo {author} {\bibfnamefont {N.}~\bibnamefont {Hatano}},\
  and\ \bibinfo {author} {\bibfnamefont {T.}~\bibnamefont {Petrosky}},\
  }\bibfield  {title} {\bibinfo {title} {{Nontrivial Eigenvalues of the
  Liouvillian of an Open Quantum System}},\ }\href
  {https://doi.org/10.1007/s10773-010-0606-9} {\bibfield  {journal} {\bibinfo
  {journal} {International Journal of Theoretical Physics}\ }\textbf {\bibinfo
  {volume} {50}},\ \bibinfo {pages} {1134} (\bibinfo {year}
  {2010})}\BibitemShut {NoStop}%
\bibitem [{\citenamefont {Petrosky}(2010)}]{Petrosky01032010}%
  \BibitemOpen
  \bibfield  {author} {\bibinfo {author} {\bibfnamefont {T.}~\bibnamefont
  {Petrosky}},\ }\bibfield  {title} {\bibinfo {title} {{Complex Spectral
  Representation of the Liouvillian and Kinetic Theory in Nonequilibrium
  Physics}},\ }\href {https://doi.org/10.1143/PTP.123.395} {\bibfield
  {journal} {\bibinfo  {journal} {Progress of Theoretical Physics}\ }\textbf
  {\bibinfo {volume} {123}},\ \bibinfo {pages} {395} (\bibinfo {year}
  {2010})}\BibitemShut {NoStop}%
\bibitem [{\citenamefont {Zwanzig}(1960)}]{Zwanzig60:1338}%
  \BibitemOpen
  \bibfield  {author} {\bibinfo {author} {\bibfnamefont {R.}~\bibnamefont
  {Zwanzig}},\ }\bibfield  {title} {\bibinfo {title} {{Ensemble Method in the
  Theory of Irreversibility}},\ }\href@noop {} {\bibfield  {journal} {\bibinfo
  {journal} {J. Chem. Phys.}\ }\textbf {\bibinfo {volume} {33}},\ \bibinfo
  {pages} {1338} (\bibinfo {year} {1960})}\BibitemShut {NoStop}%
\bibitem [{\citenamefont {Nakajima}(1958)}]{Nakajima58:948}%
  \BibitemOpen
  \bibfield  {author} {\bibinfo {author} {\bibfnamefont {S.}~\bibnamefont
  {Nakajima}},\ }\bibfield  {title} {\bibinfo {title} {{On Quantum Theory of
  Transport Phenomena Steady Diffusion}},\ }\href@noop {} {\bibfield  {journal}
  {\bibinfo  {journal} {Prog. Theor. Phys.}\ }\textbf {\bibinfo {volume}
  {20}},\ \bibinfo {pages} {948} (\bibinfo {year} {1958})}\BibitemShut
  {NoStop}%
\bibitem [{\citenamefont {Petersen}\ and\ \citenamefont
  {Pedersen}(2012)}]{IMM2012-03274}%
  \BibitemOpen
  \bibfield  {author} {\bibinfo {author} {\bibfnamefont {K.~B.}\ \bibnamefont
  {Petersen}}\ and\ \bibinfo {author} {\bibfnamefont {M.~S.}\ \bibnamefont
  {Pedersen}},\ }\href {http://localhost/pubdb/p.php?3274} {\bibinfo {title}
  {{The Matrix Cookbook}}} (\bibinfo {year} {2012}),\ \bibinfo {note} {version
  20121115}\BibitemShut {NoStop}%
\bibitem [{\citenamefont {Jonsson}\ \emph {et~al.}(2017)\citenamefont
  {Jonsson}, \citenamefont {Manolescu}, \citenamefont {Goan}, \citenamefont
  {Abdullah}, \citenamefont {Sitek}, \citenamefont {Tang},\ and\ \citenamefont
  {Gudmundsson}}]{2016arXiv161003223J}%
  \BibitemOpen
  \bibfield  {author} {\bibinfo {author} {\bibfnamefont {T.~H.}\ \bibnamefont
  {Jonsson}}, \bibinfo {author} {\bibfnamefont {A.}~\bibnamefont {Manolescu}},
  \bibinfo {author} {\bibfnamefont {H.-S.}\ \bibnamefont {Goan}}, \bibinfo
  {author} {\bibfnamefont {N.~R.}\ \bibnamefont {Abdullah}}, \bibinfo {author}
  {\bibfnamefont {A.}~\bibnamefont {Sitek}}, \bibinfo {author} {\bibfnamefont
  {C.-S.}\ \bibnamefont {Tang}},\ and\ \bibinfo {author} {\bibfnamefont
  {V.}~\bibnamefont {Gudmundsson}},\ }\bibfield  {title} {\bibinfo {title}
  {{Efficient determination of the Markovian time-evolution towards a
  steady-state of a complex open quantum system}},\ }\href
  {https://doi.org/10.1016/j.cpc.2017.06.018} {\bibfield  {journal} {\bibinfo
  {journal} {Computer Physics Communications}\ }\textbf {\bibinfo {volume}
  {220}},\ \bibinfo {pages} {81} (\bibinfo {year} {2017})}\BibitemShut
  {NoStop}%
\bibitem [{\citenamefont {Gudmundsson}\ \emph {et~al.}(2018)\citenamefont
  {Gudmundsson}, \citenamefont {Abdullah}, \citenamefont {Sitek}, \citenamefont
  {Goan}, \citenamefont {Tang},\ and\ \citenamefont
  {Manolescu}}]{GUDMUNDSSON20181672}%
  \BibitemOpen
  \bibfield  {author} {\bibinfo {author} {\bibfnamefont {V.}~\bibnamefont
  {Gudmundsson}}, \bibinfo {author} {\bibfnamefont {N.~R.}\ \bibnamefont
  {Abdullah}}, \bibinfo {author} {\bibfnamefont {A.}~\bibnamefont {Sitek}},
  \bibinfo {author} {\bibfnamefont {H.-S.}\ \bibnamefont {Goan}}, \bibinfo
  {author} {\bibfnamefont {C.-S.}\ \bibnamefont {Tang}},\ and\ \bibinfo
  {author} {\bibfnamefont {A.}~\bibnamefont {Manolescu}},\ }\bibfield  {title}
  {\bibinfo {title} {{Current correlations for the transport of interacting
  electrons through parallel quantum dots in a photon cavity}},\ }\href
  {https://doi.org/10.1016/j.physleta.2018.04.017} {\bibfield  {journal}
  {\bibinfo  {journal} {Physics Letters A}\ }\textbf {\bibinfo {volume}
  {382}},\ \bibinfo {pages} {1672} (\bibinfo {year} {2018})}\BibitemShut
  {NoStop}%
\bibitem [{\citenamefont {Santos}\ \emph {et~al.}(2019)\citenamefont {Santos},
  \citenamefont {C{\'e}leri}, \citenamefont {Landi},\ and\ \citenamefont
  {Paternostro}}]{2017arXiv170708946S}%
  \BibitemOpen
  \bibfield  {author} {\bibinfo {author} {\bibfnamefont {J.~P.}\ \bibnamefont
  {Santos}}, \bibinfo {author} {\bibfnamefont {L.~C.}\ \bibnamefont
  {C{\'e}leri}}, \bibinfo {author} {\bibfnamefont {G.~T.}\ \bibnamefont
  {Landi}},\ and\ \bibinfo {author} {\bibfnamefont {M.}~\bibnamefont
  {Paternostro}},\ }\bibfield  {title} {\bibinfo {title} {{The role of quantum
  coherence in non-equilibrium entropy production}},\ }\href
  {https://doi.org/10.1038/s41534-019-0138-y} {\bibfield  {journal} {\bibinfo
  {journal} {npj Quantum Information}\ }\textbf {\bibinfo {volume} {5}},\
  \bibinfo {pages} {23} (\bibinfo {year} {2019})}\BibitemShut {NoStop}%
\bibitem [{\citenamefont {{Batalhao}}\ \emph {et~al.}(2018)\citenamefont
  {{Batalhao}}, \citenamefont {{Gherardini}}, \citenamefont {{Santos}},
  \citenamefont {{Landi}},\ and\ \citenamefont
  {{Paternostro}}}]{2018arXiv180608441B}%
  \BibitemOpen
  \bibfield  {author} {\bibinfo {author} {\bibfnamefont {T.~B.}\ \bibnamefont
  {{Batalhao}}}, \bibinfo {author} {\bibfnamefont {S.}~\bibnamefont
  {{Gherardini}}}, \bibinfo {author} {\bibfnamefont {J.~P.}\ \bibnamefont
  {{Santos}}}, \bibinfo {author} {\bibfnamefont {G.~T.}\ \bibnamefont
  {{Landi}}},\ and\ \bibinfo {author} {\bibfnamefont {M.}~\bibnamefont
  {{Paternostro}}},\ }\bibfield  {title} {\bibinfo {title} {{Characterizing
  irreversibility in open quantum systems}},\ }\href@noop {} {\bibfield
  {journal} {\bibinfo  {journal} {ArXiv e-prints}\ } (\bibinfo {year}
  {2018})},\ \Eprint {https://arxiv.org/abs/1806.08441} {arXiv:1806.08441
  [quant-ph]} \BibitemShut {NoStop}%
\bibitem [{\citenamefont {{Baez}}(2011)}]{2011arXiv1102.2098B}%
  \BibitemOpen
  \bibfield  {author} {\bibinfo {author} {\bibfnamefont {J.~C.}\ \bibnamefont
  {{Baez}}},\ }\bibfield  {title} {\bibinfo {title} {{Renyi Entropy and Free
  Energy}},\ }\href@noop {} {\bibfield  {journal} {\bibinfo  {journal} {ArXiv
  e-prints}\ } (\bibinfo {year} {2011})},\ \Eprint
  {https://arxiv.org/abs/1102.2098} {arXiv:1102.2098 [quant-ph]} \BibitemShut
  {NoStop}%
\bibitem [{\citenamefont {Fock}(1928)}]{Fock28:446}%
  \BibitemOpen
  \bibfield  {author} {\bibinfo {author} {\bibfnamefont {V.}~\bibnamefont
  {Fock}},\ }\bibfield  {title} {\bibinfo {title} {{Bemerkung zur Quantelung
  des harmonischen Oszillators im Magnetfeld}},\ }\href@noop {} {\bibfield
  {journal} {\bibinfo  {journal} {Z. Phys.}\ }\textbf {\bibinfo {volume}
  {47}},\ \bibinfo {pages} {446} (\bibinfo {year} {1928})}\BibitemShut
  {NoStop}%
\bibitem [{\citenamefont {Gudmundsson}\ \emph {et~al.}(2013)\citenamefont
  {Gudmundsson}, \citenamefont {Jonasson}, \citenamefont {Arnold},
  \citenamefont {Tang}, \citenamefont {Goan},\ and\ \citenamefont
  {Manolescu}}]{Gudmundsson:2013.305}%
  \BibitemOpen
  \bibfield  {author} {\bibinfo {author} {\bibfnamefont {V.}~\bibnamefont
  {Gudmundsson}}, \bibinfo {author} {\bibfnamefont {O.}~\bibnamefont
  {Jonasson}}, \bibinfo {author} {\bibfnamefont {T.}~\bibnamefont {Arnold}},
  \bibinfo {author} {\bibfnamefont {C.-S.}\ \bibnamefont {Tang}}, \bibinfo
  {author} {\bibfnamefont {H.-S.}\ \bibnamefont {Goan}},\ and\ \bibinfo
  {author} {\bibfnamefont {A.}~\bibnamefont {Manolescu}},\ }\bibfield  {title}
  {\bibinfo {title} {{Stepwise introduction of model complexity in a
  generalized master equation approach to time-dependent transport}},\ }\href
  {https://doi.org/10.1002/prop.201200053} {\bibfield  {journal} {\bibinfo
  {journal} {Fortschritte der Physik}\ }\textbf {\bibinfo {volume} {61}},\
  \bibinfo {pages} {305} (\bibinfo {year} {2013})}\BibitemShut {NoStop}%
\bibitem [{\citenamefont {Gardiner}\ and\ \citenamefont
  {Collett}(1985)}]{PhysRevA.31.3761}%
  \BibitemOpen
  \bibfield  {author} {\bibinfo {author} {\bibfnamefont {C.~W.}\ \bibnamefont
  {Gardiner}}\ and\ \bibinfo {author} {\bibfnamefont {M.~J.}\ \bibnamefont
  {Collett}},\ }\bibfield  {title} {\bibinfo {title} {{Input and output in
  damped quantum systems: Quantum stochastic differential equations and the
  master equation}},\ }\href {https://doi.org/10.1103/PhysRevA.31.3761}
  {\bibfield  {journal} {\bibinfo  {journal} {Phys. Rev. A}\ }\textbf {\bibinfo
  {volume} {31}},\ \bibinfo {pages} {3761} (\bibinfo {year}
  {1985})}\BibitemShut {NoStop}%
\bibitem [{\citenamefont {{De Liberato}}\ \emph {et~al.}(2009)\citenamefont
  {{De Liberato}}, \citenamefont {Gerace}, \citenamefont {Carusotto},\ and\
  \citenamefont {Ciuti}}]{PhysRevA.80.053810}%
  \BibitemOpen
  \bibfield  {author} {\bibinfo {author} {\bibfnamefont {S.}~\bibnamefont {{De
  Liberato}}}, \bibinfo {author} {\bibfnamefont {D.}~\bibnamefont {Gerace}},
  \bibinfo {author} {\bibfnamefont {I.}~\bibnamefont {Carusotto}},\ and\
  \bibinfo {author} {\bibfnamefont {C.}~\bibnamefont {Ciuti}},\ }\bibfield
  {title} {\bibinfo {title} {{Extracavity quantum vacuum radiation from a
  single qubit}},\ }\href {https://doi.org/10.1103/PhysRevA.80.053810}
  {\bibfield  {journal} {\bibinfo  {journal} {Phys. Rev. A}\ }\textbf {\bibinfo
  {volume} {80}},\ \bibinfo {pages} {053810} (\bibinfo {year}
  {2009})}\BibitemShut {NoStop}%
\bibitem [{\citenamefont {Scala}\ \emph {et~al.}(2007)\citenamefont {Scala},
  \citenamefont {Militello}, \citenamefont {Messina}, \citenamefont {Piilo},\
  and\ \citenamefont {Maniscalco}}]{PhysRevA.75.013811}%
  \BibitemOpen
  \bibfield  {author} {\bibinfo {author} {\bibfnamefont {M.}~\bibnamefont
  {Scala}}, \bibinfo {author} {\bibfnamefont {B.}~\bibnamefont {Militello}},
  \bibinfo {author} {\bibfnamefont {A.}~\bibnamefont {Messina}}, \bibinfo
  {author} {\bibfnamefont {J.}~\bibnamefont {Piilo}},\ and\ \bibinfo {author}
  {\bibfnamefont {S.}~\bibnamefont {Maniscalco}},\ }\bibfield  {title}
  {\bibinfo {title} {{Microscopic derivation of the Jaynes-Cummings model with
  cavity losses}},\ }\href {https://doi.org/10.1103/PhysRevA.75.013811}
  {\bibfield  {journal} {\bibinfo  {journal} {Phys. Rev. A}\ }\textbf {\bibinfo
  {volume} {75}},\ \bibinfo {pages} {013811} (\bibinfo {year}
  {2007})}\BibitemShut {NoStop}%
\end{thebibliography}
%
%
%
%
\end{document}